%% file: ms.tex
\documentclass[onecolumn,iop]{emulateapj}  

\usepackage{graphicx, natbib, color, bm, hyperref, breakurl}

\input{local-commands}

\begin{document}

\title{Enzo: An Adaptive Mesh Refinement Code for Astrophysics}
\author{The Enzo Collaboration: Greg L. Bryan\altaffilmark{1}}
\author{Michael L. Norman\altaffilmark{2,3}}
\author{Brian W. O'Shea\altaffilmark{4,5}}
\author{Tom Abel\altaffilmark{6, 20}}
\author{John H. Wise\altaffilmark{7}}
\author{Matthew J. Turk\altaffilmark{1}}
\author{Daniel R. Reynolds\altaffilmark{8}}
\author{David C. Collins\altaffilmark{9}}
\author{Peng Wang\altaffilmark{6}}
\author{Samuel W. Skillman\altaffilmark{10,11}}
\author{Britton Smith\altaffilmark{4}}
\author{Robert P. Harkness\altaffilmark{12}}
\author{James Bordner\altaffilmark{2}}
\author{Ji-hoon Kim\altaffilmark{13}}
\author{Michael Kuhlen\altaffilmark{14,15}}
\author{Hao Xu\altaffilmark{2}}
\author{Nathan Goldbaum\altaffilmark{15}}
\author{Cameron Hummels\altaffilmark{16}}
\author{Alexei G. Kritsuk\altaffilmark{2}}
\author{Elizabeth Tasker\altaffilmark{17}}
\author{Stephen Skory\altaffilmark{10}}
\author{Christine M. Simpson\altaffilmark{1}}
\author{Oliver Hahn\altaffilmark{18}}
\author{Jeffrey S. Oishi\altaffilmark{19}}
\author{Geoffrey C So\altaffilmark{2}}
\author{Fen Zhao\altaffilmark{20}}
\author{Renyue Cen\altaffilmark{21}}
\author{Yuan Li\altaffilmark{1}}

\altaffiltext{1}{Columbia University, Department of Astronomy, New York, NY, 10025, USA}
\altaffiltext{2}{CASS, University of California, San Diego, 9500 Gilman Drive La Jolla, CA 92093-0424, USA}
\altaffiltext{3}{SDSC, University of California, San Diego, 9500 Gilman Drive La Jolla, CA 92093-0505, USA}
\altaffiltext{4}{Department of Physics and Astronomy, Michigan State University, East Lansing, MI 48824, USA}
\altaffiltext{5}{Lyman Briggs College and Institute for Cyber-Enabled Research, Michigan State University, East Lansing, MI 48824, USA}
\altaffiltext{6}{Kavli Institute for Particle Astrophysics and Cosmology, Stanford University, Menlo Park, CA 94025, USA}
\altaffiltext{7}{Center for Relativistic Astrophysics, School of Physics, Georgia Institute of Technology, 837 State Street, Atlanta, GA, USA}
\altaffiltext{8}{Department of Mathematics, Southern Methodist University, Box 750156, Dallas, TX 75205-0156, USA}
\altaffiltext{9}{Theoretical Division, Los Alamos National Laboratory, Los Alamos, NM 87544}
\altaffiltext{10}{Center for Astrophysics and Space Astronomy, Department of Astrophysical and Planetary Science, University of Colorado, Boulder, CO 80309, USA}
\altaffiltext{11}{DOE Computational Science Graduate Fellow}
\altaffiltext{12}{NICS, Oak Ridge National Laboratory, 1 Bethel Valley Rd, Oak Ridge, TN 37831, USA}
\altaffiltext{13}{Department of Astronomy and Astrophysics, University of California, Santa Cruz, CA 95064, USA}
\altaffiltext{14}{Theoretical Astrophysics Center, University of California Berkeley, Hearst Field Annex, Berkeley, CA 94720, USA}
\altaffiltext{15}{Department of Astronomy and Astrophysics, University of California, 1156 High Street, Santa Cruz, CA 95064, USA}
\altaffiltext{16}{Department of Astronomy/Steward Observatory, University of Arizona, 933 N. Cherry Ave., Tucson, AZ, 85721, USA}
\altaffiltext{17}{Physics Department, Faculty of Science, Hokkaido University, Kita-10 Nishi 8, Kita-ku, Sapporo 060-0810, Japan}
\altaffiltext{18}{ETH Zurich Institute for Astronomy. CH-8093 Zurich. Switzerland}
\altaffiltext{19}{Department of Astrophysics, American Museum of Natural History, Central Park West at 81st St, New York, NY 10024}
\altaffiltext{20}{Department of Physics, Stanford University, Stanford, CA 94305, USA}
\altaffiltext{21}{Department of Astrophysical Sciences, Princeton University, Peyton Hall, Princeton, NJ 08544, USA}

\begin{abstract}
This paper describes the open-source code \enzo, which uses
block-structured adaptive mesh refinement to provide high spatial and
temporal resolution for modeling astrophysical fluid flows.  The code
is Cartesian, can be run in 1, 2, and 3 dimensions, and supports a
wide variety of physics including hydrodynamics, ideal and non-ideal
magnetohydrodynamics, N-body dynamics (and, more broadly, self-gravity
of fluids and particles), primordial gas chemistry, optically-thin
radiative cooling of primordial and metal-enriched plasmas (as well as
some optically-thick cooling models), radiation transport,
cosmological expansion, and models for star formation and feedback in
a cosmological context.  In addition to explaining the algorithms
implemented, we present solutions for a wide range of test problems,
demonstrate the code's parallel performance, and discuss the \enzo\
collaboration's code development methodology.
\end{abstract}


\maketitle

\input{introduction}


\input{overview}


\input{numerical-amr}

\section{Fluid methods}
\label{sec.fluids}

\input{numerical-ppm}

\input{numerical-mhd-dedner}  
\input{numerical-mhd-ct}  
\input{numerical-zeus}

\section{Gravity and N-body}
\label{sec.allgrav}
\input{numerical-gravity}
\input{numerical-particle}

\section{Microphysics}
\label{sec.microphysics}
\input{numerical-chemistry}
\input{numerical-cooling}

\section{Radiation}
\label{sec.allrad}
\input{numerical-radiation-homogeneous} 
\input{numerical-radiation-raytracing}

\input{numerical-radiation-fld}

\section{Other physics}
\label{sec.otherphys}
\input{numerical-conduction} 
\input{numerical-starformation}

\input{numerical-timestep}

\input{numerical-analysis}

\input{code-tests}

\input{parallel}

\input{development}

\input{conclusions}

\input{acknowledgments}

\input{appendix}

\bibliographystyle{apj}
\bibliography{apj-jour,ms}  

\end{document}

%% file: local-commands.tex
\newcommand{\enzo}{{\small Enzo}}
\newcommand{\moray}{{\small Enzo+Moray}}
\newcommand{\zeus}{{\small ZEUS}}

\renewcommand{\div}{\bm{\nabla} \cdot}
\newcommand{\grad}{\bm{\nabla}}
\newcommand{\curl}{\bm{\nabla} \times}

\def\etal{{\sl et al.}}

\def\k#1 {k_{{\rm #1}}}
\def\HH{H$_2$}
\def\HHp{H$_2^+$}
\def\Hp{H$^+$}
\def\Hm{H$^-$}
\def\Hep{He$^+$}
\def\Hepp{He$^{++}$}
\def\Dp{D$^+$}

\def\mH2p{H_2^+}

\newcommand{\myvec}[1]{\mathbf{#1}}
\newcommand{\rhop}{\rho^\prime}
\newcommand{\vecvp}{\myvec{v}^\prime}
\newcommand{\vecv}{\myvec{v}}
\newcommand{\vecB}{\myvec{B}}
\newcommand{\vecx}{\myvec{x}}
\newcommand{\Ep}{E^\prime}

\newcommand{\phip}{\phi^\prime}
\newcommand{\pp}{p^\prime}

\newcommand{\Mach}{\mathcal{M}}
\newcommand{\fcond}{\myvec{F}_{\rm cond}}
\newcommand{\minmod}[2]{{\rm minmod} \left({#1},{#2} \right) }

\newcommand{\cubecm}{\ifmmode{{\rm cm^{-3}}}\else{cm$^{-3}$}\fi}
\newcommand\unit[1]{\; \textrm{#1}}
\newcommand{\mh}{m_{\rm H}}

\newcommand\abs[1]{\left|#1\right|}

%% file: introduction.tex
\section{Introduction}\label{sec.intro}

Due to the high spatial and temporal dynamical ranges involved,
astrophysical and cosmological phenomena present a taxing challenge
for simulators. To tackle such situations, a number of numerical
techniques have been developed that can be broadly split into
gridless, Lagrangian methods and grid-based, Eulerian schemes. The
most commonly used is an example of the first type known as Smoothed
Particle Hydrodynamics \citep[SPH;][]{Lucy77, SPH}. It has achieved
much success, particularly in regimes dominated by gravity. However,
its development to include an increasing number of sought-after
physical processes is still at a relatively early stage when compared
with the effort put into the latter type of Eulerian grid-based
hydrodynamic schemes \citep[e.g.,][]{laney-1998, toro-1997,
Woodward84}.

Despite this invested expertise, the Eulerian solvers in their
original form have a serious drawback: they do not provide an easy
method of adaptively increasing the spatial and temporal resolution in
small volumes of the simulation. Such flexibility is essential for
following physical processes such as gravitational instability. A
solution to this problem was first proposed by \citet{Berger89} in the
Computational Fluid Dynamics (CFD) community, and became known as
Structured Adaptive Mesh Refinement (SAMR). The principle is to
adaptively add and modify additional, finer meshes (``grids'') over
regions that require higher resolution. In addition, it is possible to
add other, more advanced physics including -- for the AMR
implementation in the astrophysics code presented in this paper --
comoving coordinates, self-gravity, radiative cooling, chemistry, heat
conduction, collisionless fluids, magnetohydrodynamics, radiation
transport, star formation and a range of other physical effects.
\newpage

There have been a number of numerical methods described in the
astronomical literature that contain elements of SAMR or have a
similar aim. For example, the N-body solver developed by
\citet{Villumsen89} used non-adaptive meshing to increase the
resolution in pre-selected regions. This static approach was later
used extensively when applied to hydrodynamics
\citep[e.g.,][]{Ruffert94, Anninos94}. Adding adaptivity is a more
recent enhancement, and there are now a number of codes that possess
this feature, both with and without hydrodynamics \citep{Couchman91,
Jessop94, Suisalu95, Splinter96, Gelato97, ART97, Truelove98,
flash_method, MLAPM01, Yahagi01, RAMSES, Quilis04, Ziegler05, Zhang06,
Astrobear09, Pluto-amr, GAMER, Nyx}. Of these, perhaps the most
comparable and widely used are FLASH \citep{flash_method}, which uses
grid blocks of fixed size, and RAMSES \citep{RAMSES} and ART
\citep{ART97}, both of which refine individual cells. It is also
possible to deform the grid to obtain high resolution
\citep[e.g.,][]{Gnedin95, Xu97, Pen98}, and more recently a few codes
have adopted an unstructured approach based on a moving Voronoi mesh
\citep{Arepo10, Tess11}.

In this paper, we present \enzo\ 2.3, a structured adaptive mesh
refinement (SAMR) code. Originally developed for cosmological
hydrodynamics, \enzo\ has since been used on a wide variety of
problems.  It has grown to become a general tool for astrophysical
fluid dynamics and is intended to be efficient, accurate and easily
extended to include new capabilities.  Although many of the components
of the \enzo\ code have been described in previous publications
\citep{1995CoPhC..89..149B, BryanThesis96, Bryan97a, Bryan97b,
Norman99, BryanCompSci99, Bryan01, Oshea04, 2007arXiv0705.1556N,
WangAbelZhang08,ReynoldsHayesPaschosNorman2009,Collins10,Wise11_Moray},
there has previously been no systematic and complete description of
the code.  In this paper we provide that description, filling in many
previous omissions and showing the code's performance for a wide
variety of test problems.

The \enzo\ code has been extensively used over the last two decades in
a wide variety of problems, resulting in the publication of more than
100 peer-reviewed papers. The variety of astrophysical systems that
\enzo\ has been used for include galaxies \citep{2003ApJ...587...13T,
2012MNRAS.425..641L, 2013MNRAS.432.1989S}, galaxy clusters
\citep{Loken02, Xu11, Skillman13}, the interstellar medium
\citep{Slyz05}, the intergalactic medium \citep{Fang01,
2011ApJ...731....6S}, the circumgalactic
medium~\citep{2013MNRAS.430.1548H}, cooling flows \citep{Li12,
2013ApJ...763...38S}, turbulence \citep{Kritsuk04,
2007ApJ...665..416K, 2009JPhCS.180a2020K, Collins11}, the formation of
the first stars \citep{ABN02, 2007ApJ...654...66O,
2009Sci...325..601T, Xu08}, and the formation of stars in our own
Galaxy \citep{Collins11, 2011ApJ...727L..20K, Collins12a}.

Numerical simulations of astronomical phenomena now play a key role,
along with observations and analytic theory, in pushing forward our
understanding of the cosmos \citep[e.g.,][]{DecadalSurvey01,
DecadalSurvey10}.  But along with this role comes responsibility.  We
believe that those developing simulation tools must fulfill two key
obligations: the first is to make those tools available to the
community as a whole, much in the way that astronomical data are now
regularly made publicly available.  The second is to document, test
and refine those methods so that they can be critically evaluated and
expanded upon by others.  Our public release of the \enzo\ code (which
can be found at \url{http://enzo-project.org}) represents our attempt
to meet the first of these obligations; this paper represents our
attempt to meet the second.

The structure of this paper is as follows.  In
Section~\ref{sec.overview}, we first provide a top-level overview of
the code method and structure.  This is designed to give a broad-brush
picture of the equations solved by \enzo\ and the methods used to
solve them.  Next, in Sections~\ref{sec.amr}
through~\ref{sec.num.analysis}, we describe the methods we use in
detail, reserving some of the longer descriptions of particular
components for the appendix in order to not interrupt the flow of the
paper.  The \enzo\ testing framework and code tests are described in
Section~\ref{sec.tests}.  The parallelism strategy and scaling results
are described in Section~\ref{sec.parallel}.  Finally, we discuss the
code's development methodology (which is, as far as we know, unique in
the astrophysics community) in Section~\ref{sec.development}.

%% file: overview.tex
\section{Physical Equations and Overview of Numerical Methodology}
\label{sec.overview}

We begin this section by first writing down the complete set of
physical equations that can be solved by \enzo, and then briefly
describe the numerical algorithms that we use to solve these
equations.  This section is intended to be an overview of \enzo's
capabilities: thus, we gather all of the equations solved into a
single place and provide a brief and high-level introduction to the
numerics of the code.  Detailed descriptions of the individual
components are then provided in
Sections~\ref{sec.amr}-\ref{sec.num.analysis}. In
Table~\ref{tab:variables} we provide a convenient summary of all
variables and physical constants used throughout this manuscript.

\input{table_of_variables_and_constants}


\subsection{Physical Equations}
\label{sec.equations}


The Eulerian equations of ideal magnetohydrodynamics (MHD) including
gravity, in a coordinate systems comoving with the cosmological
expansion, are given by

\begin{eqnarray} 
  \frac{\partial \rho}{\partial t} 
  + \frac{1}{a} \div (\rho \vecv) & = & 0,
  \label{eq:mass} \\
  \frac{\partial \rho \vecv}{\partial t}  
  + \frac{1}{a} \div \left(\rho \vecv \vecv + \myvec{I}p^* - 
    \frac{\vecB \vecB}{a} \right) & = &
  - \frac{\dot{a}}{a} \rho \vecv - \frac{1}{a} \rho \grad \phi,
  \label{eq:momentum} \\
  \frac{\partial E}{\partial t} 
  + \frac{1}{a} \div \left[ (E + p^*) \vecv - 
    \frac{1}{a} \vecB (\vecB \cdot \vecv) \right] & = &
  - \frac{\dot{a}}{a} \left( 2E - \frac{B^2}{2a} \right) - 
  \frac{\rho}{a} \vecv \cdot \grad \phi 
  - \Lambda + \Gamma + \frac{1}{a^2} \div \fcond,
  \label{eq:total_energy}  \\
  \frac{\partial \vecB}{\partial t} - 
  \frac{1}{a}  \grad \times (\vecv \times \vecB) & = & 
  0
  \label{eq:induction}
\end{eqnarray}

In these equations, $E$, $\rho$, $\vecv$, and $\vecB$ are the comoving
total fluid energy density, comoving gas density, peculiar
velocity, and comoving magnetic field strength, respectively. The
matrix 
$\myvec{I}$ is the identity matrix, and $a$ is the cosmological
expansion parameter (discussed in more detail below).  The first
equation represents conservation of mass, the second conservation of
momentum, and the third conservation of total (kinetic plus thermal
plus magnetic) fluid energy.  They are respectively, the first, second, and third
moments of the Boltzmann equation.  The fourth equation is the
magnetic induction equation.  Terms representing radiative cooling
($\Lambda$) and heating ($\Gamma$) enter on the right-hand side of the
energy equation (\ref{eq:total_energy}), as does the flux due to
thermal heat conduction ($\fcond$).

The comoving total fluid energy density $E$ is given by
\begin{equation}
E =  e + \frac{\rho v^2}{2}  + \frac{B^2}{2a}
        \label{eq:total_energy_def},
\end{equation}
where $e$ is the comoving thermal energy density. The total comoving
isotropic pressure $p^*$ is given by
\begin{equation}
p^* = p + \frac{B^2}{2a},
\end{equation}
and the quantity $p$ is the thermal pressure.  We use units such that the magnetic
permeability is unity ($\mu_0=1$).  The equations are closed by an
equation of state and Poisson's equation for the gravitational potential $\phi$:
\begin{eqnarray}
  e &=& \frac{p}{(\gamma - 1)},
  \label{eq:eq_of_state} \\
  \nabla^2 \phi &=& \frac{4 \pi G}{a} \left( \rho_{\rm total} - \rho_0 \right)
  \label{eq:potential}
\end{eqnarray}
The equation of state is shown here for an ideal gas with a ratio of
specific heats $\gamma$.  The gravitational potential $\phi$ is
sourced by the total mass density contrast, where $\rho_{\rm total} =
\rho_{\rm gas} + \rho_{\rm dm} + \rho_{\rm stars}$ and $\rho_0$ is the
mean density.  Although we write the equations including both magnetic
field terms and comoving coordinates, the code is frequently used both
in the purely hydrodynamic limit ($\vecB = 0$, referred to as HD
below) and without cosmological expansion ($a = 1$, $\dot{a} = 0$).



For completeness, we note here that we have defined several key
quantities in the comoving frame to make the previous equations more
readable.  Specifically, we define:

\begin{eqnarray}
\vecx & = & \vecx' / a, \\
\vecv & = & a \hspace{0.5mm} \dot{\vecx} = 
              \vecvp - \dot{a}\vecx, \\
\rho  & = & a^3 \rhop,   \\
p     & = & a^3 \pp, \\
E     & = & a^3 \left(\Ep - 
            \dot{a} \, \vecx \cdot \vecvp - 
            \frac{1}{2} \dot{a}^2 \vecx^2\right), \\
\phi  & = & \phip + \frac{1}{2} \, a \ddot{a} \, \vecx^2, \\
\vecB & = & a^2 \vecB^\prime,
\end{eqnarray}
where primes indicate quantities in proper coordinates (fixed frame).
We note that the definition of comoving $\vecB$ used here is such
that a uniform field is constant in a homogeneous expanding universe
and is also the quantity used in the MHD-CT solver described
below; however, it is not universal,
and is slightly different from that used both in \citet{Li08a}
and in the MHD-Dedner solver (see section~\ref{sec.num.hydro-muscl}
for more details).
The expansion parameter $a \equiv 1/(1 + z)$ follows the expansion of
a smooth, homogeneous background, where $z$, the redshift, is a
function only of $t$.  All spatial derivatives are determined with
respect to the comoving position $\vecx$, which removes the universal
expansion from the coordinate system. The evolution of $a(t)$ is
governed by the second Friedmann equation for the expansion of a
spatially homogeneous and isotropic universe
\begin{eqnarray}
\frac{\ddot{a}}{a} & = & 
      - \frac{4 \pi G }{3 a^3 } (\rho_0 
      + 3p_0/c^2) 
      + \Lambda_c c^2 / 3 .
      \label{eq:expansion} 
\end{eqnarray}
Here $\rho_0$ is the mean comoving mass density (including both
baryonic and dark matter), $p_0$ is the comoving background pressure
contribution, and $\Lambda_c$ is the cosmological constant.
This system of equations is limited to the non-relativistic regime and
assumes that curvature effects are not important --- both assumptions
are reasonable as long as the size of the simulated region is small
compared to the radius of curvature and the Hubble length $c/H$, where
$c$ is the speed of light and $H = \dot{a}/a$ is the Hubble constant.

The comoving evolution equations are equivalent to the fixed
coordinate version in the non-cosmological limit ($a = 1$, $\dot{a} =
0$).  We include the cosmological terms with the understanding that
\enzo\ is not restricted to cosmological applications -- while
historically the code was written with cosmological situations in mind
(e.g., galaxy clusters), in more recent years it has been used to
simulate a much broader range of astrophysical environments.

Any collisionless components (such as dark matter and stars) are modeled
by N-body particles, whose dynamics are governed by Newton's equations
in comoving coordinates:
\begin{eqnarray}
\frac{d \vecx}{dt} 
    & = & \frac{1}{a} \vecv, 
          \label{eq:dm_position} \\
\frac{d \vecv}{dt} 
    & = & - \frac{\dot{a}}{a} \vecv
          - \frac{1}{a} \grad \phi, 
          \label{eq:dm_velocity} 
\end{eqnarray}
The particles contribute to the gravitational potential through
Poisson's equation (Equation \ref{eq:potential}).

In addition, \enzo\ can solve the mass conservation equations for a set of
chemical species and their reactions.  For any species $i$ with comoving
number density $n_i$, these equations have the form:
\begin{equation}
  \frac{\partial n_i}{\partial t} 
  + \frac{1}{a} \div (n_i \vecv) = 
  \sum_j k_{ij}(T) \, n_i n_j + \sum_j \Gamma^{\rm ph}_j n_j 
  \label{eq:species_evolution}
\end{equation}
where $k_{ij}$ are the rate coefficients for the two-body reactions
and are usually functions of only temperature (we will
specifically note the cases where we either include three-body
reactions or have density-dependent rates).  The $\Gamma^{\rm ph}_j$
are destruction/creation rates due to photoionizations and/or
photodissociations.  Currently the species \enzo\ can follow include
H, H$^+$, He, He$^+$, He$^{++}$, and e$^-$, and through additional
options also H$^-$, H$_2$, H$_2^+$, and HD, D, and D$^+$. Lastly,
\enzo\ can also track the advection of one or more comoving metal density fields,
which can contribute to the radiative cooling and star formation
processes.

The code can include either a homogeneous radiation background 
or evolve an inhomogeneous
radiation field either by directly solving the radiative transfer
equation along rays or by solving a set of moment equations derived
from the radiative transfer equation. In comoving coordinates
\citep[e.g.,][]{Gnedin97} the radiative transfer equation reads
\begin{equation}
  \label{eq:rteqn}
  \frac{1}{c} \; \frac{\partial I_\nu}{\partial t} + 
  \frac{a_{\rm em}}{a} \, \hat{n} \cdot \grad I_\nu -
  \frac{H}{c} \; \left( \nu \frac{\partial I_\nu}{\partial \nu} -
  3 I_\nu \right) = -\kappa_\nu I_\nu + j_\nu .
\end{equation}
Here $I_\nu \equiv I(\nu, \mathbf{x}, \Omega, t)$ is specific
intensity of the radiation, with dimensions of energy per time $t$ per
solid angle $\Omega$ per unit area per frequency $\nu$.  The second
term represents the propagation of radiation, where the factor $a_{\rm
  em}/a$ accounts for cosmic expansion since the time of emission.
The third term describes both the cosmological redshift and dilution
of radiation.  On the right hand side, the first term captures
absorption ($\kappa_\nu \equiv \kappa_\nu(\mathbf{x},\nu,t)$), and the
second term emission ($j_\nu \equiv j_\nu(\mathbf{x},\nu,t)$) from any
sources of radiation (whether point or diffuse).

As an alternative to the ray-casting strategy, \enzo\ can also solve
the angle-averaged radiative transfer equation through a
flux-limited diffusion approximation, with couplings to both the gas
energy and chemical number densities. The equations solved are
\begin{eqnarray}
  \label{eq:fld_radiation}
  \frac{\partial E_r}{\partial t} + \frac1a \div\left(E_r \vecv \right) &=& 
  \div\left(D\grad E_r\right) -
  \frac{\dot{a}}{a} E_r - c\kappa E_r + \eta, \\
  \label{eq:fld_heating}
  \frac{\partial e_c}{\partial t} &=& -\frac{2\dot{a}}{a} e_c + \Gamma - \Lambda,
\end{eqnarray}
where $E_r = E_r(\vecx,t)$ is a grey radiation energy density and
$e_c$ is the internal energy correction due to photo-heating and
chemical cooling.
Here, we define $E_r$ through first assuming a fixed frequency
spectrum, i.e.  $E_{\nu}(\nu,\vecx,t) = \tilde{E}_r(\vecx,t)
\chi(\nu)$, and then defining the integrated quantity
\begin{equation}
\label{eq:grey_radiation_energy}
   E_r(\vecx,t) \equiv \int_{\nu_0}^{\infty}
   E_{\nu}(\nu,\vecx,t)\,\mathrm d\nu \  = \ 
   \tilde{E}_r(\vecx,t) \int_{\nu_0}^{\infty} \chi(\nu)\,\mathrm d\nu.
\end{equation}
The quantity $D$ in Equation (\ref{eq:fld_radiation}) is the Larsen
square-root flux-limiter \citep[see][]{Morel2000}, $\kappa$ is  the
total opacity, $\eta$ is the field of radiation sources, $\Gamma$ is
the radiation-induced photo-heating rate, and $\Lambda$ is the
chemical cooling rate.

Finally, \enzo\ implements the equations of isotropic heat conduction
in a manner similar to that of \citet{2007ApJ...664..135P}, where the isotropic heat flux is given by
\begin{equation}
\fcond = -\kappa_{\rm cond} \grad T.
\label{eq:conduction}
\end{equation}
Here $\kappa_{\rm cond} = f_{\rm sp} \, \kappa_{\rm sp}$ is the heat
conduction coefficient, given as a fraction $f_{\rm sp}$ of the
Spitzer conductivity $\kappa_{\rm sp}$ \citep{1962pfig.book.....S},
and T is the gas temperature (with fluids explicitly assumed to be
single-temperature).  Saturation of the heat flux in high temperature,
low density regimes (such as the intracluster medium in galaxy
clusters) is taken into account.

Thermal conduction in a plasma can be strongly affected by the
presence of magnetic field lines, which may strongly suppress heat
flow perpendicular to the magnetic field.  To include this effect, it
is possible to allow heat transport only along (not across) magnetic
field lines, as follows:

\begin{equation}
\fcond= -\kappa_{\rm cond} \,\myvec{b} ( \myvec{b} \cdot \grad T),
\label{eq:anisotropic_conduction}
\end{equation}
where $\myvec{b}$ is the unit vector denoting the direction of
the magnetic field.


\subsection{Overview of Numerical Methods}
\label{sec.method_overview}

In this section we briefly describe the numerical methods that are
used to solve the equations outlined in Section~\ref{sec.equations}.
We proceed through the numerical methods in the same order as will be
used in Sections~\ref{sec.amr} through~\ref{sec.num.analysis} so that
there is a one-to-one correspondence between each of the following
overviews and the complete description provided in the later sections.
The goal in this section is to introduce the reader to the basic
principles of the methods without drowning them in detail.

\subsubsection{Structured Adaptive Mesh Refinement}

The primary feature of the \enzo\ code is its Adaptive Mesh Refinement
(AMR) capability, which allows it to reach extremely large spatial and
temporal dynamical ranges with limited computational resources,
opening doors to applications otherwise closed by finite memory and
computational time. Unlike moving mesh methods
\citep{1995ApJS..100..269P,1995ApJS...97..231G} or methods that
subdivide individual cells \citep{Adjerid}, Berger \& Collela's AMR
(also referred to as \emph{structured} AMR, or SAMR) utilizes an
adaptive hierarchy of grid patches at varying levels of resolution.
Each rectangular grid patch (referred to as a ``grid'') covers some
region of space requiring higher resolution than its \emph{parent
grid}, and can itself become the parent grid to an even more highly
resolved \emph{child grid}.

The grid hierarchy begins with the root grid, which covers the entire
domain of interest with a coarse uniform Cartesian grid. Then, as the
solution evolves and interesting regions start to develop, finer grids
are placed within these coarse regions.  We restrict the ratio between
cell sizes of parent and child grids to be an integer, typically 2 or
4, and refer to a level as all grids with the same cell size.  For
simplicity, the edges of subgrids must coincide with the cell edge of
its immediate parent (coarser) grid. Additionally, the hierarchy can
be initialized with one or more static grids if a higher initial
resolution is required.

Given the hierarchy at some time $t$, we advance the solution in the
manner of a W-cycle in a multigrid solver.  First, we determine the
maximum timestep allowed for the coarsest grid based on a variety of
accuracy and stability criteria and advance the grid by that time
interval.  We then move down to the next level and advance all the
grids on that level.  If there are more levels, we repeat this
procedure until the bottom level of the hierarchy has been reached.
Once there, we continue advancing the grids on the lowest level until
they have ``caught up'' to the coarser parent grids.  This procedure
repeats itself until all grids have been advanced by the desired time.

Since interesting regions on the grid may move, the hierarchy must
adapt itself. This happens whenever a level has caught up to its
parent level, by entirely rebuilding the grids on that level and at
finer resolutions. Rebuilding is achieved by applying the grid
refinement criteria to the grids on that level and flagging zones that
require extra grids.  These criteria depend on the physical problem
being simulated; see Section~\ref{sec:refinement_criteria} for more
details.  Once a grid has a set of flagged cells, we apply a
machine-vision based algorithm \citep{Berger91} to find edges and
determine a good placement of subgrids.  Once these new subgrids have
been identified, the solution from the next coarser grid is
interpolated in order to initialize the values on the new grids.
Finally, any overlap between new subgrids and old ones is identified,
and the prior solution within the regions of overlap is copied to the
new subgrids. This procedure is then repeated on the new grids and in
this way, the entire hierarchy (from the original level examined and
continuing on to all finer levels) is rebuilt.

\subsubsection{(Magneto)-hydrodynamic solver methods}

Four different (magneto)-hydrodynamic methods are implemented in
\enzo: (i) the hydrodynamic-only piecewise parabolic method (PPM)
developed by~\citet{1984JCoPh..54..174C} and extended to cosmology
by~\citet{1995CoPhC..89..149B}; (ii) the MUSCL-like Godunov scheme
\citep{1977JCoPh..23..276V} with or without magnetic fields and
Dedner-based divergence cleaning, described in more detail in
\citet{WangAbelZhang08} and \citet{WangAbel09}; (iii) a constrained
transport staggered MHD scheme \citep{Collins10}, and (iv) the
second-order finite difference hydrodynamics method described in
\zeus~\citep{Stone92a,Stone92b}.

\paragraph{Godunov PPM method (HD only)}

We begin with the direct-Eulerian PPM implementation.  This is an
explicit, higher-order accurate version of Godunov's method for ideal
gas dynamics with the spatially third-order accurate piecewise
parabolic monotonic interpolation and a nonlinear Riemann solver for
shock capturing.  It advances the hydrodynamic equations in the
following steps: (i) Construct monotonic parabolic interpolation of
cell average data, for each fluid quantity; (ii) Compute interface
states by averaging the parabola over the domain of dependence for
each interface; (iii) Use interface data to solve the Riemann problem;
(iv) Difference the interface fluxes to update the cell average
quantities.

The PPM implementation does an excellent job of capturing strong
shocks across a few cells.  Multidimensional schemes are built up by
directional splitting and produce a method that is formally
second-order accurate in space and time and which explicitly conserves
mass, linear momentum, and energy \citep{Hawley84, Norman86}.  A
variety of Riemann solvers have been implemented.

As described in \citet{1995CoPhC..89..149B}, we modify the method for
use in hypersonic flows when the thermal energy $e$ is extremely small
compared to the total energy $E$.  This situation presents a problem
because in the total energy method the thermal energy is computed by
subtracting one large number from another (i.e. the kinetic energy
from the total energy), which tends to generate large numerical
inaccuracies. We overcome this difficulty by additionally solving a
thermal energy equation and using $e$ from this equation when we
expect the error to be large.

\paragraph{Godunov MUSCL (HD) with Dedner divergence cleaning (MHD)}
This solver was developed to attack problems in magnetic field
amplification during the formation of galaxies \citep{Wang:2009a} and
to understand the role of proto-stellar jets for the theory of star
formation \citep{Wang:2009b}. It combines the standard approach of
Godunov \citep{Godunov1959} for finite volume techniques with the
method of lines as described by \cite{leveque2002finite} and
\cite{toro-1997}. In addition, it implements the hyperbolic divergence
cleaning algorithm of \cite{2002JCoPh.175..645D}. It supports multiple
approximate Riemann solvers and non-ideal equations of
states. Consequently, this suite of solvers can be used for hydro and
magneto-hydrodynamic simulations. This class of solvers, as well as a
version of the PPM hydro solver, has been ported to nVidia's CUDA
framework, allowing \enzo\ to take advantage of modern graphics
hardware \citep{Wang:2010}.

\paragraph{Godunov MHD with Constrained Transport (MHD)} This MHD
method is second-order in time and space, and preserves the divergence
constraint, $\div \vecB = 0$, to machine precision through the
Constrained Transport (CT) method \citep{Collins10}.  CT, originally
described by \citet{Evans88}, updates the magnetic field with the curl
of an electric field, suitably formulated to preserve $\div \vecB$.
We employ the CT methods described by \citet{Balsara99} and
\citet{Gardiner05} with the second-order hyperbolic solver of
\citet{Li08a} and the divergence-free AMR scheme of \citet{Balsara01}.

\paragraph{Second-order finite difference method (HD only)}

Lastly, we briefly describe the \zeus\ method, a finite difference
algorithm originally used in the \zeus\ code \citep{Stone92a}. Note
that \enzo\ is entirely independent of the \zeus\ code, and only the
hydrodynamical algorithm of \zeus\ has been implemented in \enzo; the
MHD and radiation-hydrodynamics schemes have not. The \zeus\ method
uses a staggered mesh such that velocities are face-centered, while
density and internal energy are cell-centered.  It splits the solution
up into two steps. The first is the so-called source step, in which
the momentum and energy values are updated to reflect the pressure and
gravity forces, including an artificial viscosity required for
stability. The second step, known as the transport step, accounts for
the advection of conserved quantities (mass, momentum, and energy)
across the grid.

\subsubsection{Gravity}

The current implementation of self-gravity in \enzo\ uses a Fast
Fourier technique \citep{Hockney88} to solve Poisson's equation on the
root grid on each timestep.  The advantage of using this method is
that it is fast, accurate, and naturally allows both periodic and
isolated boundary conditions for the gravity, choices which are very
common in astrophysics and cosmology.  On subgrids, we interpolate the
boundary conditions from the parent grid (either the root grid or some
other subgrid). The Poisson equation is then solved on every timestep
using a multigrid technique on one subgrid at a time. Aside from
self-consistently calculating the gravitational potential arising from
the baryon fields and particles in the simulation, there are also a
number of options for specifying static gravitational fields.

\subsubsection{N-body Dynamics}

Collisionless matter (e.g. dark matter, stars, etc.) is modeled with
particles that interact with the baryons only via gravity.  These
particles are advanced through a single timestep using a
drift-kick-drift algorithm \citep{Hockney88} to provide second-order
accuracy even in the presence of varying timesteps.  Since the
particles follow the collapse of structure, they are not adaptively
refined.  Nor are there duplicate sets of particles for each level;
instead, each particle is associated with the highest refined level
available at its position in the domain and particles are moved
between grids as the hierarchy is rebuilt. Thus a particle has the
same timestep and feels the same gravitational force as a grid cell at
that refinement level.

Although the particles are fixed in mass once initialized (with the
exception of star particles, which can lose mass in the feedback
process), we can create them with any initial set of masses and
positions.  For example, in many cosmological simulations static
subgrids are included from the beginning in order to improve the
initial spatial and baryonic mass resolution, and these subgrids are
populated with lower-mass particles to correspondingly improve the
collisionless mass resolution.

\subsubsection{Chemistry}
\label{sec.ov.chem}

\enzo\ includes the capability of following up to 12 particle species
using a non-equilibrium chemical network.  The species can be turned
on in sets, with the simplest model including just atomic hydrogen and
helium (H, H$^+$, He, He$^+$, He$^{++}$, e$^-$), and more complete
models adding first species important for gas-phase molecular hydrogen
formation (H$^-$, H$_2$ and H$_2^+$), and then HD formation (HD, D,
D$^+$).  The cooling and heating due to these species is included (see
the next section). The solution of the rate equations is carried out
using one Jacobi iteration of an implicit Euler time discretization to
ensure stability. To ensure accuracy the rate equations are sub-cycled
within one hydrodynamic timestep, subject to the constraint that the
electron and neutral fractions do not change by more than 10\% in one
sub-cycle.

\subsubsection{Radiative Cooling and Heating}

\enzo\ can operate in a number of different modes with regard to
radiative cooling and heating. In the simplest mode, where the
multi-species flag is turned off and no individual chemical species
are tracked, the cooling rate is computed from a simple
temperature-dependent cooling function, taken from \citet{SW87}.  If
chemistry is turned on, then the code can include cooling from all
species of hydrogen and helium (including H$_2$ and Deuterium-related
species such as HD) -- and the primordial cooling rates are computed
in the same Jacobi iteration as the chemistry.  It is also possible to
include metal cooling based on a set of multi-dimensional lookup
tables computed with the \textsc{Cloudy} code
\citep{1998PASP..110..761F} as described in
\citet{2008MNRAS.385.1443S} and \citet{2011ApJ...731....6S}. Note that
the cooling and heating is most commonly treated in the optically-thin
limit, but the code can also follow radiative transfer in a limited
set of energy bins.

\subsubsection{Homogeneous radiation backgrounds}

The chemical networks and heating rates described in the previous
sections can be affected by external radiation fields, and the code
includes a number of pre-calculated meta-galactic UV radiation
backgrounds that are uniform in space but can vary in time.  These are
generally based on the redshift-dependent rates given in
\citet{1996ApJ...461...20H} and \citet{2012ApJ...746..125H}, but can
also include a uniform H$_2$ photo-dissociating background that is
either constant in time or varying as in \citet{WiseAbel05}.

\subsubsection{Radiation transport: ray tracing}

\enzo\ includes a photon-conserving radiative transfer algorithm that
is based on an adaptive ray-tracing method utilizing the HEALPix
pixelization of a sphere \citep{Abel02_RT}. Photons are integrated
outwards from sources using an adaptive timestepping scheme that
preserves accuracy in ionization fronts even in the optically-thin
limit. This has been coupled to the chemistry and cooling network to
provide ionization and heating rates on a cell-by-cell basis. The
method is described in detail (including numerous code tests) in
\citet{Wise11_Moray}.

\subsubsection{Radiation transport: Flux-limited diffusion}

A second option for radiative transfer is a moment-based method that
adds an additional field tracking the radiation energy density.  This
field is evolved using the flux-limited diffusion method, which
transitions smoothly between streaming (optically thin) and opaque
limits and is coupled to an ionization network of either purely
hydrogen, or both hydrogen and helium.  The resulting set of linear
equations is solved using the parallel HYPRE framework.  Full details
on the \enzo\ implementation of this method can be found in
\citet{ReynoldsHayesPaschosNorman2009}.

\subsubsection{Heat Conduction}

Heat conduction, both isotropic and anisotropic, can be included using
a sub-cycled, operator-split method.  The heat fluxes are computed
with simple second-order accurate finite differences, and stability is
ensured by restricting the timestep and using flux-limiters where
appropriate.

\subsubsection{Star Formation and Feedback}

A family of simple heuristic methods are used to model the formation
of stars and their feedback of metals and energy into the gas.  These
methods are based on the work of \citet{CO1992}, and involve the
identification of plausible sites of star formation based on a set of
criteria (for example, dense gas with a short cooling time, which is
both collapsing and unstable).  The local star formation rate is
computed using a range of methods, such as a density-dependent method
based on the Schmidt-Kennicutt relation \citep{K89}.  The affected gas
is converted into a star particle over a few dynamical times, and
metals and thermal energy are injected into the region surrounding the
star particle.  A related set of methods involves the simulation of
single Population III stars rather than ensembles, and is calibrated
by \textit{ab initio} simulations of primordial star formation.

\subsubsection{Timestep constraints}

All grids on a given level are advanced with the same timestep.  This
time step is determined by first calculating the largest time step
allowed for each cell and for each physical process separately (except
for chemistry and heat conduction, which are sub-cycled).  The level
is then advanced with a timestep equal to the minimum over all of
these $\Delta t$.

%% file: table_of_variables_and_constants.tex
\begin{center}


\LongTables

\begin{deluxetable}{cl}
  \tablecaption{Summary of symbols used in this manuscript\label{tab:variables}}
  \tablehead{\colhead{Symbol} & \colhead{Description}}
  
  \startdata
  \multicolumn{2}{l}{\underline{Fluid Quantities}} \\[5pt]
  $E$ & comoving total fluid energy density \\
  $e$ & comoving thermal energy density \\
  $T$ & baryon temperature \\
  $S$ & fluid entropy \\
  $\rho$ & comoving baryonic density \\
  $\rho_{\!X}$ & comoving mass density of component $X$ (gas, dark matter, stars, metals, total, etc.) \\
  $t$ & time \\
  $\vecv$ & peculiar velocity \\
  $\vecB$ & comoving magnetic field strength \\
  $p$ & comoving thermal pressure \\
  $p^*$ & comoving total (thermal + magnetic) isotropic pressure \\
  $\gamma$ & ideal gas ratio of specific heats \\
  $\phi$ & gravitational potential \\
  $\Lambda$ & radiative/chemical cooling rate \\
  $\Gamma$ & radiative heating rate \\
  $Q$ & artificial viscosity (only for \zeus\ hydro) \\[3pt]
  
  \multicolumn{2}{l}{\underline{Code Quantities}} \\[5pt]
  $\Delta x_l$ & cell size on level $l$ \\
  $\Delta t_l$ & timestep on level $l$ \\
  $d$ & dimensionality (rank) of the simulation \\
  $q$ & conserved quantity \\
  $F$ & flux of quantity $q$ \\
  $r$ & refinement factor \\
  $\epsilon_l$ & refinement criterion parameter \\
  $J$ & number of cells per Jeans length (for Jeans length based refinement) \\[3pt]
  
  \multicolumn{2}{l}{\underline{Cosmology}} \\[5pt]
  $a$ & cosmological expansion factor \\
  $\Lambda_c$ & cosmological constant \\
  $\rho_0$ & mean comoving mass density (including both baryonic and dark matter) of the universe \\
  $p_0$ & comoving background pressure \\[3pt]
  
  \multicolumn{2}{l}{\underline{Chemistry}} \\[5pt]
  $n_i$ & comoving number density of species $i$ \\
  $k_{ij}$ & two-body reaction rate coefficient \\[3pt]
  
  \multicolumn{2}{l}{\underline{Ray-tracing Radiation Transfer}} \\[5pt]
  $\Gamma_j^{ph}$ & photo-ionization/-dissociation destruction/creation rates \\
  $I_\nu$ & specific radiation intensity (energy per time per solid angle per unit area per frequency) \\
  $\kappa_\nu$ & radiation absorption coefficient \\
  $j_\nu$ & radiation emission coefficient \\
  $P$ & photon number flux along ray \\
  $\sigma_{\rm abs}$ & absorption cross section \\
  $n_{\rm abs}$ & number density of absorbing medium \\
  $dt_P$ & radiative transfer timestep \\
  $k_{ph}$ & photo-ionization rate of a single ray \\
  $\Gamma_{ph}$ & photo-heating rate of a single ray \\[3pt]
  
  \multicolumn{2}{l}{\underline{Flux Limited Diffusion Radiation Transfer}} \\[5pt]
  $E_r$ & grey radiation energy density \\
  $e_c$ & internal energy correction due to photo-heating and chemical cooling \\
  $D$ & Larsen square-root flux-limiter \\
  $\kappa$ & total opacity \\
  $\eta$ & field of radiation sources \\[3pt]
  
  \multicolumn{2}{l}{\underline{Heat Conduction}} \\[5pt]
  $\fcond$ & thermal conduction heat flux \\
  $\kappa_{\rm cond}$ & thermal conduction coefficient \\
  $\kappa_{\rm sp}$ & Spitzer conductivity \\
  $f_{\rm sp}$ & fraction of Spitzer conductivity \\[3pt]
  
  \multicolumn{2}{l}{\underline{Star Formation And Feedback}} \\[5pt]
  $t_{\rm cool}$ & cooling time \\
  $t_{\rm dyn}$ & local dynamical time \\
  $t_{\rm form}$ & star particle formation time \\
  $m_{\rm b}$ & baryonic mass of cell \\
  $m_{\rm J}$ & Jeans mass of cell \\
  $m_*$ & star particle mass \\
  $m_{\rm *min}$ & minimum star particle mass \\
  $f_{\rm *eff}$ & star formation efficiency parameter \\
  $f_{\rm Zb}$ & metallicity fraction of baryon gas in cell \\
  $f_{\rm m*}$ & fraction of star particle's mass returned to cell \\
  $f_{\rm Z*}$ & fraction of star particle's metal mass returned to cell \\
  $f_{\rm SN}$ & fraction of stellar rest mass energy deposited as feedback \\
  $\Delta m_{\rm sf}$ & mass of stars formed in current timestep \\
  $E_{\rm add}$ & feedback energy added to cell \\
  $T_{\rm dyn,min}$ & user-defined minimum dynamical time \\
  $\eta$ & overdensity threshold for star formation \\
  $\rho_{\rm SF}$ & constant proper density threshold for star formation \\
  $\tau_*$ & characteristic star formation timescale (for Schmidt-Law method) \\[3pt]
  
  \multicolumn{2}{l}{\underline{Natural Constants and Values}} \\[5pt]
  $G$ & gravitational constant \\
  $H$ & Hubble constant \\
  $k_B$ & Boltzmann constant \\
  $M_\odot$ & mass of Sun \\
  $m_H$ & mass of hydrogen atom \\
  $m_e$ & mass of electron \\
  $c$ & speed of light \\
  $c_{\rm s}$ & sound speed \\
  $z$ & redshift
  \enddata

\end{deluxetable}

\end{center}

%% file: numerical-amr.tex
\section{The Structured Adaptive Mesh Refinement (SAMR) Method}
\label{sec.amr}

The back-bone of the SAMR idea is the patch, or grid, which we take to
be rectilinear to simplify bookkeeping (in practice, the lower-level
routines such as the hydro solvers all assume Cartesian coordinates).
The single root grid covers the entire computational domain and plays
the role of the root node in the grid hierarchy.  Subgrids are placed
such that they cover (hopefully small) sub-volumes of the root grid at
higher resolution.  This structure forms a tree that may be extended
to arbitrary depths by adding additional grids to refine regions in
the subgrids.  We use the following notation to describe grid
relationships.  A grid's \textit{parent} is the coarse grid that
completely contains it.  The \textit{root grid} is the most
coarsely-resolved grid, and it has no parent.  A \textit{child grid}
is a grid completely contained by its parent grid (note that a grid
may have many children but only one parent), while sibling grids are
those that have the same resolution, but not necessarily the same
parent.  We also refer to levels of the hierarchy, where the root grid
is labeled level zero and all grids with the same resolution are on
the same level.

In order to simplify bookkeeping, a number of restrictions are imposed
on the subgrids:

\begin{itemize}
 \item The refinement factor $r$, or the ratio of the coarse cell
   width to the fine cell width, must be an integer.  In practice, we
   generally adopt $r=2$, since tests have shown this typically
   provides the best performance.
 \item Each subgrid must begin and end on the boundary of a coarse
   cell.  This means that the number of cells in a patch will be a
   multiple of the refinement factor.
 \item A subgrid must be completely enclosed by its parent.
\end{itemize}

The AMR grid patches are the primary data structure in \enzo.  Each
patch is treated as an individual object that can contain both field
variables and particle data.  Individual grids are organized into a
dynamic, distributed hierarchy of mesh patches.  

\begin{figure}
\begin{center}
\includegraphics[width=0.5\textwidth]{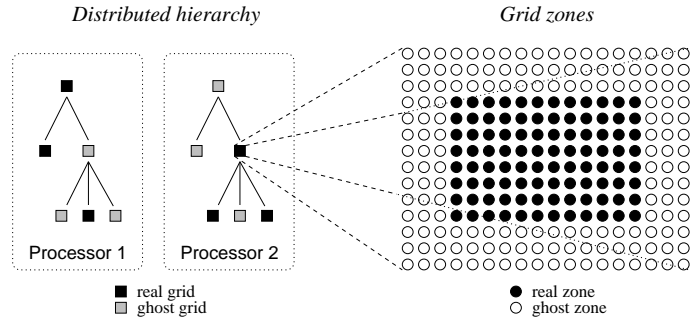}
\end{center}
\caption{\emph{Left:} Example of a simple, distributed AMR hierarchy
showing real and ghost grids.  \emph{Right:} Example 2D \enzo\ grid
showing real and ghost zones, as needed for the PPM hydro stencil. }
\label{fig.amr_hierarchy}
\end{figure}

Each grid patch in \enzo\ contains arrays of values for baryon and
particle quantities.   Grids are partitioned into a core of
\emph{active zones} and a surrounding layer of \emph{ghost zones}, as
shown in Figure~\ref{fig.amr_hierarchy}.  The active zones store field
values and ghost zones are used to temporarily store values that have
been obtained directly from neighboring grids or interpolated from a
parent grid.  These zones are necessary to accommodate the
computational stencil of the (magneto)hydrodynamics solvers
(Sections~\ref{sec.hydro.ppm} through~\ref{sec.hydro.zeus}) and the
gravity solver (Section~\ref{sec.gravity}).  The PPM and \zeus\ hydro
solvers require 3 layers of ghost zones,  and the gravity solver
requires 6 (although only for the density and potential fields).  This
can lead to significant memory and computational overhead,
particularly for smaller grid patches at high levels of refinement.

From the point of view of the hydro solver, each grid is solved as an
independent computational fluid dynamics (CFD) problem, with Dirichlet
boundary conditions stored in the ghost zones.  At the beginning of
each timestep, cells on a given level fill ghost zones by
interpolating from the parent and copying from neighbors on the same
level.  Grids then correct the flux at the interface boundary between
fine and coarse zones, and finally project their active zone data to
the parent grid. 

\begin{figure}
\begin{center}
\includegraphics[width=0.6\textwidth]{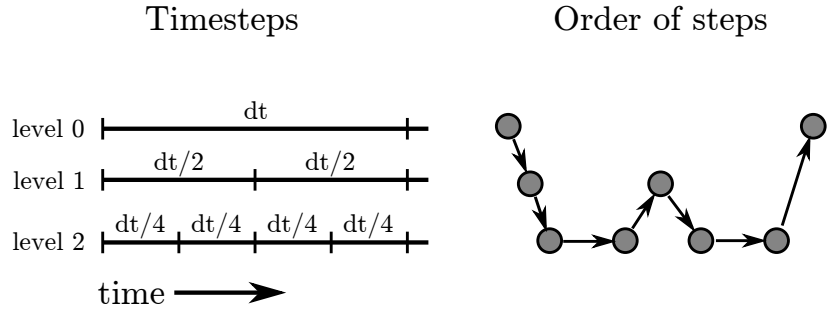}
\end{center}
\caption{\emph{Left:} Example of the timesteps on a 2-level AMR
  hierarchy.  \enzo\ does not restrict the timesteps on the finer levels
  to be a factor of $1/2^n$ smaller.  \emph{Right:} The order in which
  the AMR grids are evaluated on each level.\vspace{1ex}}
\label{fig:wcycle}
\end{figure}

Timesteps are determined on each level as described in
Section~\ref{sec.timestepping}, and the hierarchy is advanced on a
level-by-level basis in a W-Cycle.  Beginning with the coarsest level,
$l$, all grids on that level are advanced one timestep.  Then, one
timestep is taken on all grids at the next level of refinement, $l+1$,
and so on until the finest level is resolved. The finest level is then
advanced, using as many steps as it takes to reach the level
immediately above.  The finest level is then synched to the level
above it, which then proceeds forward in time one more step.  This is
shown graphically in Figure~\ref{fig:wcycle}.

At the end of every timestep on every level, each grid updates its
ghost zones by exchanging information with its neighboring grid
patches (if any exist) and/or by interpolating from a parent grid.  In
addition, cells are examined to see where refinement is required and
the entire grid hierarchy is rebuilt at that level (including all more
highly refined levels).  The timestepping and hierarchy
advancement/rebuilding process described here is repeated recursively
on every level to the specified maximum level of refinement in the
simulation.

The basic control algorithm looks very much like any used in a
grid-based code of a single resolution, and can be written
schematically as:

\begin{verbatim}
InitializeHierarchy
While (Time < StopTime):
   dt = ComputeTimeStep(0)
   EvolveLevel(0, dt)
   Time = Time + dt
   CheckForOutput(Time)
\end{verbatim}

The AMR control algorithm is contained within the recursive
EvolveLevel algorithm, which takes the level upon which to operate as
an argument.  This looks something like the following pseudo-code:

\begin{verbatim}
EvolveLevel(level):
   SetBoundaryValues
   while (Time < ParentTime):
      dt = ComputeTimeStep(level)
      SolveHydroEquations(dt)
      SolveOtherEquations(dt)
      Time = Time + dt
      SetBoundaryValues
      EvolveLevel(level+1, dt)
      FluxCorrection
      Projection
      RebuildHierarchy(level+1)
\end{verbatim}

Each function operates on all the grids on the given level.  In the
following sections, we will address each of the major elements in
turn.

\subsection{Setting the boundary values}
\label{sec:interpolation}

In order to solve the hydrodynamic equations on a patch, it is
necessary to specify the boundary conditions on that patch.  There are
two cases.  The first is that the boundary is external to the
computational domain.  This occurs for all the boundaries of the root
grid.  The second possibility is that the boundary is within the
computational domain, which is true for most, but not all, subgrids.

In the first case, we allow for four types of boundary conditions (for
simplicity, represented here as being at $x=0$, with positive $x$
interior and negative $x$ exterior to the domain):

\begin{enumerate}
  \item{\em Reflecting.} In this case, the boundary behaves like a
    mirror, with the solution on the boundary reflecting the solution
    in the computational domain ($q(-x) = q(+x)$) but with the
    velocity normal to the boundary reversed: $v_x(-x) = -v_x(+x)$.
  \item{\em Outflow.}  We approximate outflow boundary conditions by
    duplicating the solution at the edge of the computational domain:
    $q(-x) = q(0)$.
  \item{\em Inflow.} In this case, the boundary values are fixed by a
    pre-determined function ($q(-x) = q0(-x,t)$).  The code provides a
    simple way to set inflow values that are constant in both time and
    space over an entire face, or hooks are provided for special cases
    where the boundary must be time-varying.
  \item{\em Periodic.} The boundary solution is obtained from the
    other side of the grid: $q(-x) = q(x_{\rm{max}}-x)$.
\end{enumerate}

In the second case, the value must be either (1) interpolated from a
parent or (2) copied from a sibling.  In practice, we solve this
problem by first interpolating all boundaries from the grid's parent
and then performing the sibling copy wherever possible, since these
values will always be as good as, or superior to, interpolation.

We focus first on conserved, cell-centered quantities.  The
interpolation function should be conservative in the sense that the
mean over the interpolated quantity, in a given cell, must return the
cell-centered value.  It should also be monotonic, so that new extrema
are not introduced into a smoothly varying field.

In one dimension, such functions are well known; examples include
donor cell, Van Leer, and piecewise parabolic interpolation, which are
first-, second-, and third-order accurate respectively
\citep[e.g.,][]{Stone92a}.  However, we require fully
multi-dimensional interpolation functions, and so focus on
second-order schemes since they are tractable and relatively
accurate\footnote{We could, in principle, build a multi-dimensional
interpolator out of a one-dimensional interpolation function; however,
at higher order, this would not necessarily satisfy our constraint
regarding not introducing new extrema in interpolated quantities.}.
In Appendix~\ref{app:interpolation}, we explicitly write down
functions in one, two and three dimensions, but here we discuss the
general design considerations.

In two dimensions, it is relatively easy to build a linear, monotonic,
conservative function by requiring, first, that the function at the
center of the cell return the cell-centered quantity (and since it is
linear, this implies that an average over the cell will also be equal
to this value), and second, by treating the cell-diagonals as
one-dimensional Van Leer-type functions.  The reason to focus on the
cell diagonals is that for a linear function, the extrema will occur
at the cell corners, which are in turn linked by the diagonals.  This
process can be done in two dimensions since there are three
constraints (one at the center and one along each diagonal), and three
parameters in a two-dimensional linear function.

In three dimensions this process breaks down since the number of
constraints (center + four diagonals = five) exceeds the number of
parameters (four).  One way to solve this problem would be to perform
a least-squares fit under the given constraints; however, this is too
computationally expensive.  Therefore, in
Appendix~\ref{app:interpolation} we describe a heuristic approach that
satisfies four constaints analytically and then adjusts the
interpolation parameters in such a way as to satisfy the fifth.  While
the result is not necessarily optimal, it does appear to perform
adequately in our tests.

\subsection{Solving the hydrodynamic equations on a grid}
\label{sec:solve_hydro}

Given the grid and its boundary conditions, the process of solving
equations~(\ref{eq:mass})--(\ref{eq:total_energy}) can be detached
from the rest of the SAMR technique, so that it is possible to
experiment with a number of different methods.  The restrictions are
that the method should be flux-conservative, or at least produce an
accurate estimate of the flux of conserved quantities at cell
interfaces.  This is required in the flux-correction step (see
\ref{sec:flux_correction}).  There is also the issue of centering,
which refers to the position of physical quantities within a cell (the
primary types are cell-centered, face-centered, and node-centered).
The interpolation step, described in Section~\ref{sec:interpolation},
must correctly interpolate with the selected centering.

\subsection{Projection}
\label{sec:projection}

Since a refined region is simulated with (at least) two different
resolutions (one fine and one coarse), it has two or more
``solutions'' (i.e.~values of density, velocity, etc.) that represent
the same volume of space.  This is one important aspect of the SAMR
method -- some physical regions are represented by grids at multiple
resolutions.  To maintain consistency, the coarse grid quantities are
updated with the finer values, using a volume-weighted average of the
conserved quantities:

\begin{equation}
q^{\rm coarse} = r^{-d} \sum q^{\rm fine}_{i, j,k}
\end{equation}

where $r$ is the refinement factor, and the sum is over all fine cells
that ``cover" the coarse cell. The exponent, $d$, refers to the weight
of the average.  For cell-centered quantities, this is the
dimensionality of the problem.  For face-centered quantities, this is
one less than the dimensionality, and for edge-centered quantities
this is one.

\subsection{Flux correction}
\label{sec:flux_correction}

One of the significant advantages of flux-conservative methods is, of
course, their ability to maintain conserved quantities to machine
precision.  Unfortunately, the projection step upsets this flux
conservation since, at the boundary of a refined region, the coarse
and fine cells on either side of the border are updated with different
estimates of the flux across the boundary.  At one refinement level
this comes from the coarse grid computation, and in the other from the
fine grid solution.  Conservation can be restored by correcting all
coarse cells that lie outside the boundary of a fine region with the
difference between the coarse and fine estimates of the flux across
the boundary:

\begin{equation}
  q^{\rm coarse} = \tilde{q}^{\rm coarse} - \Delta t \left( F^{\rm
      coarse} - \sum_{j,k} F^{\rm fine}_{j,k} \right)
\end{equation}

where $F$ is the flux of quantity $q_i$ in the $i$-direction through
the cell interface.   We use $\tilde{q}$ to indicate the uncorrected
quantity on the coarse grid and the sum is over the $r^{d-1}$ fine
cell interfaces in the perpendicular dimensions that share a face with
the coarse cell being corrected.  This correction is carried out for
all coarse cells that share a face with a refined cell (and are not
themselves covered with more refined cells).

\subsection{Hierarchy reconstruction}
\label{sec:hierarchy_reconstruction}

Since the addition of more highly refined grids is adaptive, the
conditions for refinement must be specified. Many different criteria
can be simultaneously specified for a given simulation, and these
refinement criteria are discussed in more detail in the following
section (Section~\ref{sec:refinement_criteria}).   Once all cells in a
given grid that require additional refinement have been flagged,
rectangular boundaries are determined that minimally encompass the
flagged cells.  This is done using an iterative machine-vision
technique, as described in \cite{Berger91}.

We first find the smallest region (or ``proto-subgrid") that covers
all of the refined cells on a grid.  We compute an efficiency
$\epsilon$ for this grid, defined as the ratio of flagged cells to
total cells.  If this efficiency is sufficiently large (typically over
30\%), or if the grid is small enough, then we adopt the region as a
new grid and move on to the next region (if any).  If the region is
not acceptable, then we split it into two using the following
algorithm.  First, we determine the longest dimension and compute the
signature along that dimension.  The signature is found by summing the
number of refined cells with the same index along the longest
dimension.  Edges are typically indicated by inflection points in this
signature, so we calculate the second derivative using a simple
three-point finite difference estimator and look for the maximum in
the second derivative, with ties going to the index closest to the
center.  If no inflection points are found along the longest axis, the
next-longest axis is searched.  If no inflection points are found, the
longest axis is cut in half.  The original region is then split into
two new regions, or proto-subgrids.  This process is then re-applied
to each of the new regions.

This algorithm is not necessarily optimal, but experience shows it
does a reasonable job of generating an efficient covering of the
refined cells.  Note that some cells are unnecessarily refined with
the SAMR approach, resulting in wasted resources (unlike in
cell-splitting AMR schemes).  However, it has a number of advantages.
First, traditional grid-based hydro methods can be simply added with
only minor modifications.  Second, it is easier to accommodate larger
stencils (as in PPM, which requires three boundary zones), and third,
it can be more efficient for some machines since there are more
predictable and linear memory accesses, which more efficiently use a
CPU's cache.

\subsection{Refinement criteria}
\label{sec:refinement_criteria}

In this section, we turn to the refinement criteria themselves.
Although generalized refinement criteria are available based on
estimators for the error, such as the Richardson method
\citep{AtkinsonHan2004}, we have focused generally on methods that use
physical estimators based on the assumption that the simulator has a
specific quantity or quantities that indicate the need for refinement.
The flagging methods currently built into \enzo\ are:

\begin{enumerate}

\item{\em Slope.}  If the normalized slope $(q_{i+1} - q_{i-1})/ (2
  q_i)$ of baryon quantities between adjacent cells is larger than a
  specified value (typically 0.3), then the cell is flagged.

\item{\em Baryon mass.}  This refinement criterion is designed to
  mimic a Lagrangian method in that it tries to keep a fixed mass
  resolution.  In this method, a cell is refined if the baryonic mass
  in the cell ($M_g = \rho\, (\Delta x)^d$) is larger than a specified value:

\begin{equation}
M_g > \rho_{\rm flag}\, (\Delta x_{\rm root})^d r^{\epsilon_l l},
\end{equation}

where $\rho_{\rm flag}$ is the equivalent density on the root grid
required for refinement, $\Delta x_{\rm root}$ is the root grid cell
spacing, $r$ is the refinement factor, $l$ is the level and
$\epsilon_l$ is a parameter that can be used to make the refinement
more aggressive ($\epsilon_l < 0$) or less aggressive ($\epsilon_l >
0$) than pure Lagrangian refinement --- in other words,
super-Lagrangian or sub-Lagrangian refinement.

\item{\em Shocks.}  In this method, we identify and refine cells that
  contain shocks using the following criteria:

\begin{eqnarray}
(p_{i+1} - p_{i-1})/\min(p_{i+1}, p_{i-1}) > 0.33,  \qquad
u_{i-1} - u_{i+1} > 0, \quad\text{and} \qquad
e_i / E_i > 0.1.  \nonumber
\end{eqnarray}

We note that the parameters $0.33$ and $0.1$ in the equations above
were determined empirically, and can be modified by the user to refine
more or less aggressively, as the situation warrants.

\item{\em Particle mass.}  This refinement criterion is precisely the
  same as the baryon mass criterion except that it uses the gridded
  (cloud-in-cell, or CIC) particle density.  In general, this criteria
  is used in cosmological simulations, and thus refines on dark matter
  and stellar density.

\item{\em Jeans length.}  Following \cite{Truelove98}, the code can be
  forced to refine the Jeans length by a fixed number of cells, or,
  more specifically, whenever the following criterion is not met:

\begin{equation}
\Delta x < \left( \frac{\pi k_B T}{N_J^2 G \rho \mh} \right)^{1/2},
\end{equation}
where $J_J$ is the required number of cells per Jeans length (4 by default).

\item{\em Cooling time.}  Often in cosmological simulations, cooling
  is a rapid process and we do not necessarily want to be limited by
  the cooling time.  However, in some cases it is desirable to resolve
  the cooling process, in which case we flag cells that have a cooling
  time shorter than their sound crossing time.  The cooling time is
  given by $t_{\rm cool} = 1.5 k_BT / (n\, \Lambda(T))$ and the sound
  crossing time is $t_{\rm cross} = \Delta x / c_s$, where $c_s$ is
  the sound speed.

\item{\em Must-refine particles.}  In order to allow for refinement in
  moving regions, it is possible to generate ``must-refine" particles
  that ensure local refinement of the 8 nearest cells to a specific
  level.  The particles feel gravitational forces and their
  trajectories are followed as for any other collisionless particle.
  This feature can be used to ensure that a specific Lagrangian region
  stays refined to a specified level of resolution.

\item{\em Shear.} As described in \citet{Kritsuk06}, turbulence
  calculations can be successfully modeled with AMR if the shear is
  used as a refinement parameter; in particular, refinement occurs
  when the finite difference version of the following inequality
  holds:

\begin{equation}
\sum_i \sum_j \left( \frac{\partial v_i}{\partial x_j} \right)^2 -  \sum_i \left( \frac{\partial v_i}{\partial x_i} \right)^2
> \epsilon_s \frac{c_s^2}{\Delta x^2},
\end{equation}
where $\epsilon_s$ is a dimensionless parameter.

\item{\em Optical depth.} This refinement criterion, typically used
for simulations using radiation transport, flags cells for refinement
if they have an \ion{H}{1} optical depth lower than unity.  The
optical depth of a cell is estimated to be:

\begin{equation}
\tau = \sigma_{\rm HI} n_{\rm HI} \Delta x 
\end{equation}

where $\sigma_{\rm HI}$ is the neutral atomic hydrogen absorption
cross section, $n_{\rm HI}$ is the proper \ion{H}{1} number density,
and $\Delta x$ is the cell width (which is assumed to be the same
along each axis).  If $\tau > 1$, the cell is refined further, up to
the maximum level of refinement.

\item{\em Resistive length.}  In a way similar to the Jeans length
  criterion described above, the code can be forced to locally resolve
  the resistive length scale by a minimum of a fixed number of cells.
  This is done by ensuring that the resistive length scale, defined as

\begin{equation}
L_R = |B| / |\curl B|
\end{equation}

is always refined by at least N$_R$ cells locally (i.e.~we refine a
given cell further if $L_R \leq N_R \Delta x$).  Note that \enzo\ does
not actually solve the resistive MHD equations -- this is the
``characteristic'' local resistive length, rather than one calculated
within the MHD solver.

\item{\em Must refine region.}  This refinement criterion is used to
  ensure that all cells in a given subvolume of the simulation are
  refined to {\it at least} some minimum level of refinement.  The
  subvolume is user-specified, and may evolve over time (e.g., to
  follow a structure of interest that is traveling through the
  simulation volume).

\item{\em Metallicity.}  This refinement criterion ensures that all
  cells with a metallicity (defined as $Z \equiv \rho_{Z} / \rho_{b} /
  0.022$, where $\rho_{Z}$ is the density of metals in the cell,
  $\rho_{b}$ is the total baryon density (including metals), and 0.022
  is the solar mass fraction of metals) above a user-defined value is
  refined to {\it at least} some minimum level of refinement. 

\item{\em Second Derivative.}  This is similar to the refinement criteria used
  in Paramesh and was originally developed by \cite{1987nasa.reptQ....L}.  This
  method calculates the second derivative of a user-defined field normalized by
  the first derivative, and sets a threshold for refinement.  More background
  information can be found in the FLASH4 user guide\footnote{\url{http://flash.uchicago.edu/site/flashcode/user\_support/flash4\_ug/}},
  but we reproduce the relevant equation here for completeness. The
  multi-dimensional version of the criteria takes in to account all
  cross-derivatives, and is expressed by

 \begin{equation}
   S = \left\{ \frac{\sum_{ij}\left( \frac{\partial^2f}{\partial x_i \partial x_j} \right)^2}{\sum_{ij}\left[ \frac{1}{2\Delta x_j}\left( \abs{\frac{\partial f}{\partial x_i}}_{x_i+\Delta x_i} + \abs{\frac{\partial f}{\partial x_i}}_{x_i-\Delta x_i} \right) + \epsilon \frac{\abs{\bar{f_{ij}}}}{\Delta x_i \Delta x_j} \right]^2} \right\}^{1/2},
   \label{eq:second-deriv_analytical}
 \end{equation}

  where $f$ is the user-defined field, $\abs{\bar{f_{ij}}}$ is an average of
  $f$ over cells in the $i$ and $j$ dimensions, and $\epsilon$ is
  a user-controlled value that provides a method of ignoring small
  fluctuations, set by default to $10^{-2}$.  If $S$ is larger than a specified
  threshold (bounded by [0,1]) a cell is marked for refinement. This threshold
  is chosen at runtime and can vary for each of the fields under consideration.

\end{enumerate}

Finally, it is also possible to specify a rectangular region that is
statically refined to a certain level.  These static refined regions
are in place throughout the simulation regardless of the grid
quantities and can be used for testing purposes, or for
multiply-zoomed initial conditions (in cosmology simulations, for
example) that would not otherwise be refined.

%% file: numerical-ppm.tex
In this section, we describe the four solvers that we have implemented
for solving the fluid equations.  We describe the PPM method in
considerably more detail than the other methods in part because its
implementation in Enzo has not previously been described, but mostly
because it introduces many of the ideas and methods used for the
MUSCL-Dedner and MHDCT schemes (including expansion terms,
reconstruction, Riemann solvers, and dual energy formalism).

\subsection{Hydrodynamics: The PPM method}
\label{sec.hydro.ppm}

One (purely) hydrodynamic method used in \enzo\ is closely based on
the piecewise parabolic method (PPM) of \citet{1984JCoPh..54..174C} --
henceforth referred to as CW84 -- which has been modified for the
study of cosmological fluid flows.  CW84 describe two variants on this
method: Lagrangian Remap and Direct Eulerian.  We use the Direct
Eulerian version in \enzo, which is better suited for AMR simulations.
The Lagrangian Remap version has previously been adapted for
cosmological use as described in \citet{1995CoPhC..89..149B}, and the
implementation we use here is closely based on that work.

The first term on the right hand side of equations~(\ref{eq:momentum})
and (\ref{eq:total_energy}) comes from our choice of comoving
coordinates (a similar term also appears in
equation~(\ref{eq:dm_velocity}), the velocity relation for the dark
matter particles).  A similar term does not appear in the mass
conservation equation (\ref{eq:mass}) because of the comoving density
definition.  We note that these expansion terms could be eliminated
entirely by the proper choice of variables (including time), although
we have not done so here as they do not constitute a major source of
error.  We solve these terms using the same technique -- they are
split from the rest of the terms and solved using an (implicit)
time-centered method, which is straightforward as there are no spatial
gradients.  Note that we use this method for all four of the hydro
solvers.

The remainder of the terms in the fluid equations involve derivatives.
Because we are interested in phenomena with no special geometry, we
will restrict discussion to Cartesian coordinates.  We can
dimensionally split the equations and rewrite the one-dimensional
(Eulerian) versions of
equations~(\ref{eq:mass})--(\ref{eq:total_energy}) without expansion
terms in conservative form as,
\begin{eqnarray}
\frac{\partial \rho}{\partial t}  + \frac{1}{a} \frac{\partial \rho v }{\partial x}    & = &  
     0 \label{eq:mass1d} \\
\frac{\partial \rho v}{\partial t}  + \frac{1}{a} \frac{\partial \rho v^2}{\partial x}   + 
      \frac{1}{a} \frac{\partial p}{ \partial x} & = & 
      \rho \frac{g}{a}  \label{eq:momentum1d}  \\
\frac{\partial \rho E}{\partial t}  + \frac{1}{a} \frac{\partial \rho v E}{\partial x}  & =  &
      \rho v \frac{g}{a} \label{eq:total_energy1d}
\end{eqnarray}
Here $x$ and $v$ refer to the one dimensional comoving position and 
peculiar velocity of the baryonic gas, and $g$ is the acceleration at the cell center.
These equations are now in a form that can be solved by the split PPM scheme.


We now restrict ourselves to the solution of equations
(\ref{eq:mass1d})--(\ref{eq:total_energy1d}) in one dimension.  In the
Direct Eulerian version of PPM, this is accomplished by a three-step
process.  First, we compute effective left and right states at each
grid boundary by constructing a piecewise parabolic description of the
primative variables ($\rho$, $u$ and $E$) and then averaging over the
regions corresponding to the distance each characteristic wave can
travel ($u$, $u-c_s$ and $u+c_s$, where $c_s$ is the sound speed).
Second, a Riemann problem is solved using these effective left and
right states, and finally the fluxes are computed based on the
solution to this Riemann problem and the conserved quantities are
updated.  This is described in detail in CW84, but we will briefly
outline the procedure here both for completeness and to put the
changes we will make in context.

The Eulerian difference equations are:
\newcommand{\avp}[1]{\overline{#1}_{j+1/2}}
\newcommand{\avm}[1]{\overline{#1}_{j-1/2}}
\begin{equation}
\rho_j^{n+1} = \rho_j^{n} + \Delta t \left( 
                      \frac{ \avp{\rho}\avp{v} -  \avm{\rho}\avm{v} } {\Delta x_j}
            \right)
     \label{eq:mass_diff}
\end{equation}
\begin{equation}
\rho_j^{n+1} v_{j}^{n+1} =
       \rho_j^{n+1} v_j^n  + \Delta t \left(
          \frac{ \avp{\rho} \avp{v}^2 - \avm{\rho} \avm{v}^2 + \avp{p} - \avm{p}} {\Delta x_j}
             \right)
     + \frac{ \Delta t }{2} g_j^{n+1/2} (\rho_j^n + \rho_j^{n+1}), 
     \label{eq:momentum_diff}
\end{equation}
\begin{eqnarray}
\rho_j^{n+1} E_j^{n+1}  = 
       \rho_j^n E_j^n  & + & \Delta t  \left(
            \frac{  \avp{\rho} \avp{v} \avp{E}  - \avm{\rho} \avm{v} \avm{E}  +
                       \avp{v} \avp{p}    - \avm{v} \avm{p} } {\Delta x_j}
             \right) \nonumber \\
         & & + \frac{ \Delta t }{2} g_j^{n+1/2} (\rho_j^n v_j^n + \rho_j^{n+1} v_j^{n+1} ).
     \label{eq:energy_diff}
\end{eqnarray}

We have used subscripts to indicate zone-centered ($j$) and
face-centered ($j+1/2$) quantities, while superscripts refer to
position in time.  The cell width is $\Delta x_j$.  Although they have
been discretized in space, the accuracy of the updates depend on how
well we can compute the fluxes into and out of the cell during $\Delta
t$.  This in turn depends on our ability to compute the time-averaged
values of $p$, $\rho$ and $v$ at the cell interfaces, denoted here by
$\overline{p}_{j\pm 1/2}$, $\overline{\rho}_{j\pm 1/2}$, and
$\overline{v}_{j\pm 1/2}$.  We now describe the steps required to
compute these quantities.

We first construct monotonic piecewise parabolic (third-order)
interpolations in one dimension for each of $p$, $\rho$, and $v$.  The
pressure is determined from equation~(\ref{eq:eq_of_state}), the
equation of state.  The interpolation formula for some quantity $q$ is
given by:
\begin{eqnarray}
q_j(x) & = &  q_{L,j} + \tilde{x}(\Delta q_j + q_{6,j}(1-\tilde{x})), \\
\tilde{x}      & \equiv & {x - x_{j-1/2} \over \Delta x_j}, \qquad
             x_{j-1/2} \leq x \leq x_{j+1/2}. \nonumber
\end{eqnarray}
This is equation~(1.4) of CW84 (in the spatial rather than mass
coordinate, as is appropriate for the direct Eulerian approach).  The
quantities $q_{L,j}$, $\Delta q_j$, and $q_{6,j}$ can be viewed simply
as interpolation constants; however, they also have more intuitive
meanings.  For example, $q_{L,j}$ is the value of $q$ at the left edge
of zone j, while $\Delta q_j$ and $q_{6,j}$ are analogous to the slope
and first-order correction to the slope of $q$ (see CW84 for a
complete discussion):
\begin{equation}
\Delta q_j \equiv q_{R,j} - q_{L,j} \qquad 
q_{6,j}    \equiv 6\left[q_j - 1/2\left(q_{L,j} + q_{R,j}\right)\right].
\end{equation}

We have reduced the problem to finding $q_{L,j}$ and $q_{R,j}$.  While
this is simple in principle, it is complicated somewhat by the
requirement that these values be of sufficient accuracy and that the
resulting distribution be monotonic.  That is, no new maxima or minima
are introduced.  The resulting formulae are straightforward but
complicated and are not reproduced here, but see Equations 1.7 to 1.10
of CW84.  We also optionally allow steepening as described in that
reference.

Once we have the reconstruction, the primary quantities ($p$, $\rho$,
$v$ and $E$) are averaged over the domains corresponding to the three
characteristics $u-c$, $u$, or $u+c$ (where $c$ is the sound speed in
a cell).  The linearized gas dynamics equations are then used to
compute second-order accurate left and right states that take into
account the multiple wave families.  This process is described in CW84
and we use their equations (3.6) and (3.7).

With these effective states, an approximation to the Riemann problem
is found (see below for more detail about the Riemann solvers used),
producing estimates for $\overline{p}_{j\pm 1/2}$,
$\overline{\rho}_{j\pm1/2}$, and $\overline{v}_{j\pm 1/2}$ that are
third-order accurate in space and second-order accurate in time.
These are then used to solve the difference Equations
(\ref{eq:mass_diff})--(\ref{eq:energy_diff}) for $\rho^{n+1}$,
$v^{n+1}$, and $E^{n+1}$.

We include an optional diffusive flux (and flattening for the
parabolic curves) that can improve the solution in some cases.  Our
implementation follows that in the appendix of CW84.

In addition, as discussed earlier, the three-dimensional scheme is
achieved by operator-splitting and repeating the above procedure in
the other two orthogonal directions.  The transverse velocities and
any additional passive quantities are naturally and easily added to
this system (see Equation 3.6 of CW84).  

We note that the acceleration required in
Equation~(\ref{eq:momentum_diff}) is actually the acceleration felt by
the entire zone and not just at the zone center.  Therefore, it is
possible to find the mass-weighted average acceleration over the zone
by expanding the density and acceleration distributions and retaining
all terms up to second-order in $\Delta x$
\citep{1995CoPhC..89..149B}, although we find that the potential is so
slowly varying that this is unnecessary.


\subsubsection{The dual energy formulation for  very high Mach flows} 

The system described thus far works well for gravitating systems with
reasonable Mach numbers ($<100$) as long as the structures are well
resolved.  This section and the next detail changes that are required
to correctly account for situations where one or both of these
requirements are not met.

Large, hypersonic bulk flows appear to be very common in cosmological
simulations and they present a problem because of the high ratio of
kinetic energy $E_k$ to gas internal energy $e$, which can reach as
high as $10^8$.  Inverted, we see that the internal energy consists of
an extremely small portion of the total energy.  In such a situation,
the pressure, proportional to $E - E_{k}$, is the small difference
between two large numbers: a disastrous numerical situation.  This is
not as large a problem as it may at first appear because it only
occurs when the pressure is negligibly small.  Therefore, even if we
suffer large errors in the pressure distribution in these regions, the
dynamics and total energy budget of the flow will remain unaffected.
Nevertheless, if the temperature distribution is required for other
reasons (e.g., for calculating radiative processes), a remedy is
required.

To overcome this, we also solve the internal energy equation:
\begin{equation}
 \frac{\partial e}{\partial t} 
           + \frac{1}{a} \vecv \cdot \grad e
         = - 3 \frac{\dot{a}}{a} \frac{p}{\rho}
           - \frac{p}{a\rho} \div \vecv
\end{equation}
in comoving coordinates.  The structure is similar to the total energy
equation; the second term on the left hand side represents transport,
while the first term on the right is due to expansion of the
coordinate system.  It is differenced (again, in Eulerian form without
the expansion term) as,
\begin{equation}
\rho_j^{n+1} e_j^{n+1}  = 
       \rho_j^n e_j^n   +  \Delta t  \left(
            \frac{  \avp{\rho} \avp{v} \avp{e}  - \avm{\rho} \avm{v} \avm{e}} {\Delta x_j} \right)
            - \Delta t \ p_j^{n} \left( \frac{ \avp{v} - \avm{v}} {\Delta x_j} 
                      \right)
                  \label{eq:gasenergy_diff}
\end{equation}
Note that because of the structure of this equation, it is not in
flux-conservative form.  In particular, the pressure is evaluated at
the cell center.  Unfortunately, time-centering of this pressure has
proved difficult to do without generating large errors in the internal
energy and so we leave the pressure at the old time in this difference
equation.  This leads to some spreading of shocks; however, we note
that this equation is only used in hypersonic flows.

It is necessary, however, to conserve the total energy so that the
conversion of kinetic to thermal energy is performed properly.  We
must therefore combine the two formulations without allowing the
separately-advected internal energy $e$ to play a role in the gas
dynamics.  This is done by carrying both terms through the simulation
and using the total energy $E$ for hydrodynamic routines and the
internal energy $e$ when the temperature profile is required.  One way
to view this procedure is to treat $e$ as enhanced precision (extra
digits) for $E$ that automatically `floats' to where it is needed.  We
only require that they be kept synchronized when the two levels of
precision overlap.

When the pressure is required solely for dynamic purposes, the 
selection criterion 
operates on a cell by cell basis using,
\begin{equation}
p = \cases{ \rho(\gamma - 1)(E - \vecv^2/2),& 
                  $  (E - \vecv^2/2)/E > {\eta}_1 $; \cr
            \rho(\gamma -1)e,&
                  $  (E - \vecv^2/2)/E < {\eta}_1 $. \cr}
\end{equation}

It should be stressed that as long as the parameter ${\eta}_1$ is
small enough the dual energy method {\it will have no dynamical
effect}.  We use ${\eta}_1 = 10^{-3}$, which is consistent with the
truncation error of the scheme for grid sizes that are typically used
in our simulations.  We are now free to select the method by which the
internal energy field variable $e$ is updated so that it will not
become contaminated with errors advected by the total energy
formulation but still give the correct distribution in shocked
regions.  Since we are concerned with the advection of errors, the
selection criterion must look at each cell's local neighbourhood.  In
one dimension, this is done with,

\begin{equation}
e = \cases{ (E - \vecv^2/2),& $\rho(E - \vecv^2/2)/
    \max ({\rho}_{j-1}E_{j-1},{\rho}_j E_j,{\rho}_{j+1}E_{j+1}) > {\eta}_2$,\cr
            e,& $\rho(E - \vecv^2/2)/
    \max ({\rho}_{j-1}E_{j-1},{\rho}_j E_j,{\rho}_{j+1}E_{j+1}) < {\eta}_2$.\cr}
    \label{eq:dualselect}
\end{equation}

Thus, ${\eta}_2$ determines when the synchronization (of $e$ with $E$)
occurs.  Too high a value may mask relatively weak shocks, while
spurious heating (via contamination) may occur if it is set too low.
After some experimentation, we have chosen ${\eta}_2 = 0.1$, a
somewhat conservative value.  This scheme is optional and is generally
only required in large-scale cosmological simulations where the gas
cools due to the expansion of the universe but large bulk flows
develop due to the formation of structure.

We note that others have independently developed a similar but
distinct scheme for dealing with this problem, which is endemic to
methods adopting the total energy equation.  In \citet{TVD93}, the two
variables adopted are total energy and entropy (rather than total
energy and thermal energy), with an analogous scheme for choosing
which variable to employ.

\subsubsection{Riemann Solvers and Fallback}
\label{sec.riemann}

In this section, we describe the methods we adopt to solve the Riemann
problem, which is generally required to compute the fluxes in any
Godunov-based scheme.  This section therefore applies to all three of
our Godunov-based schemes.  The Riemann problem we are solving
involves two constant states separated by a single discontinuity at
$t=0$.  The subsequent evolution has an exact analytic solution.  This
solution is described in detail in many texts on computational fluid
dynamics \citep[e.g.,][]{toro-1997}.  In brief, there are three waves
that propagate away from the initial discontinuity.  The central wave,
characterized by a density jump but not a pressure jump, is called the
contact discontinuity.  The waves traveling to the left and right of
the contact discontinuity can be either shocks, if characteristics
converge on the wave front, or rarefaction fans if characteristics
diverge.

While there exists an exact solution to this problem, finding it is
expensive.  There are four possible combination of left- and right-
traveling shocks and rarefactions, only one of which is fully
consistent with the initial conditions.  Once the correct physical
state is determined, the pressure in the central region can only be
computed by finding the root to an algebraic equation, which is
necessarily an iterative process.  Thus a series of \emph{approximate}
Riemann solvers are typically used.  There are four approximate
Riemann solvers in \enzo: two-shock \citep{toro-1997}, Harten-Lax-van
Leer \citep[HLL,][]{toro-1997}, HLL with a contact discontinuity
\citep[HLLC,][]{toro-1997}, and HLL with multiple discontinuities
\citep[HLLD,][]{Miyoshi05}.  Two-shock is used only with the PPM
method.  HLL and HLLC are used with PPM, MUSCL (both with and without
MHD) and MHD-CT.  HLLD is exclusively an MHD solver, and works with
both the MUSCL and MHD-CT methods.

The only approximation that two-shock makes is that both left- and
right-moving waves are shocks.  This solution still requires an
iterative method for finding the pressure in between the two waves.
The HLL method alleviates this iteration by assuming that there is no
central contact discontinuity, and the signal speed in the central
region is approximated by an average over the left- and right- moving
waves.  This method is significantly faster than the two-shock method,
but it is also quite a bit more dissipative.  The HLLC is a three-wave
method that improves upon the HLL method by also including the third
wave, the contact discontinuity.

For the MHD equations, there are seven waves instead of three.  This
makes the exact solution to the Riemann problem quite a bit more
expensive.  Both the HLL and HLLC approximations can be formulated for
the MHD equations, and are employed in both the Dedner and CT solvers
in \enzo.  The HLLD solver includes two of the additional waves, the
rotational discontinuities, making it a five-wave solver.

On rare occasions, high-order solutions can cause negative densities
or energies.  Both our PPM and MUSCL solvers employ a Riemann solver
fallback mechanism \citep{Lemaster09}.  If a negative density is found
at a particular interface, the more diffusive HLL Riemann solver is
used to compute the fluxes associated with that cell, and the flux
update is repeated.

%% file: numerical-mhd-dedner.tex
\subsection{Hydrodynamics and Magnetohydrodynamics: MUSCL with Dedner cleaning}
\label{sec.num.hydro-muscl}

The second method we describe is a MUSCL-based solver than can be used
in both HD and MHD modes.  The description here will be very brief
both because the ideas are similar to those described in the previous
section, and because this implementation has previously been described
in more detail elsewhere \citep{WangAbelZhang08, WangAbel09}.

Much like the PPM solver, we have three basic steps: the first is reconstruction of the variables, the second is a solution of the Riemann problem, and the third is is updating the conserved quanties with the fluxes as written above. For the reconstruction scheme we have implemented only the simple piecewise linear reconstruction \citep{1979JCoPh..32..101V, 1985JCoPh..59..264C}, with options for both primitive and conservative variable reconstruction. The available Riemann solvers are HLL, HLLC, and HLLD, as described earlier.

To more clearly describe the Dedner divergence cleaning modifications,
we write the equations of compressible inviscid hydrodynamics in the
form of conservation laws as,

\begin{equation}
 \frac{\partial{U}}{\partial{t}} +
 \frac{\partial{F^x}}{\partial{x}} + \frac{\partial{F^y}}{\partial{y}} + \frac{\partial{F^z}}{\partial{z}}= 0, \label{hydro}
\end{equation}

The conserved variable $U$ is given by

\begin{equation}
 U = (\rho, \rho v_x, \rho v_y, \rho v_z, \rho E)^{T},
\end{equation} 

where $\rho$ is density, $v_i$ are the three components of velocity
for $i={x,y,z}$, $E=v^2/2 + e$ denotes the specific total energy and $e$ the
specific internal energy (note that in this section only, we use the specific
energy).

For the generalized Lagrange multiplier (GLM) formulation of the MHD
equations \citep{2002JCoPh.175..645D}, we consider these 
conserved variables

\begin{equation}
 U = (\rho, \rho v_x, \rho v_y, \rho v_z, \rho E+B^2/2, B_x, B_y, B_z, \psi)^{T},
\end{equation} 

where $B_i$ with $i={x,y,z}$ are the three components of magnetic
fields and $\psi$ is the additional scalar field introduced in the GLM
formulation for the divergence cleaning.  The fluxes then are

\begin{eqnarray}
 F^x &=& (\rho v_x, \rho v_x^2+p+B^2/2-B_x^2, \rho v_yv_x-B_yB_x, \cr
 && \rho v_zv_x-B_zB_x, \rho ({v^2\over2} + h)v_x+B^2v_x-B_xB\cdot v, \cr
&& \psi, v_xB_y-v_yB_x, -v_zB_x+v_xB_z, c_h^2B_x)^T,\\
 F^y &=& (\rho v_y, \rho v_xv_y-B_xB_y, \rho v_y^2+p+B^2/2-B_y^2, \cr
 && \rho v_zv_y-B_zB_y, \rho ({v^2\over2} + h)v_y+B^2v_y-B_yB\cdot v, \cr
 && v_yB_z-v_zB_y,\psi,-v_xB_y+v_yB_x, c_h^2B_y)^{T}, \\
 F^z &=& (\rho v_z, \rho v_xv_z-B_xB_z, \rho v_yv_z-B_yB_z, \rho v_z^2+p+B^2/2-B_z^2, \cr
 && \rho ({v^2\over2} + h)v_z+B^2v_z-B_zB\cdot v, \cr
    &&  -v_yB_z+v_zB_y, v_zB_x-v_xB_z,\psi, c_h^2B_z)^{T},
\end{eqnarray}

where $c_h$ is a constant controlling the propagation speed and
damping rate of $\div B$, and $h=e+p/\rho$ denotes the specific
enthalpy.  All quantities are cell-centered.

The method is dimensionally un-split in that the fluxes are computed
for all dimensions first and the conserved quantities are updated in
one step, in contrast to the Strang splitting employed in the other
fluid methods described in this paper.  Also unlike the other schemes,
time-integration is done with a second-order Runge-Kutta scheme
\citep{1988JCoPh..77..439S}.

Finally, we note that for cosmological simulations, this solver uses a
slightly different definition of the magnetic field than used in the
rest of the paper.  In particular, the field is defined as $\vecB =
a^{3/2} \vecB^{\prime}$ (where $\vecB^{\prime}$ is the proper field
strength).  This adds a source term of $-\dot{a}/(2a) \vecB$ on the
right-hand side of Equation~(\ref{eq:induction}) and removes the
factor of $a$ in the $\vecB$ term in the energy equation
(Eq.~\ref{eq:total_energy_def}) and in the definition of the isotropic
pressure $p^*$.

%% file: numerical-mhd-ct.tex
\subsection{Magnetohydrodynamics: Constrained transport}
\label{sec.num.mhd-ct}
\def\Bvec{{\bf B}}
\def\Bf{Bf}
\def\Bc{Bc}
\def\Evec{{\bf E}}
\def\Divb{\ensuremath{\div \Bvec}}

The third solver we describe is an MHD method developed by
\citet{Collins10}.  Since a full description and suite of test
problems can be found in that reference, we only describe the method
briefly here.

The divergence of the magnetic field, \Divb, is identically zero in
reality due to the fact that the evolution of the magnetic field is
the curl of a vector, and the divergence of the curl of a vector is
identically zero.  The Constrained Transport (CT) method
\citep{Evans88, Balsara99} for magnetohydrodynamic (MHD) evolution
employs this same vector property to evolve the magnetic field in a
manner that preserves $\Divb=0$.  The electric field is computed using
the fluxes from the Riemann solver.  The curl of that electric field
is then used to update the magnetic field.  The advantage of this
method is that it preserves \Divb\ to machine precision.  The primary
drawback is increased algorithm complexity.  Note also that since only
the update of the magnetic field is divergence free, any monopoles
created by other numerical sources (such as ill-chosen initial
conditions) persist.

The base Godunov method, described in \citet{Li08a}, uses spatially
and temporally second-order reconstruction (both MUSCL-Hancock and
Piecewise Linear Method), and a selection of Riemann solvers including
HLLC and HLLD \citep{Mignone07}, as described earlier.  The
constrained transport methods are the first-order method described by
\citet{Balsara99} and the second-order methods described in
\citet{Gardiner05}.  The AMR machinery is described by
\citet{Balsara01} and \citet{Collins10}.

The increased complexity of the constrained transport scheme comes in
the form of area-averaged face and length-averaged edge-centered
variables, while the rest of \enzo\ employs predominantly
volume-averaged cell-centered variables.  The magnetic field is
represented by both a face-centered field, B$_f$, and a cell-centered
field, B$_c$.  The electric field is edge-centered.  The magnetic
field is updated in four steps: first, the Riemann problem is solved
in the traditional manner, using the cell-centered field; second, an
edge-centered electric field is computed using the fluxes from the
Riemann solver; third, the curl of that electric field is used to
update the face-centered field; finally, the cell-centered magnetic
field is updated with an average of the face-centered field.

Divergence-free AMR is somewhat more complex than the AMR employed
elsewhere in \enzo.  First, the interpolation must be constrained to
be divergence free.  Thus, all three face-centered field components
are interpolated in concert.  Second, any magnetic field information
from the previous timestep must be included in the interpolation,
making the interpolation more complex than the simple parent-child
relation used for other fields.  Third, the flux correction involves
more possible grid relations than traditional AMR.  In order to
circumvent this last complexity, the electric field is projected from
fine grids to parent grids (rather than the magnetic field), and is
then used to re-update the parent magnetic field.  This is described
in detail in \citet{Balsara99} and \citet{Collins10}.

The dual energy formalism has also been incorporated in two possible
ways -- one that uses internal energy, as described in Section
\ref{sec.hydro.ppm}, and one that uses entropy \cite{TVD93,
Collins10}.

%% file: numerical-zeus.tex
\subsection{Hydrodynamics: The \zeus\ method}
\label{sec.hydro.zeus}

As an alternative to the previous Godunov methods, \enzo\ also
includes an implementation of the finite difference hydrodynamic
algorithm employed in the compressible magnetohydrodynamics code
\zeus\ \citep{Stone92a, Stone92b}.  Fluid transport is solved on a
Cartesian grid using the upwind, monotonic advection scheme of
\citet{1977JCoPh..23..276V} within a multistep (operator-split)
solution procedure that is fully explicit in time.  This method is
formally second-order accurate in space but first-order accurate in
time.  
 
As discussed in the section describing the Piecewise Parabolic Method
(Section~\ref{sec.hydro.ppm}), operator-split methods break the
solution of the hydrodynamic equations into parts, with each part
representing a single term in the equations.  Each part is evaluated
successively using the results preceding it.  In this method, in
addition to operator-splitting the expansion terms (i.e., those terms
in Eqs.~(\ref{eq:mass}) -- (\ref{eq:total_energy}) that depend on
$\dot{a}$), we divide the remaining terms into \emph{source} and
\emph{transport} steps.  The terms to be solved in the source step are
those on the right-hand side of eqs.~(\ref{eq:mass}) --
(\ref{eq:total_energy}), while the transport terms are on the
left-hand side of these equations and are responsible for the
advection of mass, momentum and energy across the grid.

The \zeus\ method uses a von Neumann-Richtmyer artificial viscosity to
smooth shock discontinuities that may appear in fluid flows and can
cause a breakdown of the finite difference equations.  The artificial
viscosity term is added in the source terms as:
\begin{eqnarray}
\rho \frac{\partial\vecv}{\partial t} &=& - \grad p - \rho \grad \phi 
- \div \textbf{Q} \\
\frac{\partial e}{\partial t} &=& -p \div \vecv - \textbf{Q} : \grad \vecv, 
\end{eqnarray}
Here \textbf{Q} is the artificial viscosity stress tensor, which we
take to be diagonal with on-axis terms given by $l^2 \rho (\partial v
/ \partial x)^2$ as proposed by von Neumann \& Richtmyer.  The length
scale $l$ determines the width of shocks and is typically a few times
the cell spacing.

In our Cartesian coordinate system, the finite difference version of
these equations is particularly simple, although there is one
important complication.  In the \zeus\ formalism, the velocity is a
face-centered quantity -- that is, the velocity is recorded on a grid
that is staggered as compared to the density, pressure and energy,
which are at the cell center.  Therefore we must remember that $v_j$
is at position $x_{j-1/2}$ (we use this notation, rather than writing
$v_{j+1/2}$ both to match the original \zeus\ paper and also to make
it easier to compare these equations to what is actually in the code).  

As in the original \zeus\ paper, the source terms are added in three
steps. First we add the pressure and gravity forces:
\begin{equation}
v_j^{n+a}  =  v_j^n - \frac{\Delta t}{\Delta x_j} \frac{p^n_j - p^n_{j-1}} {(\rho^n_j + \rho^n_{j-1})/2}.
\end{equation}
Partial updates are denoted by the $n+a$, $n+b$ notation.  We show
updates in one dimension as the extension to the multi-dimensional
case is straightforward (note that all dimensions are carried out for
each substep before progressing to the next substep). We then add the
artificial viscosity:
\begin{eqnarray}
v_j^{n+b} & = & v_j^{n+a} - \frac{\Delta t}{\Delta x_j} 
                             \frac{q^{n+a}_j - q^{n+a}_{j-1}} {(\rho^n_j + \rho^n_{j-1})/2} \\
e_j^{n+b} & = & e_j^n - \frac{\Delta t}{\Delta x_j} q^{n+a}_j (v^{n+a}_{j+1} - v^{n+a}_{j}).
\end{eqnarray}
The artificial viscosity coefficient $q_j$ is given by:
\begin{equation}
q_j = \left\{ \begin{array}{ll}
              Q_{\rm AV} \rho_j (v_{j+1} - v_j)^2 \quad & \rm{if}~(v_{j+1} - v_j) < 0 \\
               0 & \rm{otherwise}
               \end{array} \right.
\end{equation}
where $Q_{\rm AV}$ is a constant with a typical value of 2. We refer
the interested reader to \citet{Stone92a} and
\citet{1994ApJ...429..434A} for more details.  We also include the
option (turned off by default) of adding a linear artificial viscosity
as suggested in the \zeus\ paper for stagnant flow regions.  This is
given by
\begin{equation}
q_{{\rm lin},j} = Q_{\rm LIN} \rho c_j (v_{j+1} - v_{j})
\end{equation}
where $c_j^2 = \gamma p/\rho$ is the adiabatic sounds speed.

Finally, the third source step is the compression term and is given by
\begin{equation}
e^{n+c}_j = e^{n+b}_j \left( \frac{1 - (\Delta t/2) (\gamma - 1) (\div \vecv)_j }
                           {1 + (\Delta t/2) (\gamma - 1) (\div \vecv)_j } \right)
\end{equation}
We have used the notation $(\div \vecv)_j$ to indicate the
(potentially) multi-dimensional velocity divergence evaluated at the
cell center position $x_j$.  This equation differs from the previous
ones in that in the multi-dimensional case, still only one finite
difference equation is evaluated, but the divergence becomes
multi-dimensional.

We next examine the transport step, which is conservative.  Once
again, we dimensionally split the equations and present only the
one-dimensional version.  The finite difference equations actually
solved are:

\begin{equation}
\rho_j^{n+d} = \rho_j^{n} - \frac{\Delta t}{\Delta x} (v^{n+c}_{j+1/2} \rho^{*}_{j+1/2} - v^{n+c}_{j-1/2} \rho^{*}_{j-1/2} )
\end{equation}

Here $\rho^*_j$ is the correctly upwinded value of $\rho$ evaluated at
the cell-face corresponding to $v_j$, making $\rho^*_j v_j$ the mass
flux at the cell boundary and guaranteeing mass conservation.   This
requires interpolating each cell-centered quantity to the cell edge.
As recommended in \citet{Stone92a}, we use the second-order van Leer
scheme, which uses piecewise linear functions.  These are given by
Equations (48) and (49) of \citet{Stone92a}.  The transport steps for
the other variables are similar.  Note that we advect the specific
energy and specific momenta using the mass flux, as dictated by the
principle of consistent transport.  This requires appropriate
averaging for the momenta in the perpendicular directions as outlined
in equations (57)-(72) of \citet{Stone92a}.

A limitation of a technique that uses an artificial viscosity is that,
while the correct Rankine-Hugoniot jump conditions are achieved,
shocks are broadened over 3-4 mesh cells. This may cause unphysical
pre-heating of gas upstream of the shock wave, as discussed in
\citet{1994ApJ...429..434A}.  On the other hand, it is much more
robust than PPM and is easy to add additional physics.  We also note
that this method solves only the internal energy equation rather than
total energy, so the dual energy formulation discussed in
Section~\ref{sec.hydro.ppm} is unnecessary.

%% file: numerical-gravity.tex
\subsection{Gravity}
\label{sec.gravity}

Solving for the accelerations of the cells and particles on the grid
due to self-gravity involves three steps: (i) computing the total
gravitating mass, (ii) solving for the gravitational potential field
with the appropriate boundary conditions, and (iii) differencing the
potential to get the acceleration, and, if necessary, interpolating
the acceleration back to the particles. These steps are described in
detail below.

First, the massive (dark matter and star) particles are distributed
onto the grids using the second-order cloud-in-cell (CIC)
interpolation technique \citep{Hockney88} to form a
spatially-discretized density field $\rho_{\rm DM}$.  During the CIC
interpolation, particle positions are (temporarily) advanced by $0.5
v^n \Delta t$ so that we generate an estimate of the time-centered
density field.  Particles on subgrids within the grid's volume are
also added to its gravitating field using the same method. In
addition, since the gravitating field for a grid is defined beyond the
grid edges (see below), massive particles from sibling grids and sub
grids that lie within the entire gravitating field are used.  This
step can involve communication.

Next, we add the baryonic grid densities in a similar fashion.  In
particular, we treat baryonic cells as virtual CIC particles that are
are placed at the grid center but are advanced by $0.5 v^n \Delta t$
in order to approximately time-center the gravitating mass field.
Cells that are covered by further-refined grids are treated in a
similar way (i.e., we also use the subgrid cells as virtual CIC
particles).  This procedure results in a total gravitating mass field
$\rho_{\rm total}^{n+1/2}$.

To compute the potential field from this gravitating mass field on the
root grid, we use a a fast Fourier transform. For periodic boundary
conditions, we can use either a simple Greens function kernel of
$-k^{-2}$, or the finite difference equivalent \citep{Hockney88}:

\begin{equation}
G(\myvec{k}) = - \frac{\Delta x}{2 \left( \sin(k_x \Delta x/2)^2 + \sin(k_y \Delta y/2)^2 + \sin(k_z \Delta z/2)^2 \right) }
\end{equation}

where $k^2 = k_x^2 + k_y^2 + k_z^2$ is the wavenumber in Fourier space
and the potential is calculated in k-space as usual with
$\tilde{\phi}(k) = G(k) \tilde{\rho}(k)$.

For isolated boundary conditions, we use the James method
\citep{James77}.  In this case, the Greens function is generated in
real-space so as to have the correct zero-padding properties and then
transformed into the Fourier domain.  In both cases, the potential is
then transformed back into the real domain to get potential values at
the cell centers.  These are differenced with a two-point centered
difference scheme to obtain accelerations at the cell centers (except
if we are using the staggered \zeus-like solver, in which case the
accelerations are computed at the cell faces to match the velocities).
Particle accelerations are obtained using a (linear) CIC interpolation
from the grid.

In order to calculate more accurate potentials on the subgrids, \enzo\
uses a similar but slightly different technique from the root grid.
The generation of the total gravitating mass field is essentially
identical, using CIC interpolation for both the particles and baryons,
including subgrids as before.  To compute the potential on subgrids,
however, we use the standard seven-point (in three dimensions)
second-order finite difference approximation to Poisson's equation.
Boundary conditions are then interpolated from the potential values on
the parent grid.  We use either tri-linear interpolation or a natural
second-order spline for this: both methods give similar results, but
the default is the tri-linear interpolation, which empirically
provides a resonable compromise between speed and accuracy. The
potential equation on each subgrid is then solved with the given
Dirichlet boundary conditions with a multigrid relaxation technique.
This is applied to each subgrid separately.

The region immediately next to the boundary can contain unwanted
oscillations \citep[e.g.,][]{Anninos94}, and so we use an expanded
buffer zone around the grid, of size three parent grid boundary zones
(so typically six refined zones for a refinement factor of 2).  The
density is computed in this region and the potential solved, but only
the region that overlaps with the active region of the grid itself is
used to calculate accelerations.

Simply interpolating the potential without feeding it back to higher
levels leads to errors in the potential at more refined levels, due to
the build-up of errors during the interpolation of coarse boundary
values.  In addition, neighboring subgrids are not guaranteed to
generate the same potential values because of the lack of a coherent
potential solve including the whole grid hierarchy.  In an attempt to
partially alleviate this problem, we allow for an iterative procedure
across sibling grids, in which the potential values on the boundary of
grids can be updated with the potential in `active' regions of
neighboring subgrids.  To prevent overshoot, we average the potential
on the boundary and allow for (by default) 4 iterations, with the
number of iterations determined by a parameter specified at runtime.
This procedure can help in many cases, but does not necessarily
produce a coherent solution across all grids and so does not
completely solve the problem; we are working on a slower but more
accurate method that does a multigrid solve across the whole grid
(Reynolds \etal, in preparation).

At this point it is useful to emphasize that the effective force
resolution of an adaptive particle-mesh calculation is approximately
twice as coarse as the grid spacing at a given level of resolution.

%% file: numerical-particle.tex
\subsection{N-body Dynamics}
\label{sec.ov.nbody}

\enzo\ uses a particle-mesh N-body method to calculate the dynamics of
collisionless systems \citep{Hockney88}.  This method follows
trajectories of a representative sample of individual particles that
sample the phase space of the dark matter distribution, and is much
more efficient than a direct solution of the Boltzmann equation in
essentially all astrophysical situations for the levels of accuracy
that are required for simulations of structure formation.  As
described earlier, the gravitational potential is computed by solving
the elliptic Poisson's equation (Eq.~\ref{eq:potential}) and
differencing the potential to find accelerations, which are then
interpolated back to particles.  This acceleration is time-centered
(because the underlying gravitating mass field is approximately
time-centered), and so we have accelerations $\myvec{g}^{n+1/2}$ for
each particle.  These are used to update the particle positions and
velocities starting from $\myvec{x}^n$ and $\myvec{v}^n$ using a
standard drift-kick-drift technique \citep{Hockney88}:

\begin{eqnarray}
\label{eqn.driftkick}
\myvec{x}^{n+1/2} & = & \myvec{x}^n + \frac{\Delta t}{2 a^n} \myvec{v}^{n} \nonumber \\
\myvec{v}^{n+1} & = & \myvec{v}^n \left(1 - \frac{\dot{a}^{n+1/2}}{a^{n+1/2}}\right) + \frac{\Delta t}{ a^{n+1/2}} \myvec{g}^{n+1/2} \\
\myvec{x}^{n+1} & = & \myvec{x}^{n+1/2} + \frac{\Delta t}{2 a^{n+1}} \myvec{v}^{n+1} \nonumber
\end{eqnarray}

Particles are stored in the most highly refined grid patch at the
point in space where they exist, and particles that move out of a
subgrid patch are sent to the grid patch covering the adjacent volume
with the finest spatial resolution, which may be of the same spatial
resolution, coarser, or finer than the grid patch from which the
particles moved.  This takes place in a communication process at the
end of each timestep on a level.

To avoid unphysical point-mass effects, \enzo\ provides a parameter
that governs the maximum level at which particles will be regarded as
point masses.  At higher levels, contributions from particles to the
gravitating mass field will be smoothed over a spherical region
centered at each particle's position.

%% file: numerical-chemistry.tex
\subsection{Chemistry}
\label{sec.num.chemistry}

While it is often safe to assume that species (both chemical and
ionization) within a gas can be treated as being in equilibrium, in
some regimes that are found in astrophysics this assumption leads to
considerable error.  For example, the cooling and collapse of
primordial gas in Population III star formation is dominated by
molecular hydrogen, which in the absence of dust forms via an
inefficient pair of collisional processes that depend heavily on the
local, highly non-equilibrium population of free electrons.  As a
result, when modeling primordial star formation it is critical to
follow the non-equilibrium evolution of the chemical species of
hydrogen, including molecular hydrogen and deuterium.

The primordial non-equilibrium chemistry routines used in \enzo\ were
first described by \citet{abel97} and \citet{anninos97}, but have
since been extended with updated reaction rates and the inclusion of
deuterium species \citep{McGreer2008, 2009PhDT.........5T}.  These
routines follow the non-equilibrium chemistry of a gas of primordial
composition with 12 total species: H, \Hp, He, \Hep, \Hepp, \Hm, \HHp,
\HH, e$^-$, D, \Dp, and HD.  \enzo\ also computes the radiative
heating and cooling of the gas from atomic and molecular line
excitation, recombination, collisional excitation, free-free
transitions, Compton scattering of the cosmic microwave background, as
well as several models for a metagalactic UV background that heat the
gas via photoionization and photodissociation (see
Section~\ref{sec.num.cooling} for more details).  The chemical and
thermal states of the gas can be updated either at the same
hydrodynamical timestep (i.e., decoupled and operator-split) or
through the same subcycling system (i.e., a coupled chemical and
thermal system).  The default behavior of \enzo\ is to couple these
two systems at subcycles of the hydrodynamic timestep; this results in
updates to both the chemical and thermal states of the gas (which also
inform the temperature, the reaction rate coefficients and the cooling
functions of the gas) on timescales that are faster than those of the
gas dynamics.  The gamma used by \enzo\ to compute the temperature of
the gas from the energy and density characteristics utilizes a
variable gamma that includes effects of the rotational state of
molecular hydrogen, enabling it to vary from $5/3$ (fully-atomic) to
$7/5$ (fully molecular).  This further coupling of the chemical and
thermal states of the gas underscores the need for coupled chemistry
and radiative cooling solutions.

Input parameters to \enzo\ govern the chemical species that are
updated during the course of the simulation.  This can include only
the atomic species (H, \Hp, He, \Hep, \Hepp, and e$^-$), those species
relevant for molecular hydrogen formation (\HH, \HHp, and \Hm), and
can further include deuterium and its species (D, \Dp, and HD).  A
total of 9 kinetic equations are solved for the 12 species mentioned
above, including 29 kinetic and radiative processes.  See
Table~\ref{table.collisional} for the collisional processes and
Table~\ref{table.radiative} for the radiative processes solved.

The chemical reaction equation network is technically challenging to
solve due to the huge range of reaction timescales involved.  The
characteristic times for creation and destruction of the various
species and reactions can differ by many orders of magnitude and are
often very sensitive to the chemical and thermal state of the gas.
This makes a fully-implicit scheme, with convergence criteria and
error tolerance, strongly preferable for such a set of equations.
However, most implicit schemes require an iterative procedure to
converge, and for large networks (such as this one) an iterative,
fully-implicit method can be very time-consuming and computationally
costly for a relatively small increase in accuracy.  At the present
time, this makes fully-implicit methods somewhat undesirable for
large, three-dimensional simulations.

\enzo\ solves the rate equations using a method based on a
semi-implicit formulation in order to provide a stable, positive
definite and first-order accurate solution.  The update discretization
splits chemical changes into formation components and destruction
components and updates with a mixed set of time states, as described
in \citet{anninos97}.  The formation components of species $S_i$ are
computed at the current subcycle time, where the contribution of
species $S_i$ to its own destruction components are computed at the
updated time; all other contributions to the destruction component are
computed at the current time.  This mixed state improves accuracy,
ensures species values are positive definite, and is equivalent to one
Jacobi iteration of an implicit Euler solve.  This technique is
optimized by taking the chemical intermediaries \Hm and \HHp, which
have large rate coefficients and low concentrations, and grouping them
into a separate category of equations.  Due to their fast reactions,
these species are very sensitive to small changes in the more abundant
species and are (at almost all times in astrophysical calculations)
close to their equilibrium values.  Attempting to resolve their
formation and destruction times would necessitate extremely small
timesteps.  Therefore, reactions governing these two species can be
decoupled from the rest of the network and treated independently
through analytic solutions for equilibrium values.  This allows a
significant speedup in the solution speed, as the timestepping scheme
is applied only to the slower 7- or 10-species network (depending on
whether deuterium is included or not), which will be much closer to
the overall hydrodynamic timestep of the simulation.

Even so, the accuracy and stability of the scheme is maintained by
subcycling the rate solver within a single hydrodynamic timestep.
These subcycle timesteps are determined so that the estimated
fractional change in the electron concentration is limited to no more
than $10\%$ per timestep; additional criteria may be applied based on
the expected change in internal energy from radiative cooling and from
chemical heating due to the formation of molecular hydrogen.

It is important to note the regime in which this model is valid.
According to \citet{abel97} and \citet{anninos97}, the reaction
network is valid for temperatures between $10^0 - 10^8$ K.  The
original model discussed in these two references was only claimed to
be valid up to $n_H \sim 10^4$~cm$^{-3}$.  However, addition of the
3-body \HH~formation process (equation 20 in
Table~\ref{table.collisional}) allows correct modeling of the
chemistry of the gas up until the point where collisionally-induced
emission from molecular hydrogen becomes an important cooling process,
which occurs at $n_{\rm H} \sim 10^{14}$~cm$^{-3}$.  A further concern
is that the optically thin approximation for radiative cooling
eventually breaks down, which occurs before $n_{\rm H} \sim 10^{16} -
10^{17}$~cm$^{-3}$ in gas of primordial composition.  Beyond this
point, modifications to the cooling function that take into account
the non-negligible opacity in the gas must be made, as discussed by
\citet{2004MNRAS.348.1019R}, and was put into \enzo\ for the work
published in \citep{2009Sci...325..601T,2009PhDT.........5T}.  The
formation of molecular hydrogen as catalyzed by dust was recently
added to \enzo\ to enable studies of low-metallicity gas, as well as
the inclusion of appropriate timestepping criteria to account for the
input of ionizing radiation.  Even with these modifications, a
completely correct description of the cooling of primordial gas at
very high densities requires some form of radiation transport, which
will greatly increase the cost of the simulations.  Furthermore, at
very high densities, the stiffness of the molecular hydrogen reaction
rates may require better than a first-order accurate solution; as
such, the transition to this regime will likely necessitate a
fully-implicit, iterative solver.


\begin{table}
\begin{center}
{\bfseries Collisional Processes}\\[1ex]
\begin{tabular}{llllllll}
(1) & H & + & e$^-$ & $\rightarrow$ & H$^+$ &+& 2e$^-$ \\
(2) & H$^+$ &+ &e$^-$ & $\rightarrow$ & H &+ &$\gamma$ \\
(3) & He &+& e$^-$ & $\rightarrow$ & He$^+$ &+& 2e$^-$  \\
(4) & He$^+$ &+& e$^-$ & $\rightarrow$ & He &+ &$\gamma$  \\
(5) & He$^{+}$ &+& e$^-$ & $\rightarrow$ & He$^{++}$ &+& 2$e^-$  \\
(6) & He$^{++}$ &+& e$^-$ & $\rightarrow$ & He$^+$ &+& $\gamma$ \\
\hline
(7) & H &+& e$^-$ &$\rightarrow$& H$^-$ &+& $\gamma$  \\
(8) & H$^-$ &+& H &$\rightarrow$ & H$_2$ & +& e$^-$ \\
(9) & H &+ &H$^+$ &$\rightarrow$ &H$_2^+$ &+ &$\gamma$ \\
(10) & H$_2^+$ &+ &H &$\rightarrow$ &$H_2$ &+ &$H^+$ \\
(11) & H$_2$ &+ &H$^+$ &$\rightarrow$ &H$_2^+$ & +& H \\
(12) & H$_2$ &+ &e$^-$ & $\rightarrow$ & 2H & + & e$^-$  \\
(13) & H$_2$ & + & H & $\rightarrow$ & 3H &   &      \\
(14) & H$^-$ & + & e$^-$ & $\rightarrow$ & H & + & 2e$^-$ \\
(15) & H$^-$ & + & H & $\rightarrow$ & 2H & + & e$^-$ \\ 
(16) & H$^-$ & + & H$^+$ & $\rightarrow$ & 2H & & \\
(17) & H$^-$ & + & H$^+$ & $\rightarrow$ & H$_2^+$ & + & e$^-$ \\
(18) & H$_2^+$ & + & e$^-$ & $\rightarrow$ & 2H & & \\
(19) & H$_2^+$ & + & H$^-$ & $\rightarrow$ & H$_2$ & + & H  \\
(20) & 2H & + & H$_2$ & $\rightarrow$ & 2H$_2$ &  &   \\
(21) & 2H & + & H & $\rightarrow$ & H$_2$ & + & H  \\
(22) & H$_2$ & + & H$_2$ & $\rightarrow$ & H$_2$ & + & 2H  \\
(23) & 3H & & & $\rightarrow$ & H$_2$ & + & H \\
\hline
(24) & D & + & e$^-$ & $\rightarrow$ & D$^+$ &+& 2e$^-$ \\
(25) & D$^+$ &+ &e$^-$ & $\rightarrow$ & D &+ &$\gamma$ \\
(26) & H$^+$ &+ &D & $\rightarrow$ & H &+ &D$^+$ \\
(27) & H &+ &D$^+$ & $\rightarrow$ & H$^+$ &+ &D \\
(28) & H$_2$ &+ &D$^+$ & $\rightarrow$ & HD &+ &H$^+$ \\
(29) & HD &+ &H$^+$ & $\rightarrow$ & H$_2$ &+ &D$^+$ \\
(30) & H$_2$ &+ &D & $\rightarrow$ & HD &+ &H \\
(31) & HD &+ &H & $\rightarrow$ & H$_2$ &+ &D \\
(32) & H$^-$ &+ &D & $\rightarrow$ & HD &+ &e$^-$ \\

\end{tabular}
\caption[]{Collisional processes solved in the \enzo\ nonequilibrium
primordial chemistry routines.}
\label{table.collisional}
\end{center}
\end{table}

\begin{table}
\begin{center}
{\bfseries Radiative Processes}\\[1ex]
\begin{tabular}{llllllll}
(33) & H & + & $\gamma$ & $\rightarrow$ & H$^+$ & + & e$^-$ \\
(34) & He & + & $\gamma$ & $\rightarrow$ & He$^+$ & + & e$^-$ \\
(35) & He$^+$ & + & $\gamma$ & $\rightarrow$ & He$^{++}$ & + & e$^-$ \\
(36) & H$^-$ & + & $\gamma$ & $\rightarrow$ & H & + & e$^-$ \\
(37) & H$_2$ & + & $\gamma$ & $\rightarrow$ & H$_2^+$ & + & e$^-$ \\
(38) & H$_2^+$ & + & $\gamma$ & $\rightarrow$ & H & + & H$^+$ \\
(39) & H$_2^+$ & + & $\gamma$ & $\rightarrow$ & 2H$^+$ & + & e$^-$ \\
(40) & H$_2$ & + & $\gamma$ & $\rightarrow$ & H$_2^*$ & $\rightarrow$ & 2H \\
(41) & H$_2$ & + & $\gamma$ & $\rightarrow$ & 2H &  & \\
(42) & D & + & $\gamma$ & $\rightarrow$ & D$^+$ & + & e$^-$ \\
\end{tabular}
\caption[]{Radiative processes solved in the \enzo\ nonequilibrium
primordial chemistry routines.}
\label{table.radiative}
\end{center}
\end{table}

%% file: numerical-cooling.tex
\subsection{Radiative Cooling and Heating}
\label{sec.num.cooling}

\enzo\ has multiple methods for computing the energy change from radiative
cooling and heating.  All of them assume that the gas can be modeled either as
completely optically thin or with a simple, local approximation to optical
thickness.  In this section, we describe the methods for computing the cooling rates from
metal-free and metal-enriched gas.  Sample cooling curves for each of \enzo's
primary cooling methods are shown in Figure \ref{fig.cooling_rate}.

\subsubsection{Primordial Cooling}

As discussed in Section~\ref{sec.num.chemistry}, the set of reactions that
characterize a metal-free gas is simple enough to be computed in
non-equilibrium during even very large simulations.  Similarly, the radiative
cooling of metal-free gas is solved by directly computing the cooling
and heating rates from the following individual processes for atomic H
and He: collisional excitation and ionization, recombination,
free-free emission, Compton scattering off of the cosmic microwave
background (CMB), and photo-heating from a metagalactic ultraviolet background.
If the \HH~chemistry network is enabled, the following \HH~cooling 
processes are also considered: ro-vibrational transitions
\citep{2008MNRAS.388.1627G,1998A&A...335..403G}, heating and cooling from molecular
formation and destruction \citep{2009Sci...325..601T}, and
collision-induced emission \citep{2004MNRAS.348.1019R}.  If Deuterium
chemistry is enabled, then rotational transitions of HD
\citep{1998A&A...335..403G, 1983ApJ...270..578L} are treated as well.  The
radiative cooling calculation is coupled to the update of the chemistry network
such that they both occur within the same subcycling loop.  This is necessary
in regimes where rapid cooling or change in the ionization state occur, as this
will influence the chemical kinetic rate coefficients through both changes in
energy and the equation of state of the gas.  In addition to the subcycle
timestepping contraints mentioned in Section~\ref{sec.num.chemistry}, the
subcycle timestep is also not permitted to exceed 10\% of the cooling time,
$t_{cool} = e/\dot{e}$.  

A metagalactic background affects the gas through both
photo-heating and photo-ionization.  These are treated by including
redshift-dependent photo-ionization and photo-heating rate terms in the
chemistry and cooling equations for \ion{H}{1}, \ion{He}{1}, and \ion{He}{2}.
The black curves in Figure \ref{fig.cooling_rate} show cooling rates for a
metal-free gas with number density $n_{\rm H} = 10^{-4}$ cm$^{-3}$ both with
and without a radiation background.  More detail on the specific UV backgrounds
in \enzo\ will be presented in Section~\ref{sec.num.rad-homogeneous}.

\subsubsection{Metal Cooling}

\begin{figure}
  \begin{center}
    \includegraphics[width=0.8\textwidth]{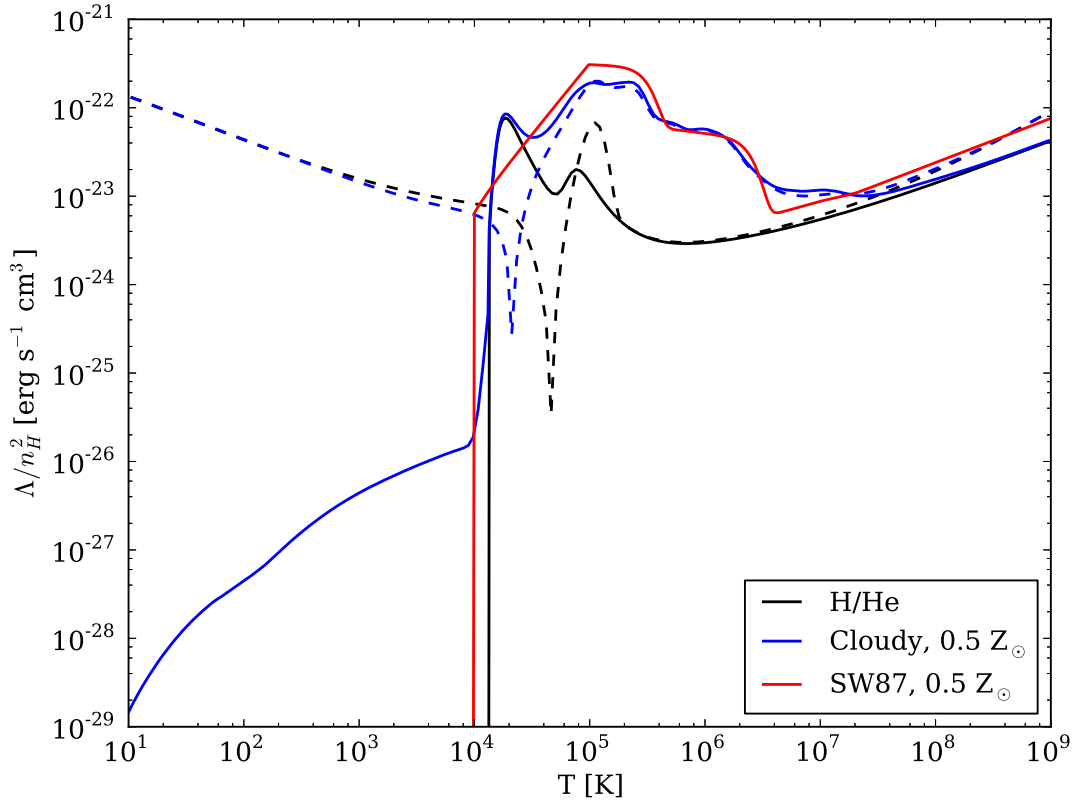}
  \end{center}
  \caption{Radiative cooling rates from the various cooling methods
    available in \enzo.  The black curves show cooling rates from a gas
    with primordial composition using the non-equilibrium chemistry
    network.  The blue curves are for a gas with metallicity of 0.5
    $Z_{\odot}$ computed with the Cloudy cooling method.  The solid
    black and blue lines assume collisional processes only while the
    dashed lines include photo-ionization and photo-heating from a UV
    metagalactic background at $z = 0$ with a gas number density of
    10$^{-4}$ cm$^{-3}$.  The rates shown by the
    dashed lines indicate a net heating below $T \sim 10^{4.5}$, where
    the rapid change in rate is evident (the curve is an absolute
    value so it can be shown on a log plot).  The red curve is the
    tabulated cooling function of \citet{SW87}, which assumes a fully
    ionized gas with metallicity of 0.5 $Z_{\odot}$.}
  \label{fig.cooling_rate}
\end{figure}

A proper treatment of the cooling from metals is significantly more
challenging due to the large number of chemical reactions and energy
transitions that must be taken into account for each element.  Because
of this, most metal cooling methods employ significant assumptions in
order to seek out a balance between accuracy and speed.  There are two
primary metal cooling methods available in \enzo.  The simpler of the
two uses the analytic cooling function of \citet{SW87}, which assumes
a fully ionized gas with a constant metallicity of 0.5 Z$_{\odot}$.
The cooling rate produced by this cooling function is shown by the red 
curve in Figure \ref{fig.cooling_rate}.

A more sophisticated method makes use of multidimensional cooling and
heating rate tables computed with the photo-ionization code
\texttt{Cloudy} \citep{1998PASP..110..761F}.  This method, detailed in
\citet{2008MNRAS.385.1443S, 2011ApJ...731....6S}, works by
using \texttt{Cloudy} to compute the cooling and heating rates from
the metal species only.  The primary assumption made is that of
ionization equilibrium.  The tables can vary along up to five dimensions:
density, metallicity, electron fraction,
redshift (for an evolving metagalactic UV background), and
temperature.  Tables can be created for any abundance pattern for
elements up to atomic number 30 (Zn) and for any incident radiation
field.  Cooling from the standard \enzo\ non-equilibrium cooling module is
applied on top of the metal contributions.  The contribution of metals
to the cooling is computed within the same subcycling loop as the
coupled primordial chemistry and cooling solver.  The blue curves in
Figure \ref{fig.cooling_rate} show cooling rates calculated with this
method for a gas with number density $n_{\rm H} = 10^{-4}$ cm$^{-3}$ and
metallicity of 0.5 Z$_{\odot}$.

%% file: numerical-radiation-homogeneous.tex
\subsection{Homogeneous radiation background}
\label{sec.num.rad-homogeneous}

\enzo\ supports the use of a set of spatially uniform (but possibly
time-varying) radiation fields that can interface with the chemistry
and cooling/heating routines described in Section~\ref{sec.microphysics}.  Many of these use
fits to the H, He and \Hep~ionizing and photo-ionization heating
rates that are of the form

\begin{equation}
\rm{rate} = k_0  (1+z)^{\alpha}  \exp{\left( \frac{\beta (z-z_0)^2 }{1 + \gamma  (z+z_1)^2} \right)}
\label{eq:homo_rad_template}
\end{equation}
where the constant coefficients $\alpha$, $\beta$, $\gamma$, $z_0$ and
$z_1$ are fits from the literature.  For radiation field types 1-3
(numbered as they appear in the \enzo\ source code and simulation
parameter values), we
give the coefficients for the photoionization (and photo-heating)
rates of H, He, and \Hep~in Table~\ref{table:homo_coefs}.  Radiation
background types 1
and 2 are based on \citet{1996ApJ...461...20H} with two different
intrinsic quasar spectra slopes, while Type 3 is from
\citet{2012ApJ...746..125H}, modified to match the normalized field
found in \citet{Kirkman05}.  

The remainder of the radiation field types either build on the fits in Types
1-3 or use a completely different form and therefore we describe them
in the text rather than in Table~\ref{table:homo_coefs}.
Type 4 is the same as Type 3 but also
adds X-ray Compton heating from \citet{MadauEfstathiou99}, using
equations (4) and (11) of that paper.

Homogeneous radiation field types 5 and 6 start with a spectral shape
that is then integrated against the appropriate H, He, and \Hep,
cross-sections to compute the ionization and photo-ioniziation heating
rates (we use 400 bins logarithmically spaced from 0.74 eV to $7.24
\times 10^9$ eV).  In particular, Type 5 has a hard, featureless
quasar-like power law spectrum $f_{\nu} = f_{\rm HI} \nu^{\alpha_0}$,
where $\alpha_0 = 1.5$ by default and the spectrum is normalized at
the HI ionization edge.  Type 6 has the same spectrum, but attenuated
by a column density of $10^{22}$ cm$^{-3}$ neutral hydrogen.  

Types 8, 9, and 14 have only a photo-dissociating Lyman-Werner flux,
with Type 8 being constant and Type 9 using the redshift-dependent
results of \citet{TrentiStiavelli09}.  Type 14 uses a fit from
\citet{WiseAbel05} for the range $6 < z < 50$, constant for $z<6$, and
zero for $z > 50$.  Other types are either undefined or currently unused.

\begin{table*}
\begin{center}
\caption{Homogeneous radiation field coefficients}
\begin{tabular*}{0.9\textwidth}{@{\extracolsep{\fill}}lrrrrrr}
\tableline\tableline
{Element} & {$k_0$} &  {$\alpha$} & {$\beta$} & {$\gamma$} & {$z_0$} & {$z_1$}  \\
\tableline
\multicolumn{7}{c}{Radiation Type 1 \citep{1996ApJ...461...20H} for the case $\alpha_q = 1.5$} \\
\tableline
H ionization & $6.7 \times 10^{-13}$ & 0.43 & 1/1.95 & 0 & 2.3 & 0 \\
He ionization & $6.3 \times 10^{-15}$ & 0.51 & 1/2.35 & 0 & 2.3 & 0 \\
He$^+$ ionization & $3.2 \times 10^{-13}$ & 0.50 & 1/2.00 & 0 & 2.3 & 0 \\
H heating & $4.7 \times 10^{-24}$ & 0.43 & 1/1.95 & 0 & 2.3 & 0 \\
He heating & $8.2 \times 10^{-24}$ & 0.50 & 1/2.00 & 0 & 2.3 & 0 \\
He$^+$ heating & $1.6 \times 10^{-25}$ & 0.51 & 1/2.35 & 0 & 2.3 & 0 \\
\tableline
\multicolumn{7}{c}{Radiation Type 2 \citep{1996ApJ...461...20H} for the case $\alpha_q = 1.8$} \\
\tableline
H ionization & $5.6 \times 10^{-13}$ & 0.43 & 1/1.95 & 0 & 2.3 & 0 \\
He ionization & $3.2 \times 10^{-15}$ & 0.30 & 1/2.60 & 0 & 2.3 & 0 \\
He$^+$ ionization & $4.8 \times 10^{-13}$ & 0.43 & 1/1.95 & 0 & 2.3 & 0 \\
H heating & $3.9 \times 10^{-24}$ & 0.43 & 1/1.95 & 0 & 2.3 & 0 \\
He heating & $6.4 \times 10^{-24}$ & 0.43 & 1/2.10 & 0 & 2.3 & 0 \\
He$^+$ heating & $8.7 \times 10^{-26}$ & 0.30 & 1/2.70 & 0 & 2.3 & 0 \\
\tableline
\multicolumn{7}{c}{Radiation Type 3 modified \citet{2012ApJ...746..125H}} \\
\tableline
H ionization &        $1.04 \times 10^{-12}$ & 0.231 & -0.6818 & 0.1646 & 1.855 & 0.3097 \\
He ionization &       $1.84 \times 10^{-14}$ & -1.038 & -1.1640 & 0.1940 & 1.973 & -0.6561 \\
He$^+$ ionization & $5.79 \times 10^{-13}$ & 0.278 & -0.8260 & 0.1730 & 1.973 & 0.2880 \\
H heating &            $8.86 \times 10^{-25}$ & -0.0290 & -0.7055 & 0.1884 & 2.003 & 0.2888 \\
He heating &         $5.86 \times 10^{-24}$ & 0.1764 & -0.8029 & 0.1732 & 2.088 & 0.1362 \\
He$^+$ heating &   $2.17 \times 10^{-25}$ & -0.2196 & -1.070 & 0.2124 & 1.1782 & -0.9213 \\
\end{tabular*}
\label{table:homo_coefs}
\end{center}
\end{table*}

%% file: numerical-radiation-raytracing.tex
\subsection{Radiation transport: ray tracing}
\label{sec.num.raytracing}

Stars and black holes strongly affect their surroundings through
radiation.  Radiation transport is a well-studied problem; however,
its treatment in multidimensional calculations is difficult because of
the dependence on seven variables -- three spatial, two angular,
frequency, and time.  The non-local nature of the thermal and
hydrodynamical response to radiation sources further adds to the
difficulty.  Here we briefly describe \enzo's ray tracing
implementation \moray, which is presented in full detail in
\citet{Wise11_Moray} with seven code tests and six applications.

We solve the radiative transfer equation in comoving coordinates
(given by Equation~\ref{eq:rteqn}).  We can make some appropriate
approximations to reduce the complexity of this equation in order to
include radiation transport in numerical calculations.  Typically
timesteps in dynamic calculations are small enough so that $\Delta a/a
\ll 1$, therefore $a_{em}/a \approx 1$ in any given timestep, reducing
the second term to $\hat{n} \partial I_\nu/\partial \mathbf{x}$.  To
determine the importance of the third term, we evaluate the ratio of
the third term to the second term.  This is $HL/c$, where $L$ is the
simulation box length.  If this ratio is $\ll 1$, we can ignore the
third term.  For example at $z=5$, this ratio is 0.1 when $L =
c/H(z=5)$ = 53 proper Mpc.  In large boxes where the light crossing
time is comparable to the Hubble time, then it becomes important to
consider cosmological redshifting and dilution of the radiation.  Thus
equation (\ref{eq:rteqn}) reduces to the non-cosmological form in this
local approximation,
\begin{equation}
  \frac{1}{c} \frac{\partial I_\nu}{\partial t} + 
  \hat{n} \frac{\partial I_\nu}{\partial \mathbf{x}} =
  -\kappa_\nu I_\nu + j_\nu .
\end{equation}
We choose to represent the source term $j_\nu$ as point sources of
radiation (e.g. stars, quasars) that emit radial rays that are
propagated along the direction $\hat{n}$.

Ray tracing is an accurate method to propagate radiation from point
sources on a computational grid as long as there are a sufficient
number of rays passing through each cell.  Along a ray, the radiation
transfer equation reduces to
\begin{equation}
\label{eqn:rtray}
\frac{1}{c} \frac{\partial P}{\partial t} + \frac{\partial P}{\partial
  r} = -\kappa P,
\end{equation}
where $P$ is the photon number flux along the ray.  To sample the
radiation field at large radii, ray tracing requires at least $N_{ray}
= 4\pi R^2 / (\Delta x)^2$ rays to sample each cell with one ray,
where $R$ is the radius from the source to the cell and $\Delta x$ is
the cell width.  If one were to trace $N_{ray}$ rays out to $R$, the
cells at a smaller radius $r$ would be sampled by, on average,
$(r/R)^2$ rays, which is computationally wasteful because only a few
rays per cell are required to provide an accurate calculation of the
radiation field (as we will show later).

We avoid this inefficiency by utilizing adaptive ray tracing
\citep{Abel02_RT}, which is based on Hierarchical Equal Area
isoLatitude Pixelation \citep[HEALPix;][]{HEALPix} and progressively
splits rays when the sampling becomes too coarse.  In this approach,
the rays are traced along normal directions of the centers of the
HEALPix pixels that evenly divide a sphere into equal areas.  The rays
are initialized at each point source with the photon luminosity
(photon s$^{-1}$) equally spread across $N_{\rm pix} = 12 \times 4^l$
rays, where $l$ is the initial HEALPix level.  We usually find $l$ = 0
or 1 is sufficient because these coarse rays will usually be split
before traversing the first cell.

The rays are traced through the grid in a typical fashion
\citep[e.g.][]{Abel99_RT}, in which we calculate the next cell
boundary crossing.  The ray segment length crossing the cell is
\begin{equation}
  \label{eqn:trace}
  dr = R_0 - \min_{i=1 \rightarrow 3} \left[(x_{\rm cell,i} - x_{\rm src,i}) /
    \hat{n}_{\rm ray,i} \right],
\end{equation}
where $R_0$, $\hat{n}_{\rm ray}$, $x_{\rm cell,i}$, and $x_{\rm
  src,i}$ are the initial distance traveled by the ray, normal
direction of the ray, the next cell boundary crossing in the $i$-th
dimension, and the position of the point source that emitted the ray,
respectively.  However before the ray travels across the cell, we
evaluate the ratio of the face area $A_{\rm cell}$ of the current cell
and the solid angle $\Omega_{\rm ray}$ of the ray,
\begin{equation}
  \label{eqn:split}
  \Phi_c = \frac{A_{\rm cell}} {\Omega_{\rm ray}} = 
  \frac{N_{\rm pix} (\Delta x)^2} {4\pi R_0^2}.
\end{equation}
If $\Phi_c$ is less than a pre-determined value (usually $>3$), the
ray is split into 4 child rays.  The pixel numbers of the child rays
$p^\prime$ are given by the ``nested'' scheme of HEALPix at the next
level, i.e. $p^\prime = 4 \times p + [0,1,2,3]$, where $p$ is the
original pixel number.  The child rays (1) acquire the new normal
vectors of the pixels, (2) retain the same radius of the parent ray,
and (3) get a quarter of the photon flux of the parent ray.
Afterward, the parent ray is discontinued.

A ray propagates and splits until at least one of the following
conditions is met: (1) the photon has traveled $c \times dt_P$, where
$dt_P$ is the radiative transfer timestep; (2) its photon flux is
almost fully absorbed ($>99.9\%$) in a single cell, which
significantly reduces the computational time if the radiation volume
filling fraction is small; (3) the photon leaves the computational
domain with isolated boundary conditions; or (4) the photon travels
$\sqrt{3}$ of the simulation box length with periodic boundary
conditions.  In the first case, the photon is halted at that position
and saved, where it will be considered in the solution of $I_\nu$ at
the next timestep.  In the next timestep, the photon will encounter a
different hydrodynamical and ionization state, hence $\kappa$, in its
path.  Furthermore any time variations of the luminosities will be
retained in the radiation field.  This is how this method retains the
time derivative of the radiative transfer equation.  The last
restriction prevents our method from considering sources external to
the computational domain. However, a uniform radiation background can
be used in conjunction with ray tracing that adds the background
intensity to the local radiation field.

The radiation field is calculated by integrating Equation
(\ref{eqn:rtray}) along each ray, which is done by considering the
discretization of the ray into segments.  In the following
description, we assume the rays are monochromatic for simplicity.  For
convenience, we express the integration in terms of optical depth
$\tau = \int \kappa(r,t) \; dr$, and for a ray segment
\begin{equation}
  \label{eqn:dtau}
  d\tau = \sigma_{\rm abs}(\nu) n_{\rm abs} dr.
\end{equation}
Here $\sigma_{\rm abs}$ and $n_{\rm abs}$ are the cross section and
number density of the absorbing medium, respectively.  In the static
case, equation (\ref{eqn:rtray}) has a simple exponential analytic
solution, and the photon flux of a ray is reduced by
\begin{equation}
  \label{eqn:flux}
  dP = P \times (1 - e^{-\tau})
\end{equation}
as it crosses a cell.  We equate the photo-ionization rate to the
absorption rate, resulting in photon conservation \citep{Abel99_RT,
Mellema06}.  Thus the photo-ionization and photo-heating rates
associated with a single ray ($k_{\rm ph}$ and $\Gamma_{\rm ph}$,
respectively) are
\begin{equation}
  \label{eqn:kph}
  k_{\rm ph} = \frac{P (1 - e^{-\tau})}{n_{\rm abs} \; V_{\rm cell} \; dt_P},
\end{equation}
\begin{equation}
  \label{eqn:gamma}
  \Gamma_{\rm ph} = k_{\rm ph} \; (E_{\rm ph} - E_i),
\end{equation}
where $V_{\rm cell}$ is the cell volume, $E_{\rm ph}$ is the photon
energy, and $E_i$ is the ionization energy of the absorbing material.
In each cell, the photo-ionization and photo-heating rates from each
ray in the calculation are summed. After the ray tracing is complete,
these rates are used as inputs to the solver described in
Section~\ref{sec.ov.chem} to update the ionization, chemical, and
energy states of the gas in each cell.

%% file: numerical-radiation-fld.tex
\subsection{Radiation transport: Flux-limited diffusion}
\label{sec.num.rad-fld}

In addition to the ray-tracing approach for radiation transport
described in Section~\ref{sec.num.raytracing}, \enzo\ currently
includes a field-based radiation transport solver for problems posed
on uniform (i.e.~non-AMR, non-static mesh refinement) grids, which has
been tuned for large-scale simulations involving many ionizing
sources.  Detailed explanations of the model and solution approach may
be found in \citet{NBHBROW2007},
\citet{ReynoldsHayesPaschosNorman2009}, \citet{NRS2009}, and
\citet{2013arXiv1306.0645N}, the salient features of which are
reproduced here.  In addition, comparisons of this solver with other
astrophysical radiation transport solvers may be found in
\cite{IlievEtAl2009}.  \enzo's field-based radiation solver focuses on
a flux-limited diffusion approximation for cosmological radiative
transfer, with couplings to both the gas energy and chemical number
densities.

The system of equations (\ref{eq:fld_radiation}-\ref{eq:fld_heating})
along with the chemical network (Equation~\ref{eq:species_evolution})
is solved independently of \enzo's hydrodynamics, gravity and
dark-matter solvers (Sections \ref{sec.hydro.ppm}-\ref{sec.ov.nbody}),
thereby allowing the advective portions of Equations
(\ref{eq:fld_radiation}) and (\ref{eq:species_evolution}) to be taken
care of by the fluid solvers.  Due to the disparate time scales
between radiation transport and chemical ionization and heating, the
remainder of these equations is solved using an operator-split
algorithm.  Within a given timestep to evolve $(E_r^n, e_c^n, {\mathrm
n}_i^n) \to (E_r^{n+1}, e_c^{n+1}, {\mathrm n}_i^{n+1})$, we first
evolve equation (\ref{eq:fld_radiation}): $(E_r^n, e_c^n, {\mathrm
n}_i^n) \to (E_r^{n+1}, e_c^{n}, {\mathrm n}_i^{n})$.  This uses an
implicit Euler time discretization, and a second-order centered finite
difference spatial discretization, resulting in a large linear system
of equations.  These are solved using a preconditioned conjugate
gradient iteration, where the preconditioner consists of a geometric
multigrid solver.  Both of these linear solvers are provided by the
HYPRE linear solver library \citep[see][]{FalgoutYang2002,
hypre-manual}.

We then evolve the heating and chemistry system,
Equations~(\ref{eq:fld_heating}) and (\ref{eq:species_evolution}):
$(E_r^{n+1}, e_c^n, {\mathrm n}_i^n) \to (E_r^{n+1}, e_c^{n+1},
{\mathrm n}_i^{n+1})$.  Due to the lack of spatial derivatives (since
advection is handled elsewhere), this system is a coupled system of
nonlinear ordinary differential equations.  This utilizes an implicit
quasi-steady-state approximation, formulated as follows.  We consider
the modified equations,
\begin{eqnarray}
  \label{eq:fld_heating_qss}
  \frac{\partial e_c}{\partial t} &=& -\frac{2\dot{a}}{a} e_c +
    \Gamma\left(\bar{E}_r,\bar{\mathrm n}_i\right) - 
    \Lambda\left(\bar{E}_r,\bar{\mathrm n}_i\right), \\
  \label{eq:fld_chemistry_qss}
  \frac{\partial {\mathrm n}_i}{\partial t} &=& k_{i,j}\left(\bar{e}_c\right)
    {\mathrm n}_e \bar{\mathrm n}_j - {\mathrm n}_i 
    \Gamma_i^{ph}\left(\bar{E}_r\right), \quad i=1,\ldots,N_s,
\end{eqnarray}
where we have defined the time-centered ``background'' states
$\bar{E}_r = \left(E_r^{n}+E_r^{n+1}\right)/2$, 
$\bar{\mathrm n}_i = \left({\mathrm n}_i^{n}+{\mathrm n}_i^{n+1}\right)/2$
and $\bar{e}_c = \left(e_c^{n}+e_c^{n+1}\right)/2$.  These equations
may each be solved analytically for their solution at the time
$t$, which we denote by
\begin{eqnarray}
  \label{eq:fld_heating_qss_sol}
  e_c(t) &=& \text{sol}_{e}\left(t,\bar{E}_r,\bar{\mathrm n}_i,e_c^n\right), \\
  \label{eq:fld_chemistry_qss_sol}
  {\mathrm n}_i(t) &=& \text{sol}_{\mathrm n_i}
  \left(t,\bar{E}_r,\bar{e_c},\mathrm n_i^n\right), \quad i=1,\ldots,N_s. 
\end{eqnarray}
We then define a nonlinear system of equations to compute the
time-evolved solutions $\left(E_r^{n+1}, e_c^{n+1}, 
{\mathrm n}_i^{n+1}\right)$ as
\begin{eqnarray}
  \label{eq:fld_heating_qss_fe}
  f_e(e_c^{n+1},{\mathrm n}_i^{n+1}) &\equiv& e_c^{n+1} -
    \text{sol}_{e}\left(t^{n+1},\bar{E}_r,\bar{\mathrm n}_i,e_c^n\right)
    = 0, \\
  \label{eq:fld_chemistry_qss_fn}
  f_{\mathrm n_i}(e_c^{n+1},{\mathrm n}_i^{n+1}) &\equiv& 
    {\mathrm n}_i(t) - \text{sol}_{\mathrm n_i}
    \left(t,\bar{E}_r,\bar{e_c},\mathrm n_i^n\right)=0, \quad
    i=1,\ldots,N_s.  
\end{eqnarray}
This system of $N_s+1$ nonlinear equations is solved using a damped
fixed point iteration, 
\[
   U_i = U_i - \lambda f_i(U), \quad i=1,\ldots,N_s+1,
\]
where $U$ is a vector containing the solutions to equations
(\ref{eq:fld_heating_qss_fe}-\ref{eq:fld_chemistry_qss_fn}).  In this
iteration, for the first 50 sweeps we use $\lambda=1$. For more
challenging problems where this does not converge, we switch to a
damping parameter of $\lambda=0.1$.

%% file: numerical-conduction.tex
\subsection{Thermal conduction}
\label{sec.num.conductions}

\enzo\ implements the equations of isotropic heat conduction in a
manner similar to that of \citet{2007ApJ...664..135P}.  The isotropic
flux of heat is given by equation~(\ref{eq:conduction}) and we use a
value for the Spitzer conduction coefficient, $\kappa_{\rm sp} = 4.6
\times 10^{-7}$~T$^{5/2}$ erg s$^{-1}$~cm$^{-1}$~K$^{-1}$
\citep{1962pfig.book.....S}.  In this situation we are using a value
for the Coulomb logarithm, $\log \Lambda = 37.8$, that is appropriate
for the intracluster medium \citep{1988xrec.book.....S} -- in
astrophysically-relevant, fully ionized plasmas this value varies by
no more than 50\% (see, e.g., Smith et al. 2013, submitted).  It is
quite possible that the local heat flux computed in this way can
become unphysically large in the high-temperature, low-density cluster
regime when using this formulation; therefore, we take into account
the saturation of the heat flux \citep{1977ApJ...211..135C} at a
maximum level of

\begin{equation}
F_{sat} \simeq 0.4 n_e k_b T \left( \frac{2 k_b T}{\pi m_e} \right)^{1/2}.
\end{equation}

To ensure a smooth transition between the Spitzer and saturated
regimes, we define an effective conductivity using the formalism of
\citet{1988xrec.book.....S}

\begin{equation}
\kappa_{eff} = \frac{\kappa_{\rm cond}}{1 + 4.2 \lambda_e / \ell_T} \; ,
\end{equation}

where $\lambda_e$ is the electron mean free path and $\ell_T \equiv T
/ |\grad T|$ is the characteristic length scale of the local
temperature gradient.  We also assume that the conductivity of the
plasma can be described in terms of an effective conductivity, which
can be expressed as a fraction f$_{\rm sp}$ of the Spitzer
conductivity (where f$_{\rm sp} \leq 1.0$ are considered physically
realistic values).  This takes into account physical processes below
the resolution limit of the simulation, such as tangled magnetic
fields, that can suppress heat transport.

Thermal conduction in a plasma can be strongly affected by the
presence of magnetic field lines, which may suppress heat flow
perpendicular to the magnetic field.  In that case, we allow for heat
transport only parallel to the magnetic field lines in
magnetohydrodynamic simulations.  Mathematically, this is given by
equation~(\ref{eq:anisotropic_conduction}.  As with the isotropic
thermal conduction, we allow a multiplicative factor f$_{\rm sp}$ to
take into account the possible suppression of magnetic fields below
the resolution limit of the simulation.

Both isotropic and anisotropic thermal conduction in \enzo\ are
treated in an operator-split manner.  Furthermore, within the heat
transport module, transport along the x, y, and z directions are
computed in a directionally-split fashion, with heat flux along each
direction calculated at the + and - faces of the cell using the
arithmetic mean of the cell-centered temperature in cells $n$ and
$n+1$ or $n-1$ and $n$, respectively (empirically, this is more stable
than taking the geometric mean of the cell-centered temperatures).
The addition of transport along magnetic field lines requires the
calculation of cross-terms in the temperature derivatives at cell
faces, which can result in spurious oscillations in the temperature
field in regions where the temperature gradient is strong in more than
one spatial direction.  Controlling these oscillations requires the
addition of a flux limiter for calculations of the temperature field.
In this case, we choose the monotonized central difference flux
limiter \citep{1977JCoPh..23..263V}, which serves to maintain
numerical stability without sacrificing substantial speed or accuracy.

%% file: numerical-starformation.tex
\subsection{Star formation and feedback}
\label{sec.ov.star}

\subsubsection{Overview}

Due to the computationally unfeasible number of stars in a galaxy
($10^{11}$) and the lack of detailed understanding of star formation, a
number of phenomenological star formation models are included in
\enzo.
Broadly speaking, these methods all work in similar
ways: at a specified time interval, all grid cells that are at the
highest local level of refinement (i.e., that have no child cells) are
examined to see if they meet a set of criteria for star formation.
This may simply be a baryon density threshold, but can also include more
complex tests, such as an examination of local cooling and dynamical
time scales, molecular hydrogen fraction, metallicity, and converging
gas flows.  If the star formation criteria are met, some mass of gas
is taken away from the cell in question and a ``star particle'' with
the same mass is placed in the center of that cell with the same
velocity as the removed gas.  This star particle is then
allowed to inject mass, momentum, thermal energy, metals, and possibly
magnetic fields and/or cosmic ray populations into its local
environment.  In general, the particle is treated as an ensemble of
stars, with feedback properties occuring over time according to the
assumed initial mass function of the stellar population.

In the following sections, we describe some of the more widely-used
star formation and feedback methods used in \enzo.  We note that
similar methods have been employed in many other codes used for galaxy
formation, with comparable implementations for both star formation and
feedback in other grid-based codes.  With regards to particle-based
codes, star formation is broadly similar in implementation, though
feedback is typically implemented in a very different way due to the
Lagrangian nature of the method \citep[see,
e.g.,][]{sh03a,sh03b,hs03}.

\subsubsection{Cen \& Ostriker}
\label{sec:starform_cen}

The Cen \& Ostriker method is a heuristic model of star formation on
galactic scales.  This method, first described by \citet{CO1992},
assumes that stars form in substantially overdense, converging, and
gravitationally unstable gas.  Algorithmically, each cell at the
locally highest level of refinement is examined at each timestep to
see if it meets the critiera for star formation.  Star particles
are allowed to form in a cell if the following criteria are met:
\begin{eqnarray}
\rho_b/\bar{\rho}_b & \geq & \eta,  \\[2mm]
\label{cendens}
\div \myvec{v}_b & < & 0, \\
\label{cencont}
t_{\rm cool} & < & t_{\rm dyn} \equiv \sqrt{3 \pi / 32G \rho_{\rm tot}}, \\
m_{b} & > & m_{J} \equiv G^{-3/2} \rho_{b}^{-1/2}c^{3} 
\left[ 1 + \frac{\delta\rho_{d}}{\delta\rho_{b}} \right]^{-3/2}
\end{eqnarray}
where 
$\eta$ is the user-defined
overdensity threshold, 
and $m_{b}$ and $m_{j}$ are the baryonic mass in the
cell and the Jeans mass of the cell, and c is the isothermal sound speed
in the cell.  If all of these criteria are met, the mass of a star
particle is calculated as \(m_{*} = m_{b} \frac{ \Delta t}{ t_{\rm
    dyn} } f_{\rm *eff} \), where $f_{\rm *eff}$ is the star formation
efficiency parameter.

If $m_{*}$ is greater than a minimum star mass $m_{\rm *min}$, a particle
is created and given several attributes: mass, a unique index number,
the time of formation $t_{\rm form}$, the local dynamical free-fall time
$t_{\rm dyn}$ and the metallicity fraction of the baryon gas in the cell
$f_{\rm Zb}$.  There is a user-defined minimum dynamical time
$T_{\rm dyn,min}$, which is observationally motivated and affects the
feedback rates (see below).  The particle is placed in the center of
the cell and given the same peculiar velocity as the gas in the cell,
and is then treated in the same manner as the dark matter particles.
An amount of gas corresponding to the new particle's mass is
removed from the cell.

The star formation algorithm creates each star particle
instantaneously.  However, feedback should take place over a
significant timescale, as all of the stars contained within the ``star
particle'' would in reality form (and massive stars would die) over a
substantial period of time.  Therefore, we assume that for the
purposes of feedback that the mass of stars formed at a time $t$ with
timestep $\Delta t$ is:
\begin{equation}
\Delta m_{\rm sf} = \int_t^{t+\Delta t} \frac{dM}{dt} dt =
 \int_{\tau_0}^{\tau_1} m_* \tau e^{-\tau} d\tau =
 m_* \left[(1+\tau_0) e^{-\tau_0} - (1 + \tau_1) e^{-\tau_1} \right]
 \end{equation}
where $\tau_0 = (t-t_{\rm form})/t_{\rm dyn}$ and $\tau_1 = (t+\Delta t-t_{\rm form})/t_{\rm dyn}$.

During this timestep, the star particle returns
metal-enriched gas and thermal energy from supernovae and from stellar
winds.  Since massive stars have very short lifetimes, we assume that
there is an immediate feedback of some fraction $f_{\rm SN}$ of the rest
energy from the stars into the gas, such
that $E_{\rm add} = f_{\rm SN} \Delta m_{\rm sf} c^2$, where c is the speed of
light.  In addition, a fraction $f_{Z*}$ of the stellar mass is fed back 
in the form of metals.  Finally, a fraction of the mass $f_{m*}$
is added back into the gas
along with momentum in order to simulate the mass ejection from
all stars (not just supernovae).

There are six user-defined parameters in this algorithm: three deal
with star formation ($\eta$, $m_{\rm *min}$ and $t_{\rm
  dyn,min}$), and three deal with feedback ($f_{\rm SN}$, $f_{Z*}$ and
$f_{m*}$).  Some of these parameters are completely free, while others
can be guided by observation or theory.  For example, the supernova
feedback parameter, $f_{\rm SN}$, can be constrained by assuming that, for
every $200 M_\odot$ of stars created, one supernova occurs, and this
event feeds back approximately $10^{51}$ ergs of thermal energy,
giving:

\begin{equation}
f_{\rm SN} = \frac{10^{51}\, \mathrm{erg}}{200~M_\odot\, c^2} \simeq 3 \times 10^{-6}
\end{equation}

The metal yield $f_{Z*}$, defined as the mass in metals produced per
unit mass of stars created, can be constrained by, e.g., the
theoretical model of \citet{1995ApJS..101..181W}.  This model suggests
that $f_{Z*} = 0.02$ is an appropriate number.  The minimum dynamical
time is set to be $t_{\rm dyn,min} = 10^7$~years to agree with the SN timescales
seen in nearby OB associations.

The other parameters, such as the overdensity threshold $\eta$,
minimum star mass $m_{\rm *min}$, and mass ejection fraction $f_{m*}$ are
not well constrained either theoretically or observationally.  Indeed,
$m_{\rm *min}$ is a purely numerical parameter designed to keep the code
from producing too many star particles, and thus has no observational
or theoretical counterpart.  The $\eta$ parameter, on the other hand,
nominally has physical meaning (i.e., the density above which star
formation must occur on a very short timescale in a self-gravitating
cloud); however, in the vast majority of simulations the densities
that are reachable are nowhere near the densities of protostellar
clouds, and thus this parameter becomes a rough proxy for finding the
densest environments in a given simulation.  

\subsubsection{Schmidt-Law method}
\label{sec:starform_kravtsov}

This method of star particle creation is designed to reproduce
the global Schmidt law of star formation~\citep{2003ApJ...590L...1K,
  1959ApJ...129..243S}.  This algorithm is deliberately minimal, and
is explicitly geared towards modeling star formation in a
phenomenological way on kiloparsec scales.  Stars are assumed to form
with a characteristic gas timescale $\tau_*$ such that $\dot{\rho}_* =
\rho_{\rm gas}/\tau_*$. 
This ``constant
efficiency'' model on the scale of star formation regions is well
motivated observationally
\citep{1996AJ....112.1903Y,2002ApJ...569..157W}.  Star formation is
only allowed to take place in very dense regions with $\rho_{\rm gas}
\geq \rho_{\rm SF}$, where $\rho_{\rm SF}$ is a constant proper (as
opposed to comoving) density threshold.  No other criteria are
imposed.  Typical choices for $\tau_*$ and $\rho_{\rm SF}$ are $\tau_*
= 4$ Gyr and $\rho_{\rm SF} = 1.64~$M$_\odot$~pc$^{-3}$~$(n_{\rm H}
\sim 50$~cm$^{-3})$.  The adopted timescale is derived from the
observationally-determined normalization of the Schmidt law, and the
density threshold is determined by observations of star forming
regions on $\sim 100$ pc scales.

Algorithmically, the star formation events in the Schmidt-law algorithm are
assumed to occur once every global timestep (with the constraint $\Delta t_0 \leq
10^7$ years).
In cells where star formation is determined to occur (i.e. $\rho_{\rm
  gas} \geq \rho_{\rm SF}$), star particles with a mass of $m_* =
\dot{\rho}_* V_{\rm cell} \Delta t_0$ (where $V_{\rm cell}$ is the
volume of the mesh cell) are assumed to form instantaneously in a
manner similar to that described in Section~\ref{sec:starform_cen}.
The \enzo\ implementation of this algorithm is similar, except that
instead of forming stars only at the root grid timestep, we allow
stars to form at the timestep of the highest level of resolution at
any particular point in space.  As can be seen from the equation for
$m_*$ above, this can result in very small stellar masses.  To avoid
memory and processor time issues related to having very large numbers
of star particles we impose a threshold mass $M_{\rm *,min}$ such that
a star particle only forms if $m_* \geq M_{\rm *,min}$.  An
appropriate choice of $M_{\rm *,min}$ does not significantly change
the star overall star formation history of a simulation, though it may
delay the onset of star formation in a given cell relative to a
simulation without a particle mass threshold.

Each ``star particle'' is assumed to represent an ensemble of stars
and is treated as a single-age stellar population (as in the previous
section).  Kravtsov assumes that the stellar initial mass function (IMF) is described by a Miller \&
Scalo functional form with stellar masses between $0.1$ and
$100~M_\odot$ \citep{1979ApJS...41..513M}.  All stars in this IMF with
$M_* > 8~M_\odot$ deposit $2 \times 10^{51}$ erg of thermal energy
and a mass $f_z M_*$ of metals into the cell in which they form
without delay, with $f_z \equiv \min(0.2, 0.01~M_*-0.06)$
(i.e. instantaneous deposition of metals).  The definition of $f_z$ is
a rough approximation of the results of \citet{1995ApJS..101..181W}.

\subsubsection{\HH-regulated method}
\label{sec:starform_H2reg}

The methods described in the previous two sections are generally used
in simulations that have relatively poor resolution, $\Delta x \ga 1$~kpc.
At this physical scale, individual star forming regions are not
resolved, so all uncertainty about the behavior of molecular clouds is
folded into a density threshold for star formation.  In calculations
with much higher resolution, however (on the order of a few pc), individual
molecular clouds can be resolved, thus rendering these approximations
invalid.  To that end, \citet{2012ApJ...749...36K} have implemented
a star formation algorithm that is specifically geared to
high-resolution cosmological simulations of galaxy formation, where
the formation of molecular hydrogen is followed directly and stars
are allowed to form at the highest level of refinement when the local
H$_2$ fraction exceeds a pre-determined threshold.  Cells are examined
every root grid timestep, $\Delta t_0$, and cells at the highest level
that exceed the H$_2$ limit form star particles with masses
proportional to the inferred mass of molecular hydrogen in the
star-forming region
\citep{2008ApJ...689..865K,2009ApJ...693..216K,2010ApJ...709..308M}.
The mass of the particle is calculated as:
\begin{equation} 
m_p = \epsilon \rho_{\rm gas} (\Delta x_{m})^3 \frac{\Delta t_0}{t_*}
\end{equation}
where $\Delta x_{m}$ is the resolution of the maximum level of
refinement, $\epsilon$ is an efficiency parameter with a standard
value of 0.01 \citep[as motivated by][]{2007ApJ...654..304K}, and
t$_*$ is the local free-fall timescale.

The feedback method used in this method is identical to that described
in Section~\ref{sec:starform_cen}.

\subsubsection{Population III star formation}
\label{sec:starform_pop3}

Unlike in the previous sections, under some circumstances it is both
possible and desirable to simulate stars individually, rather than
treating particles as ensembles.  One particular example of this is
Population III star formation
\citep{ABN02,2007ApJ...654...66O,2008ApJ...685...40W,2009Sci...325..601T},
where a given halo may only form one star of primordial composition.
To accommodate this, \enzo\ contains a star formation algorithm that
forms individual Population III stars directly \citep{2007ApJ...659L..87A,
  2008ApJ...685...40W, 2012MNRAS.427..311W}.  Using criteria similar
to~\citet{CO1992}, a star particle forms when a cell meets all of the
following conditions:

\begin{enumerate}
\item A baryon overdensity of $5 \times 10^5$ (corresponding to a
  hydrogen number density of 
  roughly $10^3$ cm$^{-3}$ at $z=10$),

\item A converging gas flow ($\div \myvec{v} < 0$), and

\item A molecular hydrogen mass fraction f$_{\rm H2} > 5 \times 10^{-4}$,
\end{enumerate}

These are comparable to the conditions typical of a collapsing metal-free molecular cloud roughly
10 million years before the birth of a Pop III main-sequence star.  If
multiple neighboring cells are flagged as being able to form stars, a
single star is created instead.  This star has a mass that is randomly
sampled from a stellar IMF with a functional form of

\begin{equation}
f(\log M) dM = M^{\alpha} \exp \Big[ -\Big( \frac{M_{\rm char}}{M}
\Big)^{\beta} \Big]
\end{equation}

Feedback from Population III stars created using this algorithm comes in multiple forms.
Radiative feedback using the Moray radiation transport algorithm
\citep{Wise11_Moray} is available, using the mass-dependent hydrogen
ionizing and Lyman-Werner photon luminosities and lifetimes of the
Population III stars from \citet{2002A&A...382...28S}.  At the end of
their main-sequence lifetimes, explosion energies, ejected gas mass,
and ejected metal are calculated using a variety of sources that
depend on the mass of the star, and are described in detail in Section
3.2.1 of \citet{2012MNRAS.427..311W}.

%% file: numerical-timestep.tex
\section{Timestepping}
\label{sec.timestepping}

In \enzo, the integration of the equations being solved is generally
adaptive in time as well as in space.  The timestep $\Delta t$ is set
on a level-by-level basis by finding the largest timestep such that
all of the criteria listed below (that are relevant for the simulation
in question) are satisfied.  The timestep criteria are given by the
following expressions, showing the one-dimensional case for clarity:

\begin{equation}
\Delta t_{\rm hydro} = \min \left( \kappa_{\rm hydro} \frac{a \Delta x}{c_{s} + |v_x|} \right)_L ,
\label{eqn:dthydro}
\end{equation}

\begin{equation}
\Delta t_{\rm MHD} = \min \left( \kappa_{\rm MHD} \frac{a \Delta x}{v_{f} + |v_x|} \right)_L ,
\label{eqn:dtMHD}
\end{equation}

\begin{equation}
\Delta t_{\rm dm} = \min \left(\kappa_{\rm dm} \frac{a \Delta x}{v_{dm,x}} \right)_L ,
\label{eqn:dtdarkmatter}
\end{equation}

\begin{equation}
\Delta t_{\rm accel} = \min \left( \sqrt{\frac{\Delta x}{|\myvec{g}|}} \right)_L ,
\label{eqn:dtaccel}
\end{equation}

\begin{equation}
\Delta t_{\rm rad} = \min \left(  \sqrt{\frac{\Delta x}{|\myvec{a}_{rad}|}} \right)_L,
\label{eqn:dtrad}
\end{equation}

\begin{equation}
\Delta t_{\rm cond} = \min \left(  \frac{ k_{\rm cond}}{f_{\rm sp}} \frac{\Delta x^2
    n_b}{\kappa_{\rm sp}(T)} \right)_L,
\label{eqn:dtcond}
\end{equation}

\begin{equation}
\Delta t_{\rm exp} = f_{\rm exp} \left( \frac{a}{\dot{a}} \right) ,
\label{eqn:dtexpand}
\end{equation}

In equations~(\ref{eqn:dthydro})-(\ref{eqn:dtcond}), the $\min (
\ldots)_L$ formalism means that this value is calculated for all cells
or particles on a given level L and the minimum overall value is taken
as the timestep.

Equation~(\ref{eqn:dthydro}) ensures that all cells satisfy the
Courant-Freidrichs-Levy (CFL) condition for accuracy and stability of
an explicit finite difference discretization of the Euler equations.
In this equation, $\kappa_{\rm hydro}$ is a dimensionless numerical
constant with a value of $0 < \kappa_{\rm hydro} \leq 1$ (with a
typical value of $\kappa_{\rm hydro} \sim 0.3-0.5$) that ensures that
the CFL condition is always met, and $c_s$ and $v_x$ are the sound
speed and peculiar baryon velocity in a given cell.

Equation~(\ref{eqn:dthydro}) is valid for one dimension, and is used
when the equations of hydrodynamics are being solved.  For 2 or 3
dimensions, it was shown by \cite{Godunov1959} that using the harmonic
average of the timestep found along each of the coordinate axes yields
a maximum $\kappa_{\rm hydro} = 0.8$.  So letting $\Delta t_x$,
$\Delta t_y$, and $\Delta t_z$ be the analogues of
equation~(\ref{eqn:dthydro}) along the $x,y$ and $z$ axes,

\begin{equation}
  \Delta t_{\rm hydro} = \min \left( \frac{\kappa_{\rm hydro}} {1/\Delta t_x
  +1/\Delta t_y + 1/\Delta t_z} \right)_L
\end{equation}

For all other criteria except for equation~(\ref{eqn:dtMHD}), multiple
dimensions are accounted for by repeating the one dimensional
criterion along each axis, and taking the minimum.

Equation~(\ref{eqn:dtMHD}) is only enforced when the equations of
magnetohydrodynamics are being solved, and is directly analogous to
equation~(\ref{eqn:dthydro}) in that it ensures that the CFL condition
is being enforced at all times.  In this equation, $\kappa_{\rm MHD}$
is a dimensionless numerical constant with a value of $0 < \kappa_{\rm
MHD} \leq 1$ (with a typical value of $\kappa_{\rm MHD} \sim 0.5$)
that ensures that the CFL condition is always met, and $v_f$ and $v_x$
are the ``fast wave speed'' and peculiar baryon velocity in a given
cell.  The fast wave speed comes from a stability analysis of the MHD
equations, and is given by:
\begin{equation}
v_f = \sqrt{ \frac{1}{2} \left(  v_A^2 + c_s^2 + \sqrt{(v_A^2 +
      c_s^2)^2 - 4 v_A^2 c_s^2}  \right)  },
\label{eqn:vfastmhd}
\end{equation}
where $c_s$ is the sound speed and $v_A$ is the Alfven speed,
calculated as $v_A = \sqrt{B^2/\rho_B}$ in units where $\mu_0 = 1$.

Equation~(\ref{eqn:dtdarkmatter}) is analogous to
equation~(\ref{eqn:dthydro}) and helps to ensure accuracy in the
N-body solver by requiring that no particle travels more than one cell
width.  The parameter $\kappa_{\rm dm}$ is like $\kappa_{\rm hydro}$,
with a similar range of values.  This criterion is used when massive
particles are included in a simulation.

Equations~(\ref{eqn:dtaccel}) and~(\ref{eqn:dtrad}) are supplementary
to equation~(\ref{eqn:dthydro}) in that they take into account the
possibility of large accelerations due to either gravity
(equation~\ref{eqn:dtaccel}) or radiation pressure
(equation~\ref{eqn:dtrad}).  In equation~(\ref{eqn:dtaccel}),
$\myvec{g}$ is the gravitational acceleration in each cell on level l.
In equation~(\ref{eqn:dtrad}), $\myvec{a}_{\rm rad}$ is the estimated
acceleration due to radiation pressure in each cell on level l,
defined as

\begin{equation}
\myvec{a}_{rad} = \frac{ \sum_i \frac{\dot{E_i}}{c} \hat{r_i} }{m_b} 
\end{equation}

where the sum calculates the energy deposited in a cell during the
previous timestep due to \textit{all photon packets} that crossed that
cell, with $\hat{r_i}$ being a unit vector that accounts for the
packet direction.

Equation~(\ref{eqn:dtcond}) is the stability condition for an explicit
solution to the equation of heat conduction.  In this expression,
$n_b$ is the baryon number density, $\kappa_{\rm sp}(T)$ is the
Spitzer thermal conductivity, and f$_{\rm sp}$ is a user-defined,
dimensionless conduction suppression factor whose value must be
f$_{\rm sp} \leq 1$.  $k_{\rm cond}$ is a dimensionless prefactor
whose value must be $0 < k_{\rm cond} \leq 0.5$, and is exactly 0.5
for the implemention in \enzo.  From a practical perspective, it is
useful to note that, unlike other timestep criteria discussed above
(which effectively scale as $\Delta x / \sqrt{T}$ with the grid cell
size $\Delta x$ and temperature $T$), the timestep criterion due to
thermal conduction scales as $\Delta x^2 / T^{2.5}$, which can result
in a rapid decrease in timestep in regions of high resolution and/or
temperature.

Finally, equation~(\ref{eqn:dtexpand}) is a cosmological constraint
that limits the timestep so that the simulated universe only expands
by some fractional amount, f$_{\rm exp}$, during a single step.  In
this equation, $a$ and $\dot{a}$ refer to the scale factor of the
universe and its rate of change, respectively.  This criteria is
necessary because the expansion of the universe and its first
derivative with respect to time both appear in the equations of
cosmological (magneto)hydrodynamics and particle motion, and some
limit is required for the stability of the PPM algorithm in comoving
coordinates.  This criterion typically limits the simulation's
timestep only during the earliest phases of a cosmological simulation,
before substantial structure has formed.

%% file: numerical-analysis.tex
\section{Analysis}
\label{sec.num.analysis}

\subsection{Inline analysis with yt}

Detailed analysis of simulation results requires both the tools to ask
sophisticated questions of the data and the ability to process vast
quantities of data at high time-cadence.  As simulations grow in size
and complexity, storing data for post-processing simply becomes
intractable.  To cope with this, we have instrumented \enzo\ with the
ability to conduct analysis during the course of a simulation.  This
enables analysis with extremely high time cadence (as often as every
subcycle of the finest refinement level), without attempting to write
an entire checkpoint output to disk.  The current mechanism for
conducting analysis in \enzo\ during the course of the simulation
utilizes the same computional resources as are used by the simulation
itself by transferring their usage from \enzo\ to the analysis
routines; this is often referred to as \textit{in situ} analysis or
visualization.  Utilizing a dynamically-scheduled second set of
computing resources, often referred to as \textit{co-scheduled}
analysis or visualization, provides greater flexibility and overall
throughput at the expense of simplicity.

We expose \enzo's mesh geometry and fluid quantities to the analysis
platform \texttt{yt} \citep{2011ApJS..192....9T, 2011arXiv1112.4482T}.
At compile time, \enzo\ is (dynamically or statically) linked against
the Python and NumPy libraries necessary to create proxy objects
exposing the mesh geometry, fluid quantities and particle arrays as
NumPy arrays.  This information is then passed to a special handler
inside \texttt{yt}.  \texttt{yt} interprets the mesh and fluid
information and, without saving data to disk, constructs a native
representation of the in-memory state of the simulation that appears
identical to an on-disk simulation output.  \enzo\ then executes a
user-provided analysis script, which is able to access the in-memory
simulation object.  Once the analysis script has returned control to
\enzo, the simulation proceeds.  This process can occur at either the
top of the main ``EvolveHierarchy'' loop or at the end of a timestep
at the finest level, and the frequency with which it is called is
adjustable by a run-time parameter.  During the course of conducting
analysis, the simulation is halted until the conclusion of the
analysis.

Most analysis operations that can be performed on data sets that
reside on disk can be performed on in-memory data sets in \texttt{yt}.
This includes projections (i.e., line integrals, both on- and
off-axis), slices, 1-, 2-, and 3-D fluid phase distributions,
calculation of derived quantities and arbitrary data selection.  As of
version 2.5 of \texttt{yt}, the Rockstar phase-space halo finder
\citep{2013ApJ...762..109B} can be executed through \texttt{yt} on
in-memory \enzo\ data, and so can the Parallel HOP halo finder
\citep{1998ApJ...498..137E, 2010ApJS..191...43S}.  Operations that
currently cannot be conducted on in-memory datasets are those that
require spatial decomposition of data.  For instance, calculating
marching cubes on a data object with \texttt{yt} is a fully local
operation and can be conducted \textit{in situ}.  However, calculating
topologically-connected sets requires a spatial decomposition of data
and thus cannot be conducted \textit{in situ}.  This prohibition
extends to halo finding operations other than Rockstar, most
multi-level parallelism operations, and volume rendering.

Where microphysical solvers or other operator-split physics
calculations can be done in Python, \texttt{yt} can serve as a driver
for these calculations.  A major feature set that is currently being
developed is to pass structured (i.e., non-fluid) information back
from \texttt{yt} into \enzo.  For instance, this could be the result
of semi-analytic models of the growth and evolution of star clusters,
galaxy particle feedback parameters that have been influenced by
merger-tree analysis, or even spectral energy distributions that are
calculated within \texttt{yt} and provided to \enzo\ as input into
radiative transfer calculations.  Future versions will include this,
as well as the ability to dynamically allocate computional resources
to \texttt{yt} such that the simulation may proceed asynchronously
with analysis (\textit{co-scheduled} analysis).  With this
functionality will also come the ability to dynamically partition
data, such that spatially-decomposed operations such as volume
rendering become feasible during the course of a simulation.

\subsection{Tracer particles}

One of the inherent drawbacks of a grid-based fluid method is the
inability to follow the evolution of a single parcel of fluid as it
travels through the simulation volume.  To address this, \enzo\ has
the capability to introduce Lagrangian ``tracer particles'' into a
calculation either at the beginning of the simulation or when
restarting the calculation.  These tracer particles are put into the
simulation in a rectangular solid volume with uniform, user-specified
spacing.  Each particle's position and velocity is updated over the
course of a single timestep $\Delta t$ as follows:

\begin{eqnarray}
\label{eqn.drifttrace}
x^{n+1/2} & = & x^n + (\Delta t/2) v^{\rm interp,n} \nonumber \\
v^{n+1} & = & v^{\rm interp,n+1} \\
x^{n+1} & = & x^{n+1/2} + (\Delta t/2) v^{n+1} \nonumber
\end{eqnarray}

This is essentially a drift-kick-drift particle update from time $n$
to time $n+1$ -- however, instead of computing an acceleration at the
half-timestep $t^n + \Delta t/2$ (as is done for massive particles --
see Equation~\ref{eqn.driftkick}), the particle's velocity is updated
both at the beginning of each timestep and at the half-timestep by
linearly interpolating the cell-centered baryon velocity to the
position of the particle ($v^{\rm interp}$), and assigning it to that
value.

\enzo\ saves tracer particle data at user-specified intervals,
independent of the intervals at which regular data sets are written.
The data written out typically includes the tracer particle's unique
ID and position, as well as the velocity, density, and temperature of
the gas at its location, but is easily extensible to output any
grid-based quantity that a user requires.  This capability has been
used quite effectively in several papers, including
\citet{2010ApJ...715.1575S} and \citet{2012ApJ...748...12S}. It is
crucial to keep in mind that tracer particles model a fixed number of
Lagrangian fluid trajectories. The fluid on the grid, however, models
the motion of all the mass and represents the average quantity in a
grid cell's volume. Consequently, after a period of evolution, tracer
particles -- even if they initially had the same density distribution
as the gas -- will not have the same density distribution as the
fluid. For example, they tend to accumulate at stagnation points of
the flow, and care has to be taken in using these particles
appropriately. Tracer particles are very useful in studies such as the
variety of histories of the hydrodynamic quantities in Lagrangian
fluid elements and when evaluating complex chemical and cooling models
in regard to the simpler ones used in the actual numerical evolution.

\subsection{Shock finding}

Identification of shocks and their pre- and post-shock conditions can
be accomplished through a combination of either temperature or
velocity jumps with dimensionally split or unsplit search methods.
The primary method used in \enzo~is the dimensionally unsplit
temperature jump method described in detail in
\citet{2008ApJ...689.1063S}.  We briefly outline the method here.

For every cell, we first determine whether it satisfies the following
conditions necessary to be flagged as a shock:

\begin{eqnarray}
\div \myvec{v} < 0,\\
\grad T \cdot \grad S > 0,\\
T_2 > T_1,\\
\rho_2 > \rho_1,
\end{eqnarray}

where $\myvec{v}$ is the velocity field, $T$ is the temperature,
$\rho$ is the density, and $S=T/\rho^{\gamma-1}$ is the entropy.
$T_2$ and $T_1$ are the post-shock~(downstream) and
pre-shock~(upstream) temperatures, respectively. An optional
temperature floor may be chosen, which is useful for situations such
as cosmological simulations without a radiation background where
underdense gas in the intergalactic medium (IGM) can cool
adiabatically to unphysically low temperatures.

Once a cell is flagged, the local temperature gradient is calculated,
which is then used to traverse cells parallel to the gradient to
search for the first pre- and post-shock cell that do not satisfy the
above conditions.  If during the search a cell satisfying the
conditions is found to have a more convergent flow, that cell is
marked as the center, and the search is started again. Using the
temperature values from each of these cells, the Mach number is then
solved using the Rankine-Hugoniot temperature jump conditions:

\begin{equation}
\frac{T_2}{T_1} = \frac{(5\Mach^2 - 1)(\Mach^2 + 3)}{16\Mach^2},
\end{equation}

where $\Mach$ is the upstream Mach number.

Shock finding in the context of AMR is applied grid-by-grid.  If a
search for pre/post-shock cells goes outside the bounds of the grid
ghost zones, the search is stopped for that particular shocked
cell. In most situations this is adequate since the hydrodynamic shock
is captured in fewer than the number of ghost zones.  Shock finding
can be run either upon data output or at each step in the evolution of
an AMR level if the Mach and pre/post-shock quantities are needed for
additional physics modules.

%% file: code-tests.tex
\section{Code tests}
\label{sec.tests}

Ensuring that a complex piece of software is behaving correctly is a
non-trivial task.  While there are a range of techniques that one can
apply to ensure correctness, the \enzo\ code uses two primary methods:
a suite of test problems that can be compared to previous versions of
the code, used to ensure that \enzo\ is running correctly on a new
computer, with a new compiler, or after substantial changes have been
made to the code base; and by direct comparisons to other
astrophysical fluid dynamics codes.  We describe the \enzo\ test
methodology in Section~\ref{sec.tests.suite}, discuss code comparisons
involving \enzo\ in Section~\ref{sec.tests.compare}, and show a small
set of representative test problems in
Section~\ref{sec.tests.problems}.  We further note that the \enzo\
test suite (Section~\ref{sec.tests.suite}) contains hundreds of test
problems, as well as the ability to compare to a ``gold standard''
solution, and thus all of the tests shown here are easily reproducible
by the reader.  To facilitate this, all of the test problems included
in this paper, as well as scripts to run the test problems and
generate the figures found in Section~\ref{sec.tests.problems}, can be
found on the \enzo\ website.

\subsection{Verifying and validating the \enzo\ code}
\label{sec.tests.vandv}

\input{tests-vandv}

\subsection{Representative test problems}
\label{sec.tests.problems}

The \enzo\ test suite (described in Section~\ref{sec.tests.suite})
contains hundreds of test problems that probe the code's behavior in a
wide range of physical circumstances, and explicitly tests each physics
package in the \enzo\ code, both individually and in combination.  It
is impractical to include a substantial fraction of these problems in
a method paper; as a result, we have chosen to publish the results of
only a small subset of particularly crucial test problems here.  If
the interested reader desires, they can download \enzo\ and
\texttt{yt}, run the test suite, and see the results of any other test
problem and its comparison to an analytic solution (if available), or
to a ``gold standard'' solution from a stable version of \enzo.

The general structure of each test problem description is as follows.
We will describe the construction of the test problem (including its
initial and boundary conditions), the analytical or expected
solutions, and motivate why we have included it in the paper.  After
that, we will show and describe \enzo's solution to the test problem.
We remind the user that, as discussed at the end of
Section~\ref{sec.tests.suite}, they can download a Mercurial
repository containing this paper, the \enzo\ test problem parameter
files that produced the simulation data used to generate the figures
in this section, and the \texttt{yt} scripts necessary to create the
figures themselves.

\input{tests-sodshocktube}

\input{tests-wavepool}

\input{tests-shockpool}

\input{tests-doublemach}

\input{tests-sedov}

\input{tests-pointsourcegravity}

\input{tests-orbit}

\input{tests-selfsiminfall.tex}

\input{tests-zeldovichpancake}

\input{tests-orszagtang}

\input{tests-onezonecollapse}

\input{tests-photoevaporation}

\input{tests-fld-ifront}

\input{tests-aniso-conduct}


%% file: tests-vandv.tex
\subsubsection{The \enzo\ test suite}
\label{sec.tests.suite}

\enzo\ is capable of simulating a large variety of problems types,
with all but a few of these types requiring only a parameter file as
an input.  The most notable exception is the cosmology simulation,
which takes as input initial conditions created by other codes.  The
suite of test problems spans a wide range in complexity.  At one end
of this spectrum are simple problems that utilize only a single
component of \enzo\ and for which analytic solutions exist for
comparison with the simulation results.  At the opposite end are
problems that exercise a large portion of \enzo's machinery in
concert.  Together, the available problem types fully cover all of
\enzo's functionality.  This enables them to serve as a vehicle for
verifying that the code's behavior remains stable over time, on new
computing platforms, and after modification of the codebase.

\enzo\ uses an automated testing framework that allows a user to run,
with a single command, a set of test problems and compare the results
against results produced by any other version of the code.  Within the
\enzo\ source distribution, the test problem parameter files are
stored in a nested directory structure organized according to the
primary functionality tested (e.g., hydrodynamics, gravity, cooling,
etc.).  Each parameter file is accompanied by a text file containing
various descriptive keywords, such as the machinery tested, the
dimensionality, and the approximate run time.  A test runner script is
responsible for taking as input from the user a series of keywords
that are used to select a subset of all available test problems.  The
test problems are also grouped into three suites: the quick, push, and
full suites, each a superset of the ones named before.  The quick
suite is considered to minimally cover the primary functionality in
\enzo\ and is designed to be run repeatedly during the development
process.  The push suite has slightly increased feature coverage and
is mandated to be run before code changes are accepted into the main
repository.  The full suite consists of all test problems that can be
run with no additional input.  Approximate run times for the quick,
push, and full suite are 15 minutes, 1 hour, and 60 hours,
respectively, on a relatively new desktop computer (circa 2013).

After the test problems are selected by the test runner script, they
are run in succession using either the \enzo\ executable contained
within that distribution or an external executable built from another
\enzo\ version and specified by the user.  This allows for any version
of the code to be tested with an identical set of test problems.
After running the test problem simulations, the test runner then
performs a series of basic analysis tasks using the \texttt{yt}
analysis toolkit \citep{2011ApJS..192....9T, 2011arXiv1112.4482T}.
The default analysis performed on all test problems includes
calculation of various statistics (such as extrema, mean, and
variance) on the fields present in the output data.  Custom analysis
provided by scripts that accompany the test problem parameter files is
run for special cases, such as when an analytical solution exists that
can be compared against the simulation data.  After the analysis is
performed, the results are compared against a set of gold standard
results that are maintained on a website and downloaded on the fly by
the test runner script.  Alternately, results from any version of
\enzo\ can be stored locally and compared to any other version of the
code.  In Section~\ref{sec.tests.problems}, we describe some of the
key test problems that are used to verify proper behavior.  All of
these test problems, as well as the scripts to generate the figures
from them, are available at
\url{http://bitbucket.org/enzo/enzo-method-paper}, the Bitbucket
repository for the \enzo\ method paper.  The \texttt{test\_problems}
subdirectory in this repository contains all of the files necessary to
regenerate all of the figures in Section~\ref{sec.tests.problems}.  We
note for completeness that the figures in
Section~\ref{sec.tests.problems} were generated using \texttt{yt}
version 2.5.3 and the \url{https://bitbucket.org/enzo/enzo-dev}
repository of \enzo\ with changeset \texttt{5d8c412}, which
corresponds to \enzo\ 2.3.

\subsubsection{Code comparisons}
\label{sec.tests.compare}

Over the course of its existence, \enzo\ has been involved in numerous
comparisons with other astrophysical codes used for self-gravitating
fluid dynamics.  In general, \enzo\ behaves in a manner similar to
other grid-based (and particularly AMR-based) codes, as we will
summarize below.

\enzo\ has been involved in multiple cosmological code comparisons,
including the Santa Barbara Cluster code comparison project
\citep{SantaBarbara}, a large comparison of N-body simulations
\citep{2008CS&D....1a5003H}, as well as several direct comparisons
between \enzo\ and the GADGET SPH code in a variety of set-ups,
including N-body and adiabatic hydrodynamics
\citep{2005ApJS..160....1O,2005MNRAS.364..909V, 2011MNRAS.418..960V}
and simulations of the Lyman-alpha forest \citep{2007MNRAS.374..196R}.
Compared to the other codes involved in these projects, the \enzo\
code typically has a more difficult time resolving small-scale
self-gravitating structures (for an equivalent dark matter particle
count and nominally equivalent force resolution), but does comparably
well as a tree-based code for larger structure, and is typically
superior in terms of resolving fluid features due to its higher-order
(and artificial viscosity-free) PPM hydrodynamics solver.  When
examining classical test cases such as the Santa Barbara project
\citep{SantaBarbara}, \enzo\ forms galaxy clusters with very similar
density, temperature, and entropy profiles to other grid-based codes
that use Godunov-type hydro methods, which systematically differed
from particle-based codes using SPH in this code comparison.
Similarly, in galaxy cluster simulations that look specifically at the
properties of cosmological shocks \citep[e.g.][]{2011MNRAS.418..960V},
\enzo\ produces results that are similar to other high-order
grid-based hydrodynamics codes, and a far superior performance of
\enzo\ is observed (in terms of resolution of fluid features and shock
detection) in low-density regions when compared to a particle-based
code.  In tests of the Lyman-alpha forest that include radiative
cooling and a uniform metagalactic ultraviolet background, \enzo\ and
GADGET provide results on metrics such as the matter power spectrum
that are comparable to within 5\% \citep{2007MNRAS.374..196R}.

Several comparisons have been made that focus specifically on
hydrodynamics solvers and fluid behavior.  In particular, the work of
\citet{2007MNRAS.380..963A} and \citet{Tasker2008} perform direct
comparisons between several grid- and particle-based codes for a
variety of fluid-centric test problems (including shocked gas clouds,
self-gravitating, translating clouds, and Sedov-Taylor explosions),
and show that \enzo\ is comparable or superior in behavior to the
other grid-based hydrodynamics codes involved in the comparison, and
provide useful information on the sort of practical challenges that a
user of an AMR code such as \enzo\ may experience.  More specific
comparisons, including one testing the linear and nonlinear growth of
the Kelvin-Helmholz instability \citep{2012ApJS..201...18M}, as well
as a comparison that more broadly examines Galilean invariance in
grid-based codes \citep{2010MNRAS.401.2463R}, show that \enzo, and in
particular its implementation of the PPM hydrodynamics solver,
converge to the correct solution as expected, and generally provide
less diffusive solutions than lower-order codes, including those that
use artificial viscosity.  Finally, there have been two code
comparison projects that focus on turbulence simulations.  The first
studied the behavior of decaying isothermal supersonic turbulence
\citep{2009A&A...508..541K}, and the second examined supersonic
magnetohydrodynamical turbulence \citep{2011ApJ...737...13K}.  Both
included the \enzo\ code, with the former testing the PPM
hydrodynamics and the latter both the constrained transport MHD
implementation of \citet{Collins10} and the Dedner method of
\citet{WangAbelZhang08}.  In both cases, \enzo\ performed similarly to
other grid-based codes that use Godunov-based fluid solvers, and
typically had better effective resolution than particle-based codes
when using the same number of particles as the number of grid cells in
the \enzo\ simulation.

Three other comparisons between the \enzo\ code and other simulation
tools have been performed that focus on physics other than gravity and
fluid flow.  The flux-limited diffusion radiation transport scheme was
measured against several test problems by \citet{IlievEtAl2009}, which
involved tests with and without analytical solutions.  \enzo\ produced
results similar to both the analytical solutions and results obtained
by other codes.  We note, however, that there were minor differences
throughout the comparison between various codes, and the majority of
the codes differed in at least a subset of the tests.
\citet{2011ApJ...726...55T} show the result of varying reaction rate
coefficients for the formation of molecular hydrogen via the 3-body
process in both the \enzo\ and GADGET-2 codes, observing similar
trends between the two codes. However, at nominally equivalent
resolution where the particle and cell gas masses are comparable,
\enzo\ simulations typically displayed a substantially higher level of
gas structure.  This is unsurprising due to \enzo's higher-order hydro
solver.  Finally, \citet{2012ApJ...744...52P} show the results of
comparing \enzo\ in its non-AMR mode to a smoothed-particle
hydrodynamics code, SNSPH, in the context of common-envelope binary
stellar interactions.  The authors show that the codes display
reasonable convergence properties as a function of simulation
resolution, and also agree quite well with each other. However, the
observed mass-loss rates do not agree particularly well with
observations.

%% file: tests-sodshocktube.tex
\subsubsection{Sod Shock Tube}
\label{sec.tests.sodshock}

\begin{figure}
\begin{center}
\includegraphics[width=\textwidth]{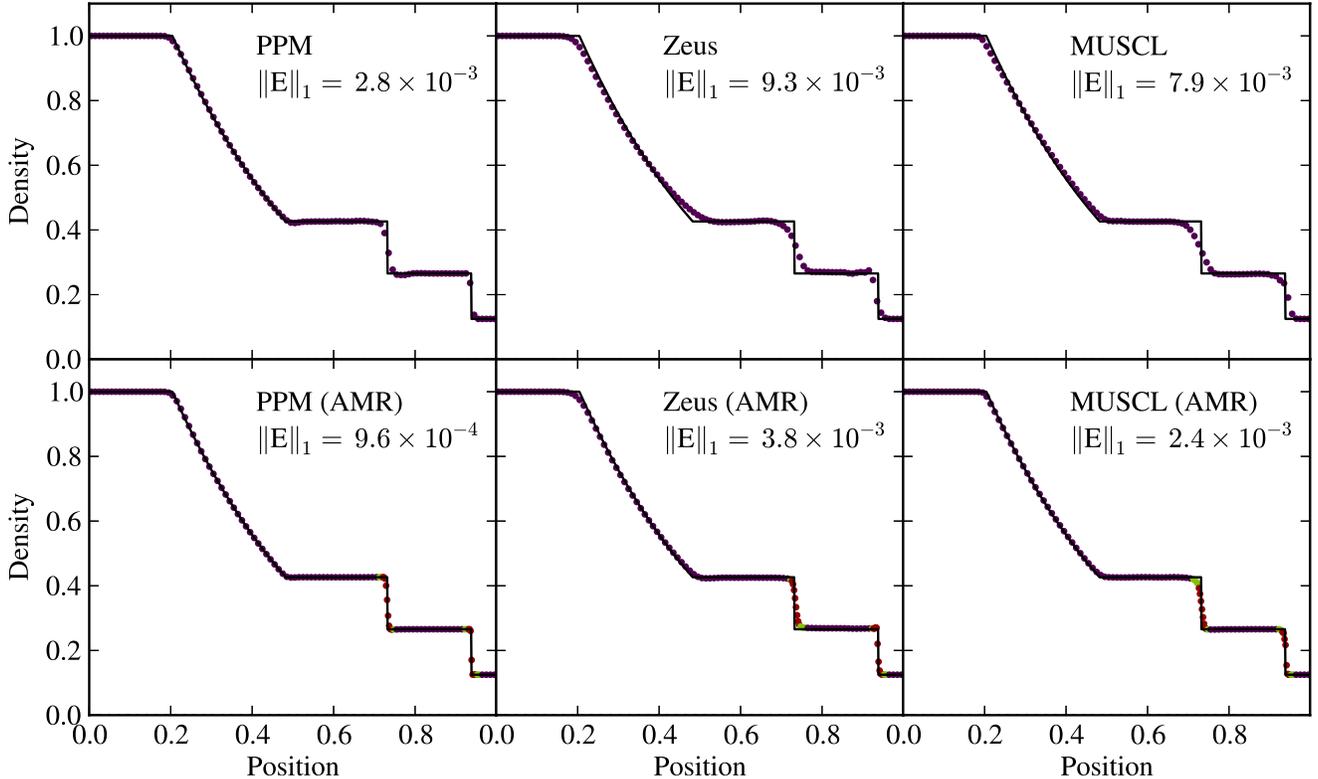}
\caption{The density distribution of the classic Sod Shock Tube for
three different solvers (from left to right column) and with (bottom
row) and without (top row) AMR.  In each case 100 zones are used on
the root level and the results are shown at $t=0.25$.  All cells are
plotted and color-coded by level with purple indicating level 0, green
level 1 and red level 2 (at the time shown, only small region
surrounding the contact discontinuity and the shock are refined).  In
each panel, we show the analytic solution as a solid line and the
$E_1$ error norm in the upper right.}
\label{fig.sodshocktube}
\end{center}
\end{figure}

We begin the paper test suite with the classic one-dimensional Sod
shock tube problem \citep{Sod78}, which provides a good test of a
hydrodynamical solver's ability to resolve a clean Riemann problem
with clear separation between the three resultant waves.  These waves
consist of a rarefaction fan, a contact discontinuity, and a moderate
shock.  The initial state is $\rho_{\rm L}, P_{\rm L} = 1.0, 1.0$ on
the left of the boundary at $x=0.5$ and $\rho_{\rm R}, P_{\rm R} =
0.125, 0.1$ on the right.  All velocities are initially zero.  In
Figure~\ref{fig.sodshocktube}, we show the density solution at the
final time ($t=0.25$) for three of our hydro solvers -- the spatially
third-order PPM, as well as the two second-order \zeus\ and MUSCL
schemes.  We use 100 cells across the domain, which is a relatively
standard choice in code method papers, and show solutions both with
and without adaptive mesh refinement.  Without AMR (top row in Figure
\ref{fig.sodshocktube}), it is clear that the PPM scheme produces by
far the cleanest solution with all wave families crisply reproduced
(in particular, the contact discontinuity and shock).  \zeus\ and
MUSCL produce similar results, with MUSCL doing a slightly better job
on the rarefaction fan.  In Figure \ref{fig.sodshocktube}, we also
show the integrated absolute deviation from the exact solution,
$||E_1|| = \sum_i \Delta x_i |F(x_i)-{\rm Exact}(x_i)|$.  These
numbers confirm the qualitative differences noted previously.

We also run the same simulation but with two levels of AMR (using a
refinement factor of 2), triggered based on a normalized slope greater
than 10\% in the density.  This refinement criterion results in only
the refinement of strong gradients, and does not include the
rarefaction fan at late times.  The results are shown in the bottom
row of Figure~\ref{fig.sodshocktube}.  Using AMR, the results are much
better for all three methods, with much sharper shocks and contact
discontinuities and even a better representation of the rarefaction
wave, which is only refined beyond the root grid at early times.
Although the results are improved for all methods, PPM still produces
the best result, as is shown clearly by the computed error norms
(displayed again in each of the individual panels).

%% file: tests-wavepool.tex
\subsubsection{Wave pool}
\label{sec.tests.wavepool}

In this one-dimensional test we pass a short wavelength linear wave
through a static singly-refined region.  The full domain is from 0 to
1 and the refined region is set from 0.25 to 0.75 at all times.  We
modify the left boundary consistent with a single linear sound wave
with density given by $\rho(x,t) = \rho_0 (1 + A \sin(kx - \omega
t))$, where the amplitude is $A = 0.01$ and the wavelength $\lambda =
k/2\pi = 0.1$.  Similar expressions exist for the pressure and
velocity.  The initial, unperturbed density and pressure are set to
unity and we adopt $\gamma = 1.4$.  The (unrefined) root grid is
covered with 100 cells, resulting in a wavelength for the linear wave
of only 10 cells.  This is, therefore, a challenging problem for hydro
methods.  We are particularly interested in any reflection or
artifacts introduced by the wave entering and exiting the refined
regions.

\begin{figure}
\begin{center}
\includegraphics[width=0.8\textwidth]{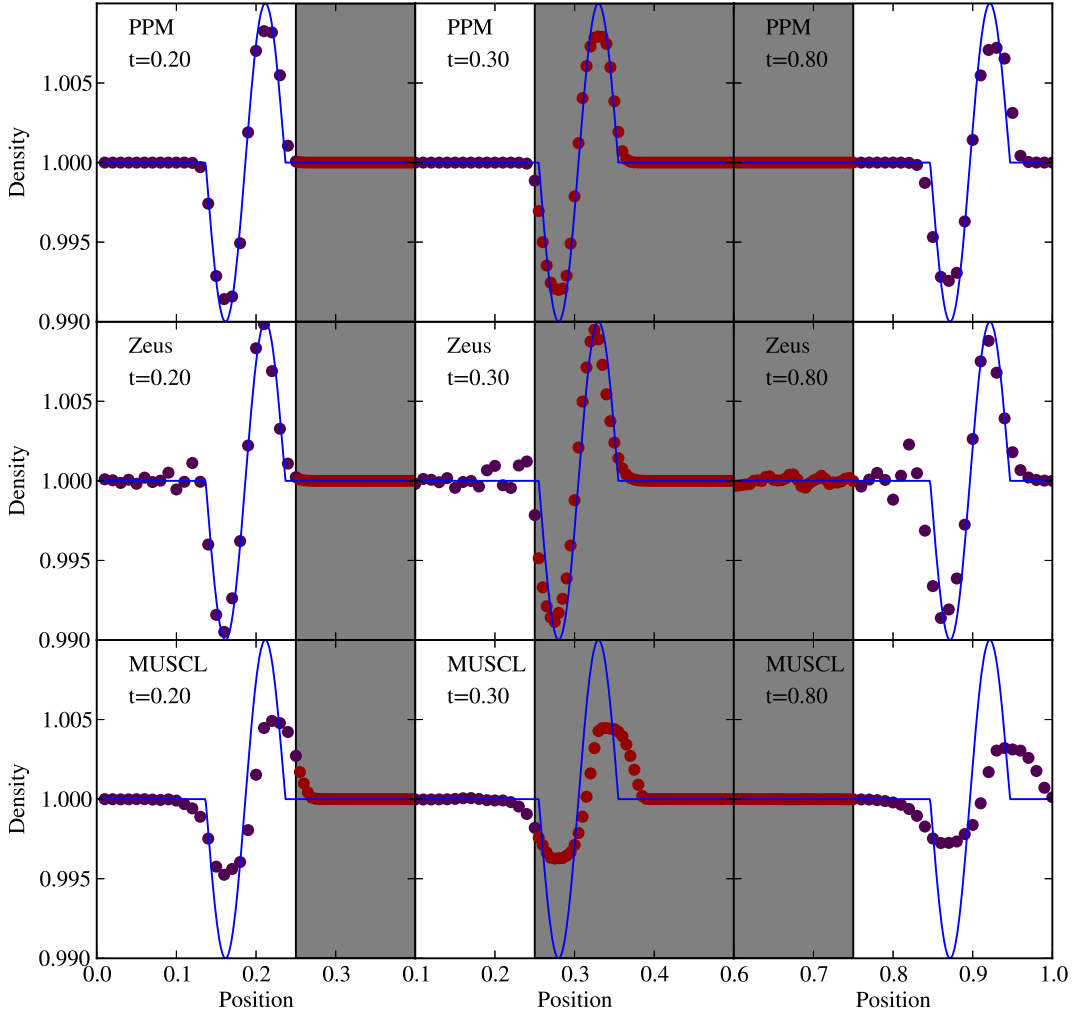}
\caption{This plot shows, for each column, three snapshots at $t=0.2,
0.3$ and 0.8 of a linear wave as it propagates through the domain.  A
static refined region extends from $x = 0.25$ to 0.75 and is shown in
grey in each panel.  The individual cells are also color-coded by
level: blue indicates the root grid, and red is for the refined
region.  Each row shows the result for a different hydrodynamic solver
(top is PPM, middle is \zeus, and the bottom is MUSCL).  The solid
line shows the analytic solution for a linear, undamped wave.  Note
that we focus each panel on a small region of the entire domain to
better show the wave itself.}
\label{fig.wavepool}
\end{center}
\end{figure}

Figure~\ref{fig.wavepool} shows the evolution of the wave at three
different times ($t = 0.2, 0.3$ and 0.8) for three of our solvers
(PPM, \zeus, and MUSCL).  In all panels, the shaded regions denote the
statically refined region. The leftmost column shows the wave just
before it enters the refined region so that we can gauge how the
solver is operating in the absence of AMR.  The center column shows
the wave after it has fully entered the refined region.  The rightmost
column shows the wave well after it has exited the refined region.

With the PPM solver, we notice that even before the wave reaches the
refined region it has been slightly damped. This is not unexpected for
such a short wavelength mode, even for higher-order solvers such as
PPM.  The remaining panels demonstrate that the wave cleanly enters
and exits the refined region.  No significant reflection is seen on
either entry or exit, and the amount of damping is mild.

For the \zeus\ solver, we see that even before it has entered the
refined region there are small oscillations excited behind the wave,
although it is also worth noting that the wave itself is beautifully
propagated without significant damping or phase errors.  In this test
we use only our standard, low amount of quadratic artificial
viscosity.  The trailing oscillations could be damped by additional
viscosity, but we do not add any in order to be sensitive to any
artifacts at the grid boundary.  The remaining panels show that the
trailing oscillations continue, but do not generate any additional
noise. We note that the end result is similar to the case without any
refined region.

Finally, the MUSCL solver also shows a very clean entry and exit from
the refined region without any oscillations, although with mild
damping on exit. However, because we use a piecewise linear
reconstruction, the wave is spread more than with the other methods.

%% file: tests-shockpool.tex
\subsubsection{Shock pool}
\label{sec.tests.shockpool}

The next problem is similar to the last one but instead examines how a
shock with a Mach number of 2 passes in and out of a static refined
region (again defined from $z=0.25$ to 0.75).  The density and
pressure in the domain are initially set equal to 1.0 with zero
velocity.  At $t=0$ the left boundary is set with the density,
pressure and velocity appropriate for a $\mathcal{M}=2$ shock wave.

\begin{figure}
\begin{center}
\includegraphics[width=0.8\textwidth]{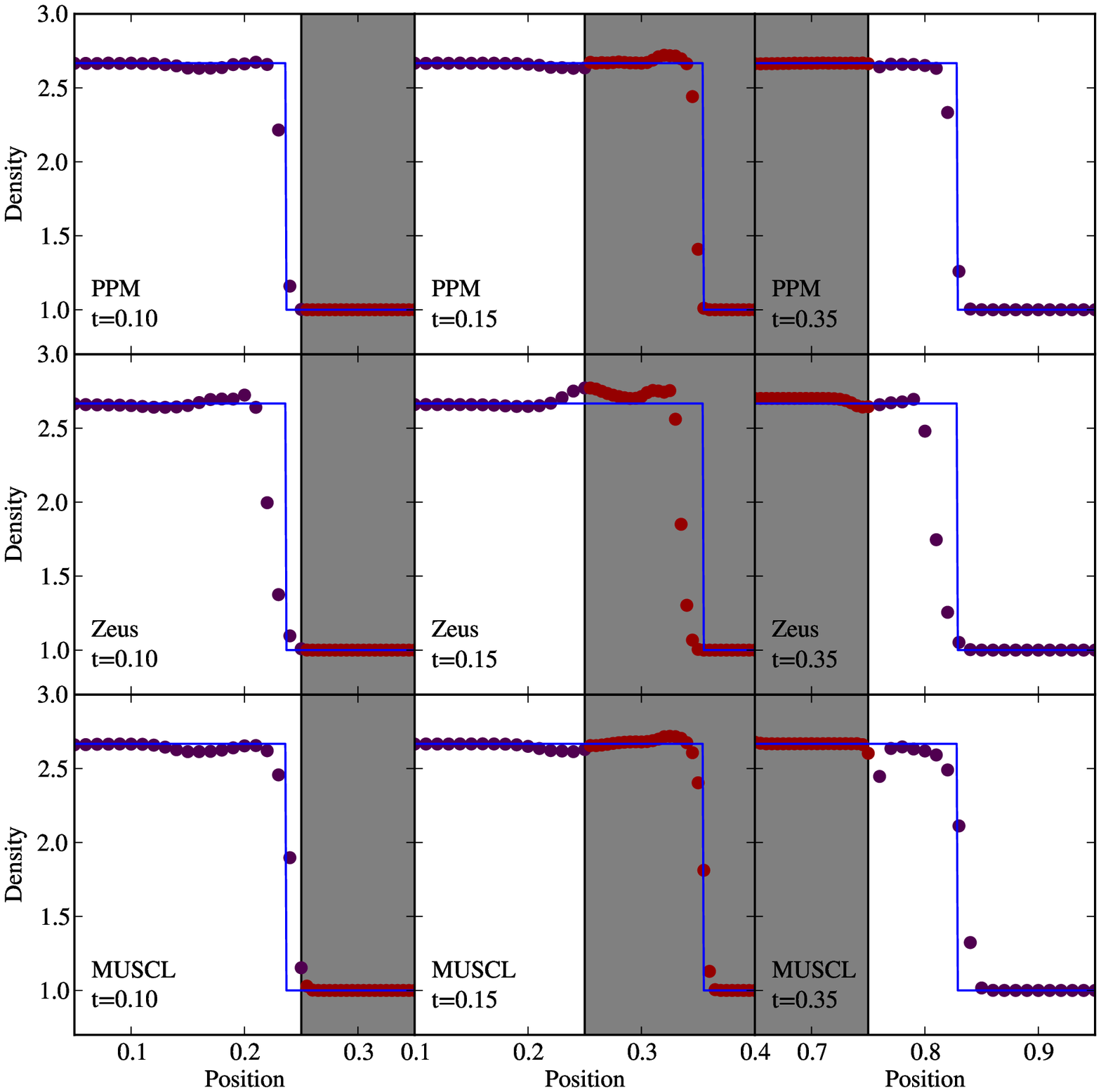}
\caption{This plot shows, for each column, three snapshots at $t=0.1,
0.15$ and 0.35 of a $\mathcal{M}=2$ shock as it propagates through the
grid with a static refined region extends from $x = 0.25$ to 0.75
(shown in grey in each panel).  Each row shows the result for a
different solver (PPM/\zeus/MUSCL from top to bottom).  The solid line
shows the analytic solution.  Note that we focus each panel on a small
region of the entire domain to better show the shock front.}
\label{fig.shockpool}
\end{center}
\end{figure}

In Figure~\ref{fig.shockpool}, we show the evolution of the shock wave
at three times, again corresponding to just before entering the refine
region (left column), after entering the refined region (center), and
after exiting the refined region (right).  We examine the same set of
three solvers as in the previous test problem.

Beginning with PPM (top row of Figure \ref{fig.shockpool}), we see
that this method captures the shock in a small number of zones with
only a small amount of oscillation.  Upon entering the refined region,
the shock finds itself broader than the natural width of the scheme
(since the cell spacing is decreased by a factor of 2), and so the
shock front contracts, causing a slight entropy perturbation in the
post-shock gas.

For \zeus\ (middle row of Figure~\ref{fig.shockpool}), the shock is
broadened because of artificial viscosity and there are slightly more
post-shock oscillations, although again quite mild.  The impact of
entering and exiting the refined region is somewhat larger than in
PPM; however, the most noticeable difference is the incorrect position
of the shock front. This is due to the fact that the scheme is not
energy-conserving (see also the Sedov problem in
Section~\ref{sec.tests.sedov}), and has very little to do with the
passage through the refined region.

Finally, the MUSCL scheme produces a shock that is intermediate in
width between the two previous cases.  This method is
energy-conserving, and thus correctly reproduces the shock speed.  The
oscillations are mild except for the cell immediately outside of the
refined region upon exiting.

%% file: tests-doublemach.tex
\subsubsection{Double Mach reflection}
\label{sec.tests.doublemach}

\begin{figure}
\begin{center}
\includegraphics[width=0.9\textwidth]{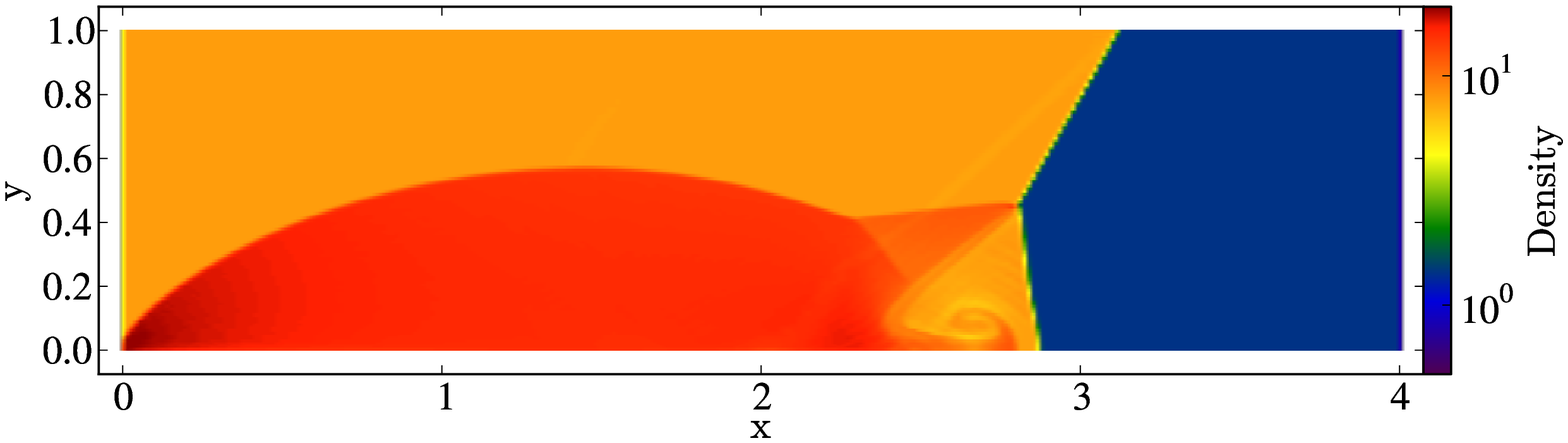}
\caption{Density field at the final time in the Double Mach test.  A
Mach 10 shock is injected into the domain with a shock normal that is
30$^\circ$ from the x-axis with a time-dependent +y-boundary condition
that mimics a shock of infinite length.  The solution is self-similar,
with the jet and whorls at $x \simeq 2.5-3.0$ in this figure being
very sensitive to numerical diffusion.  Our results compare favorably
to the higher-order images from \citet{1984JCoPh..54..115W}.}
\label{fig.doublemach}
\end{center}
\end{figure}

The double Mach reflection test is a classic two-dimensional test of
hydrodynamic algorithms originally described in
\citet{1984JCoPh..54..115W} \citep[and more recently
in][]{2008ApJS..178..137S}.  In this problem, shown in
Figure~\ref{fig.doublemach}, a shock is injected at an angle to a
reflecting surface (the -y-boundary), and a jet appears along the
reflecting surface.  The ideal solution is self-similar, and the
appearance of this solution is highly sensitive to numerical
diffusion.  If numerical noise is present, a Kelvin-Helmholz
instability develops along this jet and breaks the self-similarity.

In the test problem shown in Figure~\ref{fig.doublemach}, a 2D
simulation with $960 \times 240$ cells was created with a domain of $x
= [0, 4]$ and $y = [0, 1]$.  We use an ideal gas equation of state of
$\gamma = 1.4$, a pre-shock density of 1.4, and a pre-shock specific
internal energy of $2.5/1.4$ (all in arbitrary units).  A Mach 10
shock is initialized with a shock normal that is 30$^\circ$ from the
x-axis and an initial position on the lower boundary of x$ = 1/6$.
The lower y-boundary and right x-boundary are reflecting; the left x
boundary is inflowing, and the upper y-boundary has a time-dependent
boundary condition that allows the shock to propagate into the domain
as if it extends to infinity.  The simulation starts at t$ = 0$ and
runs until t$ = 0.205$ (arbitrary units), at which point the rightmost
extent of the shock should be at roughly x$ = 3$.  In this simulation
we use the direct Eulerian implementation of the piecewise parabolic
hydrodynamic method with the diffusion, flattening, and shock
steepening all enabled.

It is instructive to compare our Figure~\ref{fig.doublemach} to Figure
9 in~\citet{1984JCoPh..54..115W}.  By the end of the simulation, a
dense jet is apparent at the leading edge of the shock, propagating
along the x-axis.  The shape of this jet is sensitive to numerical
diffusion, and our figure compares favorably to those shown in
Woodward \& Colella.

%% file: tests-sedov.tex
\subsubsection{Sedov Explosion}
\label{sec.tests.sedov}

\begin{figure}
\begin{center}
\includegraphics[width=0.4\textwidth]{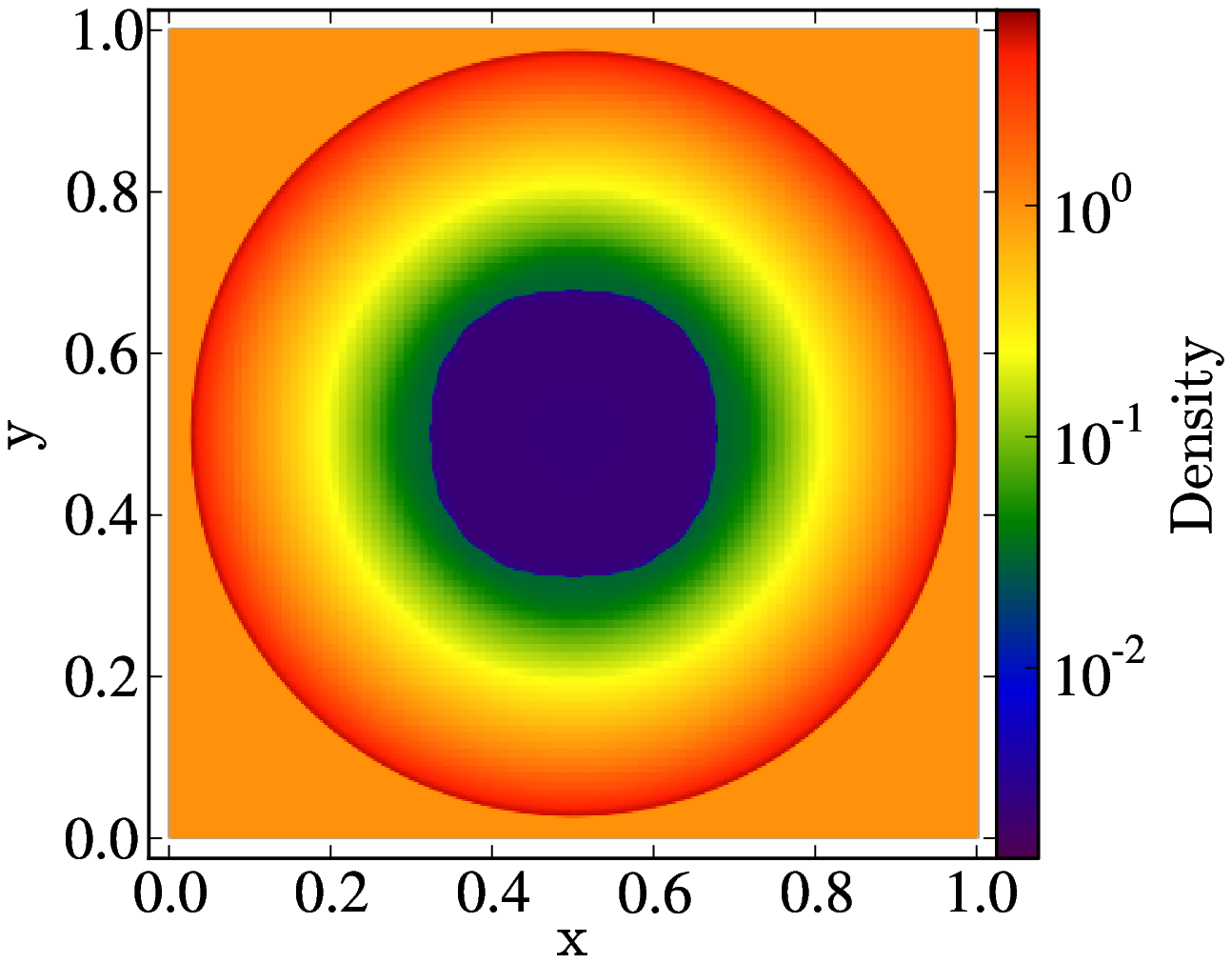}
\includegraphics[width=0.4\textwidth]{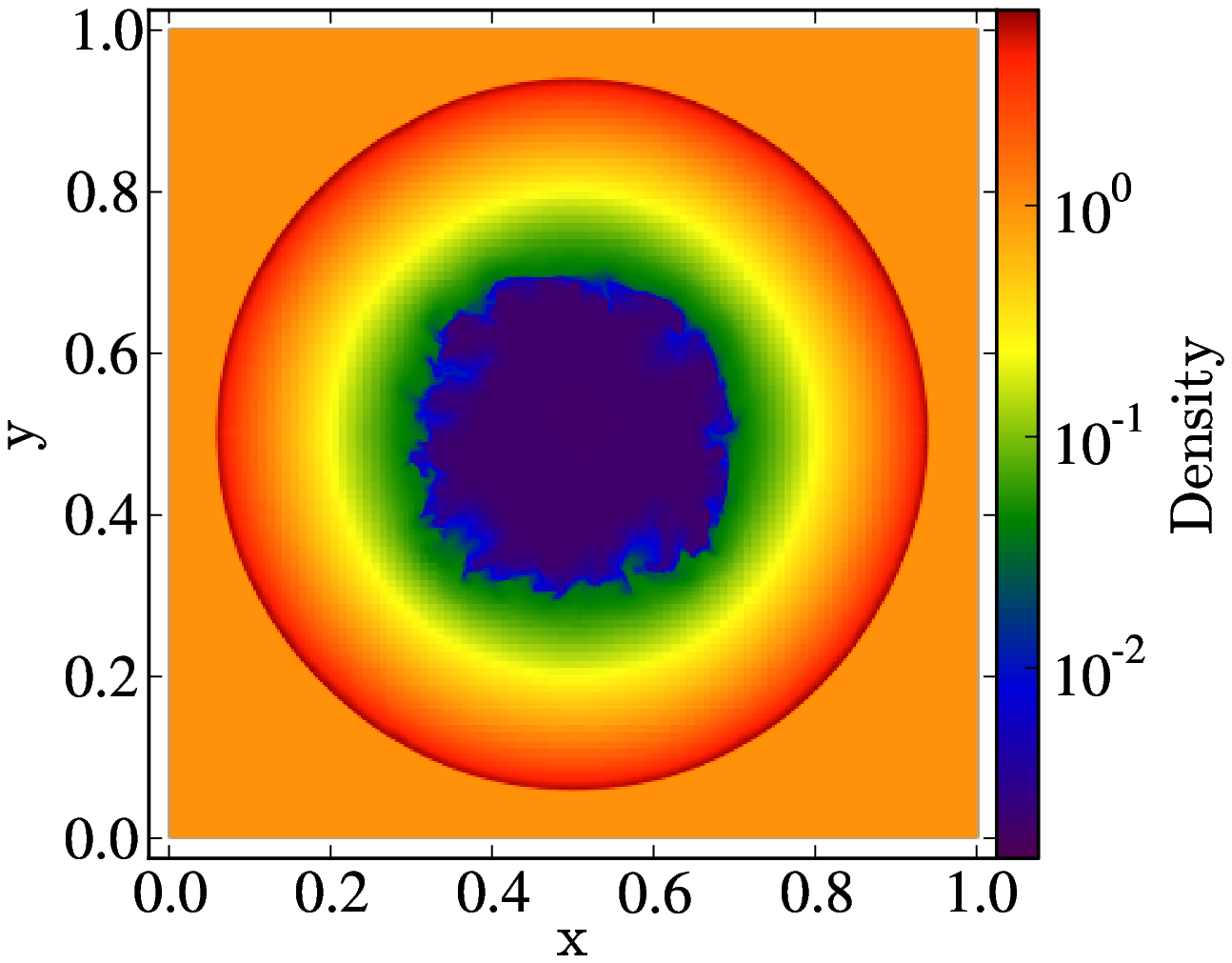}
\caption{Density slices from the Sedov Blast test at $t = 0.07$. Left:
results using the Piecewise Parabolic Method hydro scheme.  Right:
results using the \zeus\ hydro method. Notably, the \zeus\ shock front
has progressed less far than in the PPM run. This is due to energy
loss when conserving only internal, and not total, energy.}
\label{fig.sedov1}
\end{center}
\end{figure}

\begin{figure}
\begin{center}
\includegraphics[width=\textwidth]{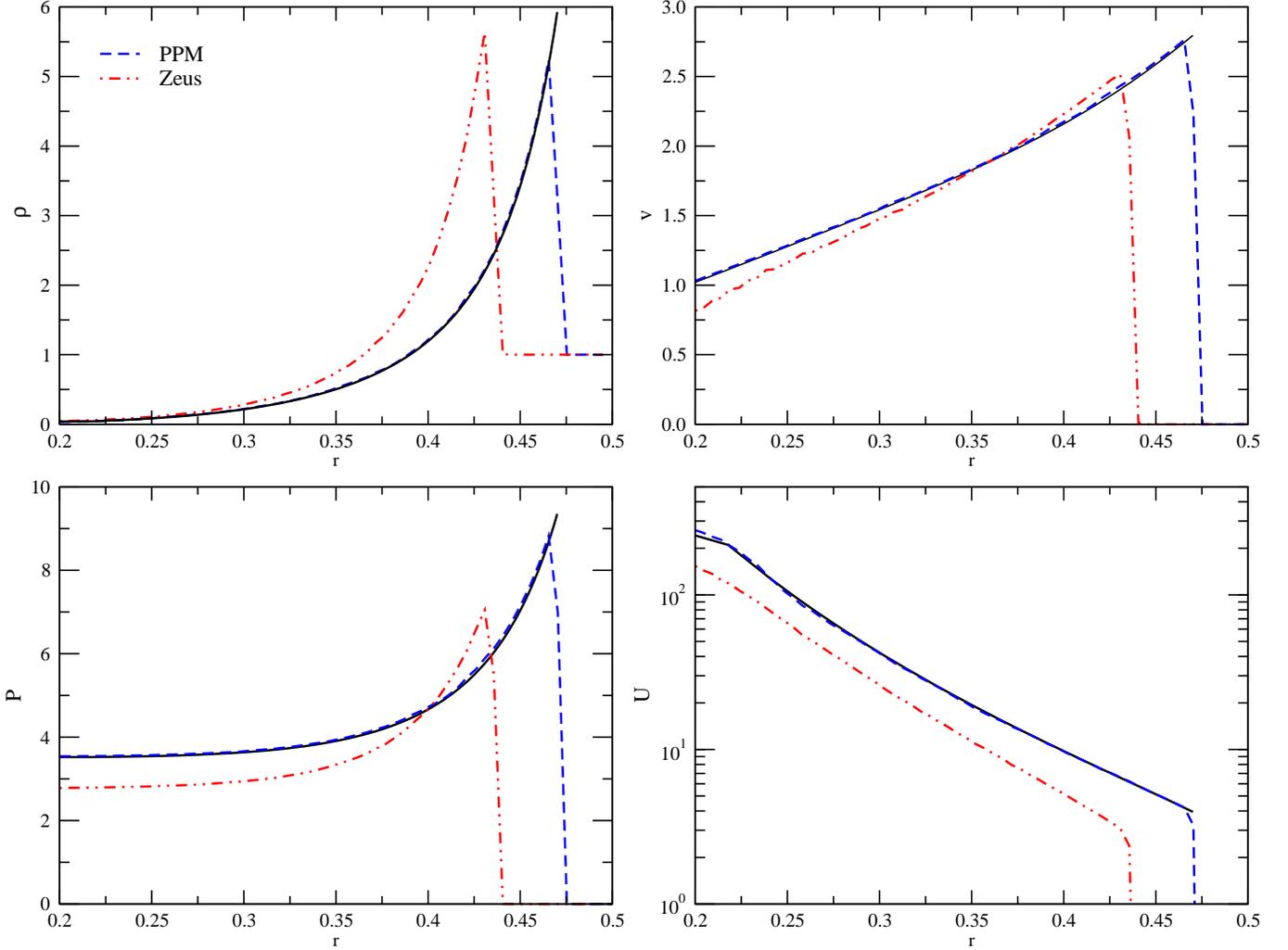}
\caption{Radial profiles for the Sedov Blast test at $t =
0.07$. Clockwise from top-left shows density, velocity, internal
energy and pressure.  The black solid line shows the analytic
solution.  The blue dashed line shows the simulation using the PPM
method, and the red dot-dashed line using \zeus.  The \zeus\ result
substantially lags the true result due to total energy not being
explicitly converged.}
\label{fig.sedov2}
\end{center}
\end{figure}

The Sedov Blast test \citep{Sedov1959} models an intense explosion,
initiated by depositing thermal energy into a homogenous distribution
of gas. The result is a strong spherical shock wave centered on the
point of energy injection.  This problem is a popular test of
astrophysical simulation codes for three reasons.  First, this problem
is representative of the astrophysical phenomenon of a supernova
explosion.  Second, an analytical solution to the evolution of the
Sedov blast wave exists, allowing for a direct test of the accuracy of
the code.  The radius of the shock front as a function of time is
given by

\begin{equation} r(t) =
\left(\frac{E_0}{\alpha\rho_0}\right)^{1/5}t^{2/5}
\end{equation}

\noindent where $E_0$ is the initial energy injection, $\rho_0$ is the
background density and $\alpha = 1.0$ for cylindrical symmetry and an
ideal gas with $\gamma = 1.4$; for the full derivation see
\citet{Sedov1959}.  Third, as the spherical shock expands, its
symmetry, or lack thereof, serves to highlight any directional
preferences of the hydrodynamics solvers.  The test presented here is
the two-dimensional version that is included in the \enzo\
distribution. The three-dimensional results from this test, both for
\enzo\ and three other leading astrophysics codes, can be found in
\citet{Tasker2008}.

In the initial state, the box contains a homogenous distribution of
gas at a density of 1 (note that, in the absence of gravity or
radiative cooling, this problem is scale-free and thus without
units). Thermal energy is deposited into a single cell at the center
of the box with $E_0 = 10.0$. The problem is run in two dimensions
with reflecting boundary conditions and a box having a length of $1$
in both directions.  We selecte a top grid of $100 \times 100$ cells
and a maximum of four levels of adaptive refinement, refining by
factors of two on shocks and the slopes of the density and total
energy fields. The exception to this scheme is in the initial
conditions, where additional refinement is placed directly around the
injection point to better resolve spherical geometry. The results are
assessed at $t = 0.07$, which corresponds to a time just before the
shock reaches the box edge (see Figure~\ref{fig.sedov1}).

Figure~\ref{fig.sedov2} shows the radial profiles for the simulation
run with the PPM hydro-solver (blue dashed line) and the \zeus\
hydro-solver (red dash-dot-dot line) together with the analytical
solution (black solid line). Clockwise from the top-left are density,
velocity, internal energy and pressure. PPM matches the analytical
solution extremely well for all quantities. However, the shock front
in the \zeus\ simulation lags behind the analytical position. This can
also be seen in the slices shown in Figure~\ref{fig.sedov1}. The cause
of this discrepancy is that \zeus\ shows a substantial energy loss
during the first few timesteps and produces a diamond-shaped, rather
than spherical, shockfront during this time. After this, the code
correctly conserves energy but this intial energy loss remains clearly
visible in the position of the shock at $t = 0.07$. This problem was
addressed directly by \citet{Clarke2010}, who attributed the source of
the issue to this version of \zeus\ solving the internal, rather than
total, energy equation. In situations with strong energy gradients,
this choice caused an energy loss and the artificial viscosity
produces the direction-dependent shockfront shape. In their paper,
\citet{Clarke2010} present results from an alternative version of
\zeus\ that conserves total energy. This problem is less marked for
smaller energy gradients and it should be noted that the \zeus\ hydro
algorithm's stability and speed make it a highly competitive choice,
despite the disagreements in this test.

%% file: tests-pointsourcegravity.tex
\subsubsection{Point source gravity test}
\label{sec.test.gravitypointsource}

\begin{figure}
\begin{center}
\includegraphics[width=0.6\textwidth]{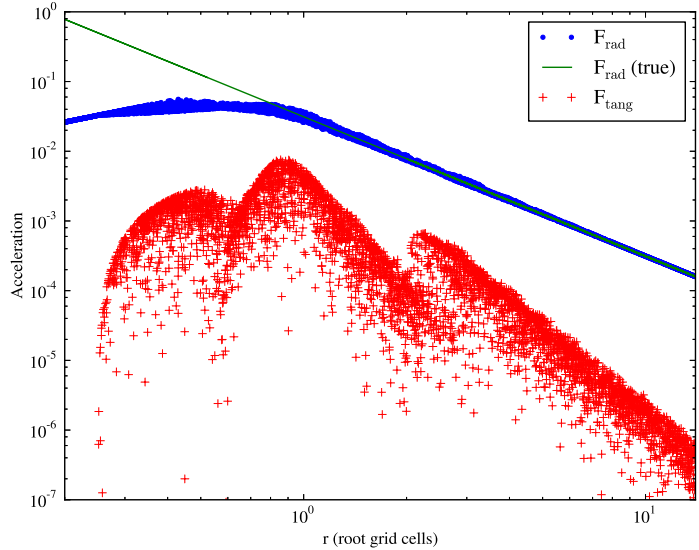}
\caption{The radial and tangential accelerations measured around a
  point source placed in a static AMR grid.  The solid line shows the
  analytic radial profile.}
\label{fig.gravitytest}
\end{center}
\end{figure}

This is a simple test of the ability of the Poisson solver to
correctly reproduce the gravitational acceleration around a point
source without any dynamics.  We place a single particle in the center
of the unit domain with root grid dimensions $32^3$ and add a static,
nested grid refined by a factor of two that covers the central $1/8^3$
of the domain (i.e. so the refined grid is only $8^3$).  We then place
5000 particles throughout the domain distributed uniformly in angle
and in the logarithm of the radius (from the central point).  The
measured accelerations are shown in Figure~\ref{fig.gravitytest}.  As
is typical for Particle-Mesh based methods, the errors decrease at
large distances and peak on roughly the grid scale.  The accelerations
are relatively smooth across the top grid/fine grid boundary with the
most noticeable impact being a drop in the amplitude of tangential
errors on the fine grid.  The force law rolls over at about 1.6 fine
cell widths (equivalent to 0.8 root grid widths in this figure),
consistent with the CIC-deposition and interpolation used.

%% file: tests-orbit.tex
\subsubsection{Orbit Test}
\label{sec.test.testorbit}

\begin{figure}
\begin{center}
\includegraphics[width=0.45\textwidth]{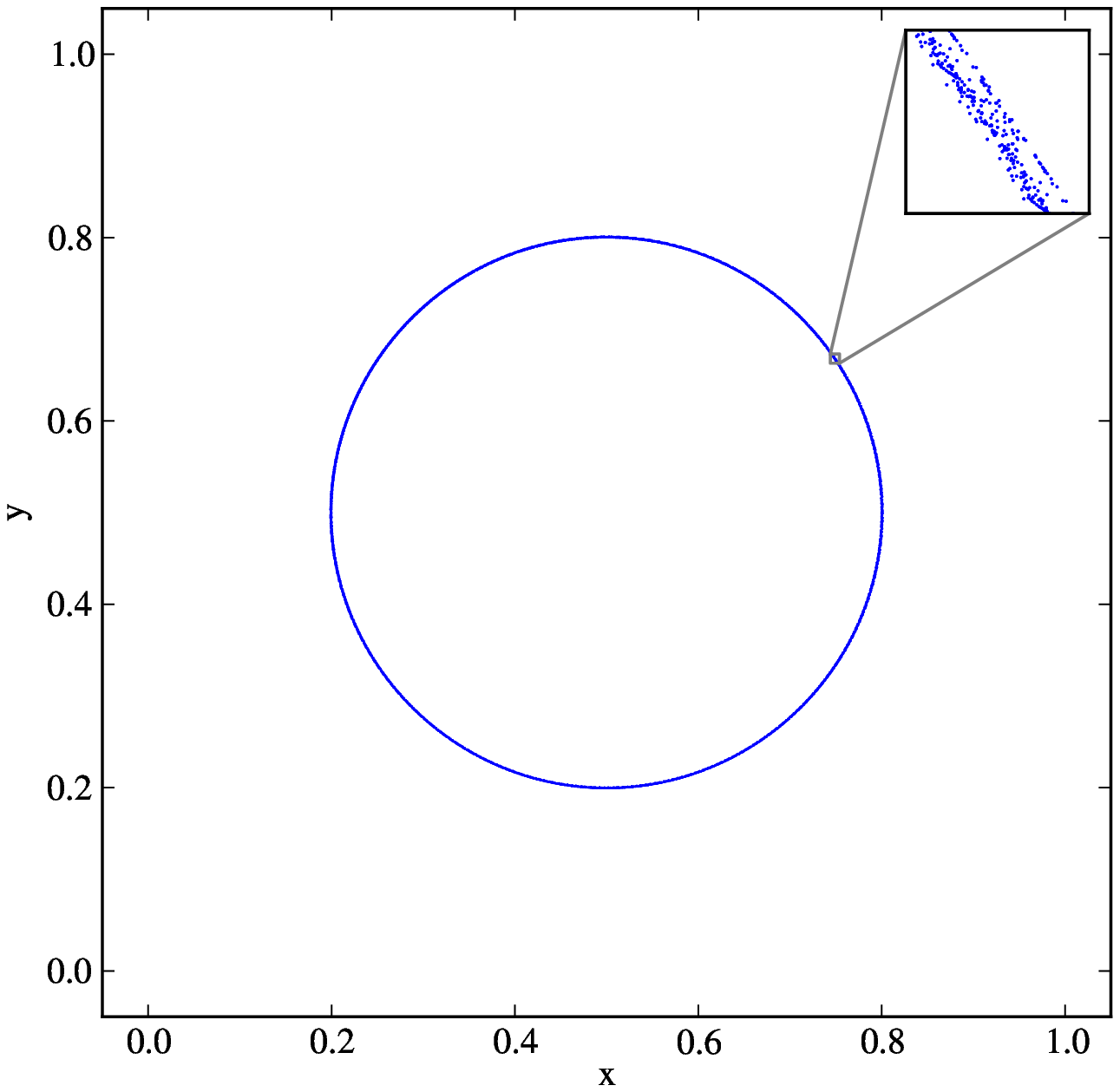}
\includegraphics[width=0.45\textwidth]{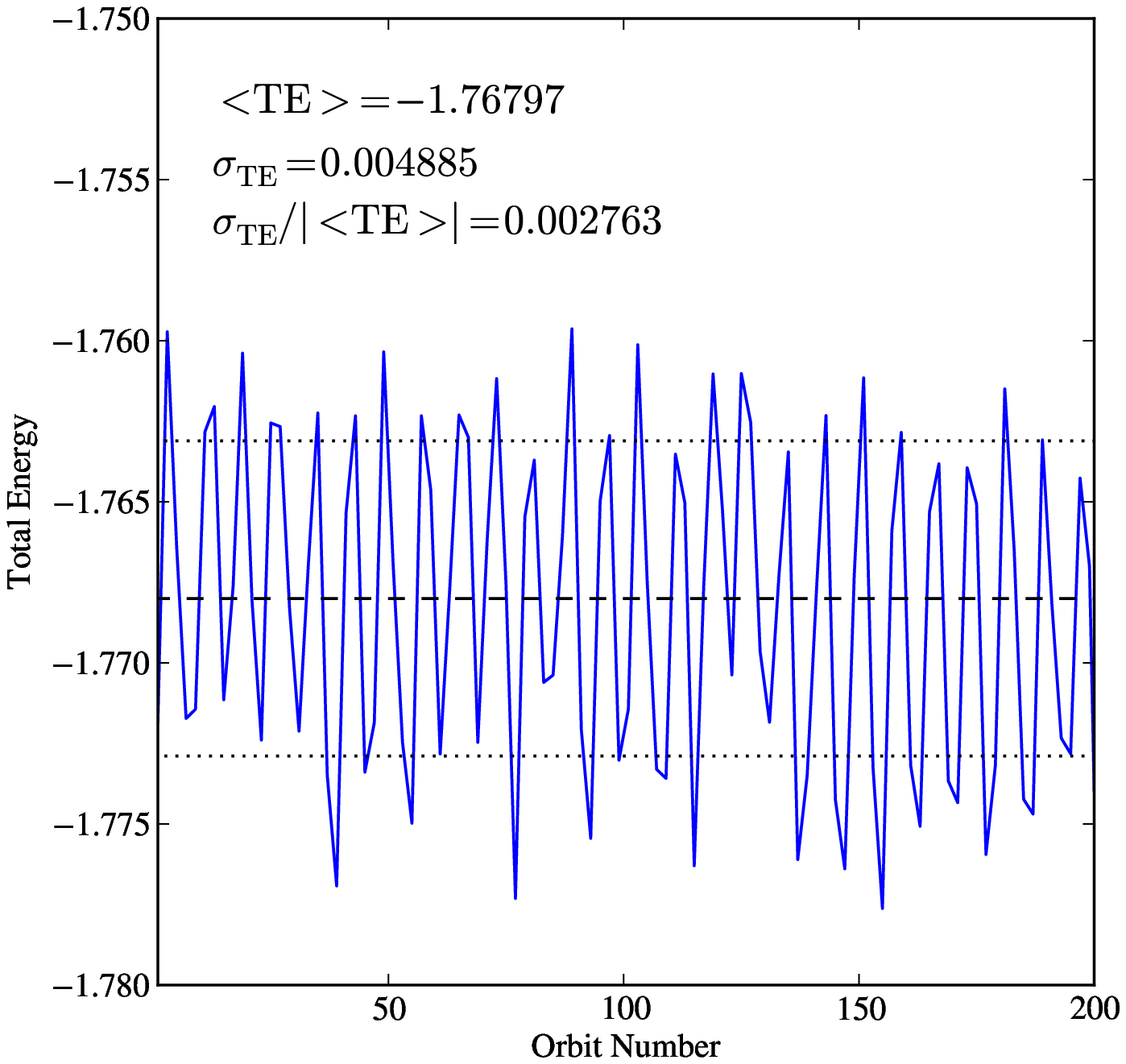}
\caption{Test particle behavior in the Orbit problem, which is used to
test the gravity solver and particle integration.  An effectively
massless test particle is set in circular orbit around a massive
central particle, with the gravitational potential calculated on a 3D,
$32^3$ grid, and evolved forward in time for 200 orbits.  Left panel:
path of the test particle in the orbital plane plotted for the
duration of the simulation, with a zoomed box showing a small portion
of the orbit to highlight deviations from perfect circularity.  Right
panel: total specific energy (kinetic plus potential) for the test
particle for the duration of the simulation.  The particle's mean
energy (integrated over the length of the simulation) is represented
by a horizontal dashed line, and +/- one standard deviation by
horizontal dotted lines.  All quantities are in arbitrary units.}
\label{fig.orbittest}
\end{center}
\end{figure}

The ``orbit test'' simulates the orbit of an effectively massless
particle around a massive primary particle and is used to test the
accuracy of both the gravity solver and particle integration.  In this
problem (shown in Figure~\ref{fig.orbittest}), a 3D box with unit size
and a $32^3$ grid is created, and a particle with mass 1 is placed at
the center of the box (both box length and particle mass are in
arbitrary units).  A second ``test'' particle, with mass of $10^{-6}$,
is placed at a distance of 0.3 L$_{box}$ from the primary particle in
the x-y plane (at $z=0.5$~L$_{box}$).  Both particles are given
velocities such that they will orbit their common center of mass in
the x-y plane.  The simulation is run for 200 orbits, writing out a
dataset approximately once per orbit (with the particle positions
being written out several hundred times per orbit).

The left panel of Figure~\ref{fig.orbittest} shows the path of the
test particle in the x-y plane (at z=$0.5$), plotted over the duration
of the simulation.  The test particle maintains a path that has
effectively unnoticeable deviation from circularity, and with no
discernable drift in either net radius or orbit center over time.  The
right panel displays the total specific energy of the test particle
(defined as $\frac{1}{2} {\bf v}^2 + \phi$, where $\phi$ is the
gravitational potential at the position of the test particle) as a
function of time, plotted once per orbit.  The mean total specific
energy is $-1.76797$ in arbitrary units, with a standard deviation of
$0.004885$ ($0.276\%$ of the mean value), and with a maximum deviation
from the mean of $0.546\%$ of the mean value.  Given that the
potential is calculated on $32^3$ cells, the two particles are
separated by between 10 and 14 cells depending on their relative
positioning in the grid, and the particles are integrated using a
second-order leapfrog method, this level of accuracy is expected and
acceptable.

%% file: tests-selfsiminfall.tex
\subsubsection{Self-Similar infall test}
\label{sec.tests.infall}

This test problem is based on the self-similar solutions to the
cosmological spherical infall problem found by
\citet{Bertschinger1985}.  It features a small density perturbation in
a homogeneous $\Omega = 1$ universe and is a strong test of the
cosmological evolution, gravity, and hydrodynamic portions of the
code.  It is a close analogue to cosmological halo formation.

For the initial conditions, we adopt a $32^3$ top grid with a single,
initial subgrid (covering $1/8^3$ of the domain) with a refinement
factor of 2.  An overdensity $\delta = 40$ is placed in a single cell
near the center of the domain.  We begin at $z=199$ and evolve to
$z=0$, which is a sufficient time for the evolution to largely forget
its initial conditions and approach the self-similar result.  An
overdensity refinement criteria ($\delta_{\rm crit} = 1.1$ on the top
grid) is used to add additional grids, going up to 5 additional levels
beyond the root grid.  We use the PPM solver without radiative
cooling; only baryons are used for this problem.  An ideal gas law
with $\gamma = 5/3$ is adopted.

\begin{figure}
\begin{center}
\includegraphics[width=0.7\textwidth]{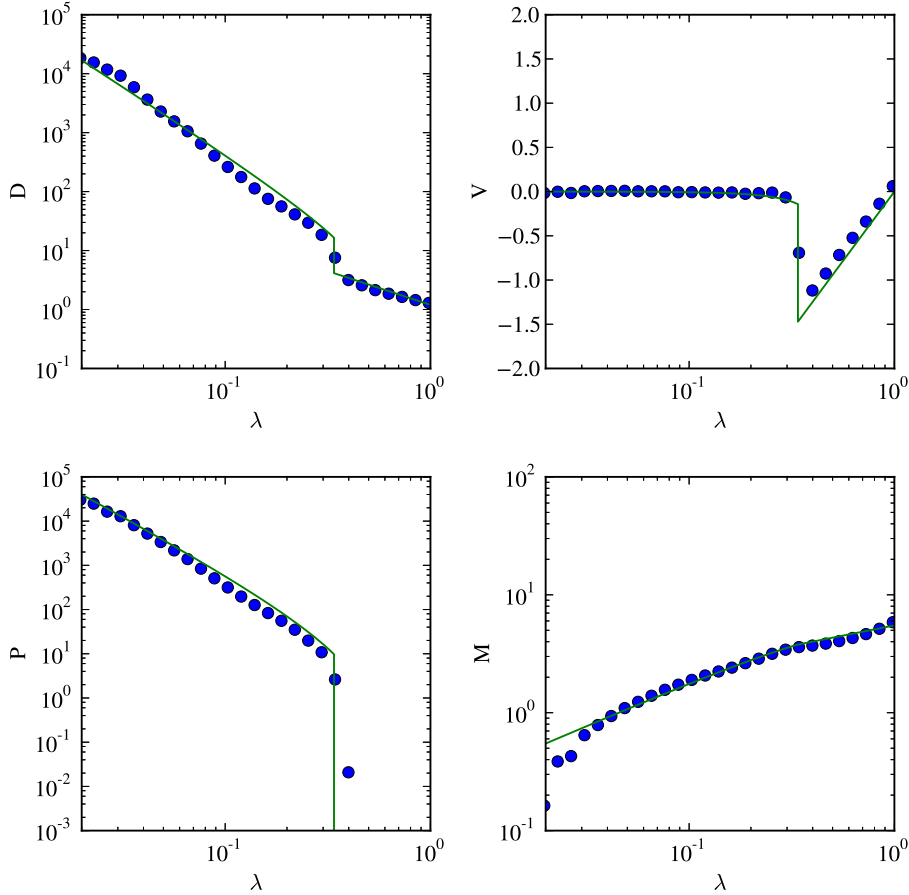}
\caption{The results of the self-similar spherical infall test.  The
panels show, as a function of the dimensionless radius $\lambda$, the
dimensionless density $D$ (top left), velocity $V$ (top right),
pressure $P$ (bottom left), and enclosed mass $M$ (bottom right).  In
each case we show azimuthally averaged profiles from the simulation as
circles and the analytic solution for $\gamma = 5/3$ as solid lines.}
\label{fig.sphericalinfall}
\end{center}
\end{figure}

In Figure~\ref{fig.sphericalinfall}, we show the results, scaled
according to the dimensionless variables as defined in equations (2.9)
and (3.2) of \citet{Bertschinger1985}.  This demonstrates that we can
quickly and easily obtain a good solution with only a fairly modest
initial grid.  The shock is sharply resolved and the asymptotic
profiles at small $\lambda$ are recovered.  At very small values of
$\lambda$, the initial conditions have not been fully forgotten and
the self-similar solution is not recovered.  This can be seen most
clearly in the dimensionless mass.  However, this is simply because of
the limited amount of time for which we evolve the solution.

%% file: tests-zeldovichpancake.tex
\subsubsection{Zel'dovich Pancake}
\label{sec.tests.pancake}

The Zel'dovich pancake \citep{1970A&A.....5...84Z} is particularly
relevant to cosmological simulations, because it includes many
features that are critical to structure formation: hydrodynamics,
expansion, and self gravity.  This problem represents the formation of an ideal,
isolated caustic, and is thus a useful proxy for much more
complicated structures in full 3-dimensional cosmological simulations,
such as the collapse of gas onto a cosmological halo or filament.

The initial conditions are simple, and we follow the prescription of
\citet{Anninos94}.  Assuming a geometrically flat cosmology, the density
perturbation is given by
\begin{equation}
\rho(x_l) = \rho_0 \left[ 1 - \frac{1+z_c}{1+z} \cos(k x_l) \right]
\end{equation}
with the internal energy of the gas set so that the entropy 
is constant throughout.  The velocity perturbation is given by
\begin{equation}
v(x_l) = -H_0 \frac{1 + z_c}{(1+z)^{1/2}} \frac{\sin(k x_l)}{k}
\end{equation}

In the equations above, $\rho_0$ is the background density, z$_c$ is a
free parameter and is the redshift where the sheet forms a caustic
(i.e., where it `pancakes'), z is the redshift of initialization, x$_l$ is the
Lagrangian mass coordinate, $k = 2 \pi / \lambda$ (where $\lambda$ is
the perturbation wavelength), and H$_0$ is the value of the Hubble
constant at $z = 0$.  Note that this solution is expressed in terms of 
Lagrangian positions, so one needs to convert this into the Eulerian
coordinates, x$_e$, that are more useful to a grid-based calculation:
\begin{equation}
x_e = x_l - \frac{1 + z_c}{1 + z} \frac{\sin(k x_l)}{k}
\end{equation}

\begin{figure}
\begin{center}
\includegraphics[width=0.8\textwidth]{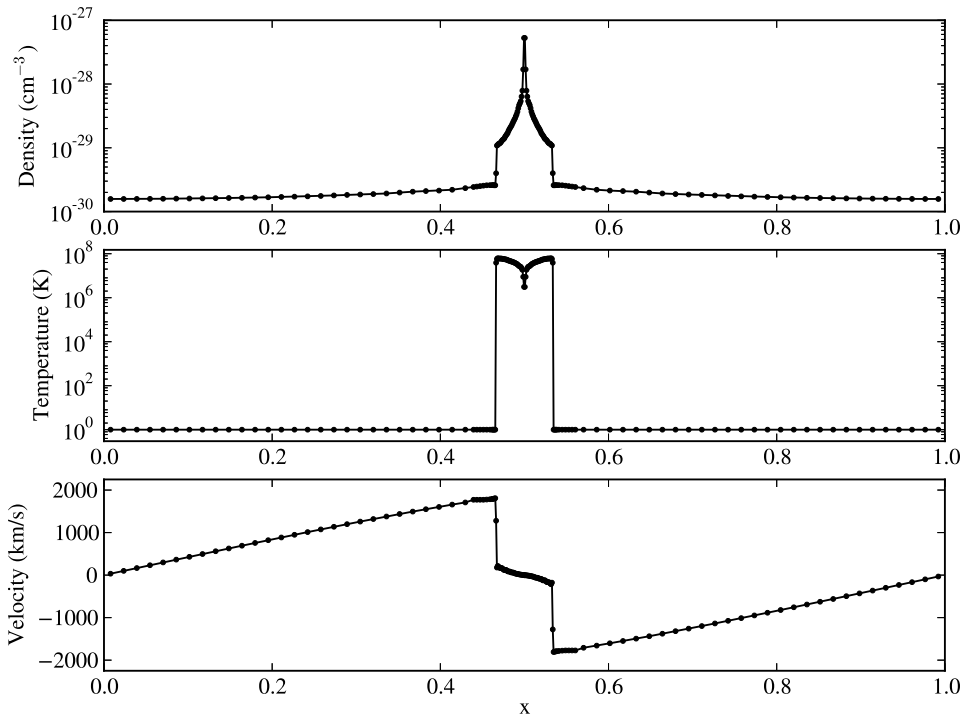}
\caption{Zel`Dovich Pancake test shown at z$ = 0$, initialized in a
$\Omega_m = 1$ universe at z$ = 20$ on a one-dimensional grid having
64 cells, and further refined by factors of four based on cell mass
and the presence of shocks for up to two additional levels of mesh,
having a maximum effective resolution of 1024 grid cells. The top,
middle, and bottom rows show density, temperature, and velocity of the
gas, respectively; all are a function of position in units of the box
size.  The central region ($x \simeq 0.45-0.55$) has been adaptively
refined, as can be seen by the locations of grid points.  Shocks and
the central density peak are clearly resolved, with well-delineated
jumps at the appropriate locations.}
\label{fig.pancake}
\end{center}
\end{figure}

We note that the solution described above is exact up to the point of
caustic formation.  In Figure~\ref{fig.pancake}, we show the results
of a test of the adaptive mesh version of \enzo's Zel'dovich Pancake
test.  A one-dimensional box of length $64$~Mpc/h is initialized at $z
= 20$ in an $\Omega_M = 1$ universe with $h = 0.5$ and a background
temperature of 100~K, with a background density of $\rho_0 = \rho_c$.
The simulation is initialized with 64 grid cells, refining by factors
of four using criteria based on cell mass and the presence of shocks,
for a maximum of 2 levels (i.e., an equivalent maximum resolution of
1024 grid cells).  The simulation is evolved to z$ = 0$ using the PPM
hydro method.

The final output of the calculation, with the key features of this
test problem, is shown in Figure~\ref{fig.pancake}.
The strong shocks and large density gradients are well-resolved, with
density and velocity jumps being well-delineated and at the correct
locations.  The key features of 
this test problem can be resolved with far fewer cells -- simulations
including a mere 8 cells resolve the key features, as shown in
Sections 3.3.4-3.3.5 of~\citet{BryanThesis96} -- but we choose a
higher resolution here for illustrative purposes.

%% file: tests-orszagtang.tex
\subsubsection{MHD: Orszag-Tang Vortex}
\label{sec.tests.mhd}

\begin{figure}
\begin{center}
\includegraphics[width=0.4\textwidth]{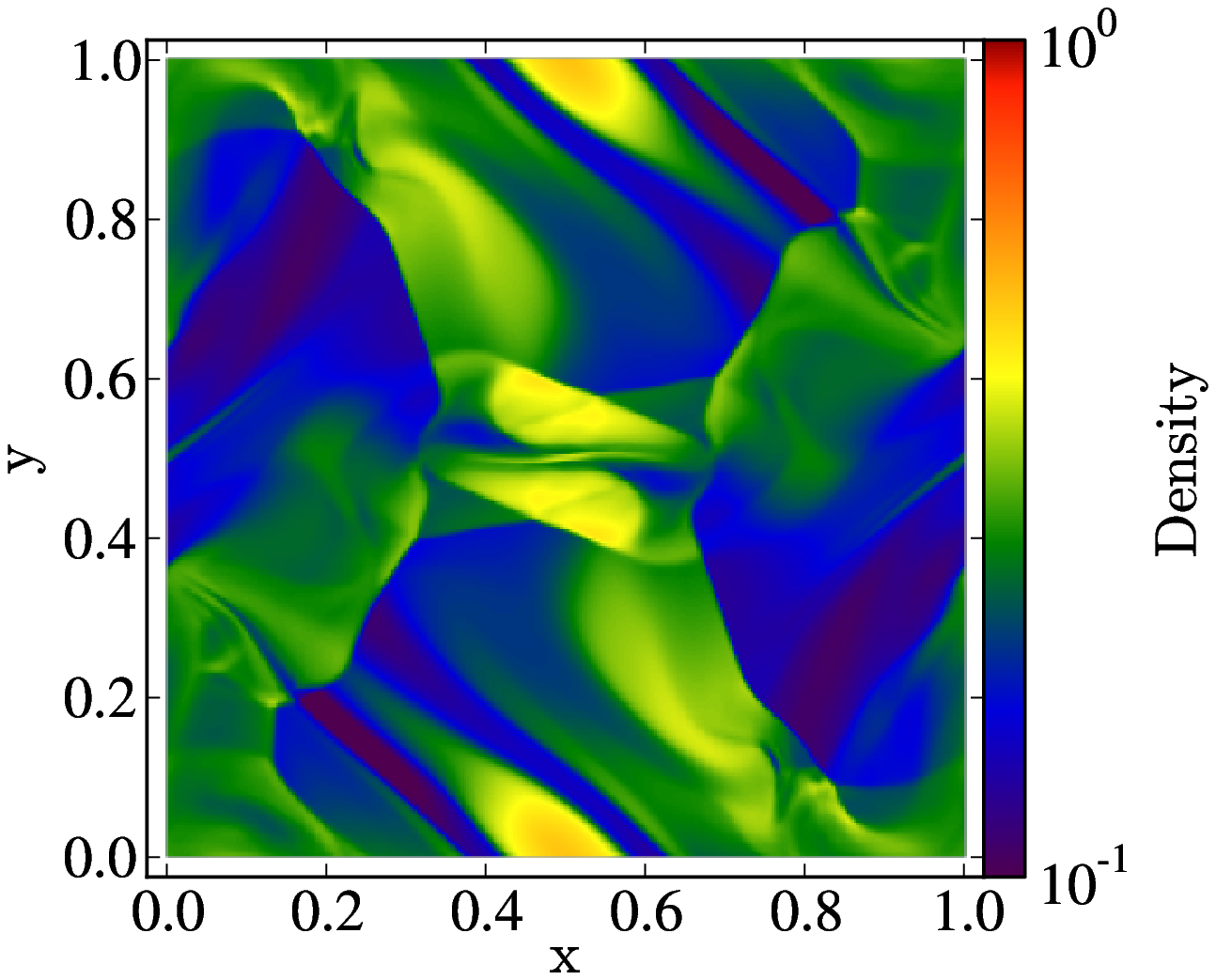}
\includegraphics[width=0.4\textwidth]{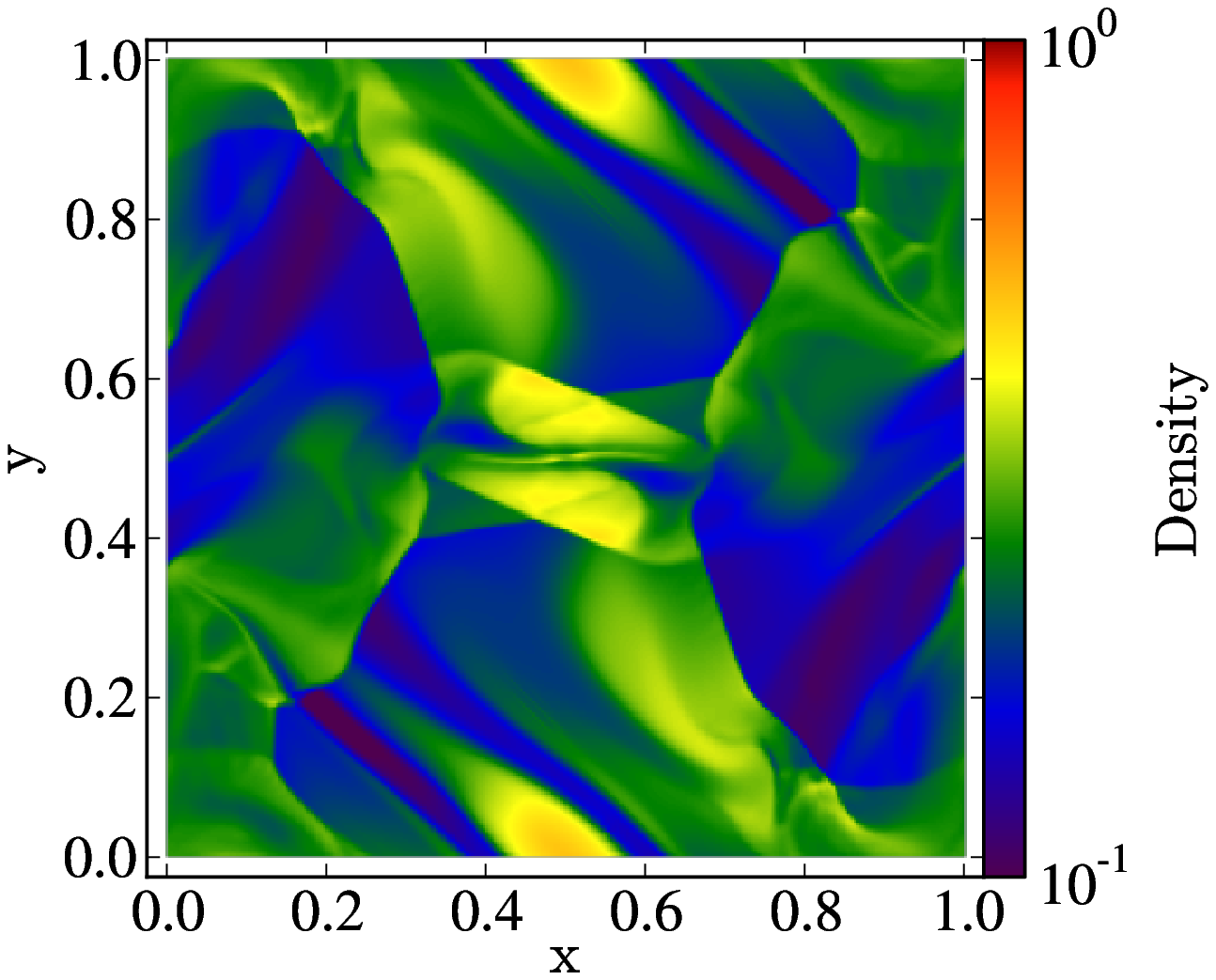}
\caption{Density field from the Orszag-Tang vortex test, at $t=0.48$.
Left: solution using constrained transport MHD.  Right: solution using
Dedner MHD. The initial conditions are uniform density, with a single
rotating velocity structure and two circular magnetic structures.
These initial conditions generate significant small-scale structure in
both the CT and Dedner schemes, which have approximately equal
effective resolution.}
\label{fig.orszag}
\end{center}
\end{figure}

Figure~\ref{fig.orszag} shows the Orszag-Tang vortex problem
\citep{Orszag79}.  The left panel shows the result using \enzo's
constrained transport MHD method, while the right panel shows the
result using \enzo's implementation of Dedner MHD.  The Orszag-Tang
vortex test is a classig MHD test problem, and shows that significant
small scale structure can be generated in MHD from large scale initial
perturbations.  It is often used to compare the effective resolution
of different MHD schemes.  The test begins with uniform density,
$\rho_0=25/(36\pi)$, and pressure, $P_0=5/(12\pi)$ (as with other
tests in this section, in the absence of gravity or chemistry/cooling
we use dimensionless units).  There is a single rotational mode in the
velocity, and two in the magnetic field: ${\bf v}_0 = (-\sin(2\pi y)
\hat{x},\, \sin(2\pi x) \hat{y})$, ${\bf B}_0 = (-\sin(2\pi y)
\hat{x},\, \sin(4\pi x) \hat{y})$.  The simulation is evolved to
$t=0.48$.  One can see that the structures are accurately represented
as compared to, for example, \citet{Toth00}, and that the resolution
of shocks is comparable in both methods.

%% file: tests-onezonecollapse.tex
\subsubsection{One-zone collapse test}
\label{sec.tests.1-zone}

The one-zone collapse test simulates the collapse of a
self-gravitating gas cloud using a semi-analytic model for the
evolution of gas density and adiabatic heat input as a function of
time.  It is designed to test the chemistry and cooling modules over a
wide range in densities and over physically-motivated timescales.
Because this test disables the hydrodynamic and gravity solvers and
uses a simple model for the density evolution, it is far faster than
running a true collapse simulation.  The density evolution is based on
the self-similar Larson-Penston solution for isothermal collapse
\citep{1969MNRAS.145..271L, 1969MNRAS.144..425P} with a modification
to account for the efficiency with which the heat introduced by
compression can be radiated away \citep{1983ApJ...265.1047Y}.  Our
implementation, described briefly here, follows the work of
\citet{2005ApJ...626..627O}, and we direct the interested reader to
this paper for further details.  The evolution of the gas density,
$\rho$, is given by

\begin{equation}
\frac{d\rho}{dt} = \frac{\rho}{t_{\rm col}},
\end{equation}
where the collapse timescale, $t_{\rm col}$, is

\begin{equation} \label{eqn.tcol}
t_{\rm col} = \frac{t_{\rm dyn}}{\sqrt{1 - f}},
\end{equation}
In this equation, $t_{\rm dyn}$ is the dynamical time for the collapse
of a spherical cloud, and is expressed as
\begin{equation}
t_{\rm dyn} = \sqrt{\frac{3 \pi}{32 G \rho}}.
\end{equation}

 The collapse timescale is altered from the dynamical time by a factor
$1/\sqrt{1-f}$ in Equation \ref{eqn.tcol}, which is an approximation
of the ratio of the gas pressure to the force of gravity.  The value
of $f$ depends on the effective adiabatic index, $\gamma_{\rm ef}
\equiv (\partial \ln p / \partial \ln \rho)$, which we linearly
extrapolate from derivative values at the two previous timesteps.  For
the value of $f$ in this test problem, we use the piecewise function
of \citet{2005ApJ...626..627O}, given by

\begin{equation}
f = \left\{
  \begin{array}{ll}
  0, & \gamma_{\rm ef} < 0.83,\\
  0.6 + 2.5 (\gamma_{\rm ef} - 1) - 6.0 (\gamma_{\rm ef} - 1)^{2}, & 0.83 <
  \gamma_{\rm ef} < 1,\\
  1.0 + 0.2 (\gamma_{\rm ef} - 4/3) - 2.9 (\gamma_{\rm ef} - 4/3)^{2}, & \gamma_{\rm ef} > 1.
\end{array} \right.
\end{equation}

\begin{figure}
  \begin{center}
    \includegraphics[width=0.8\textwidth]{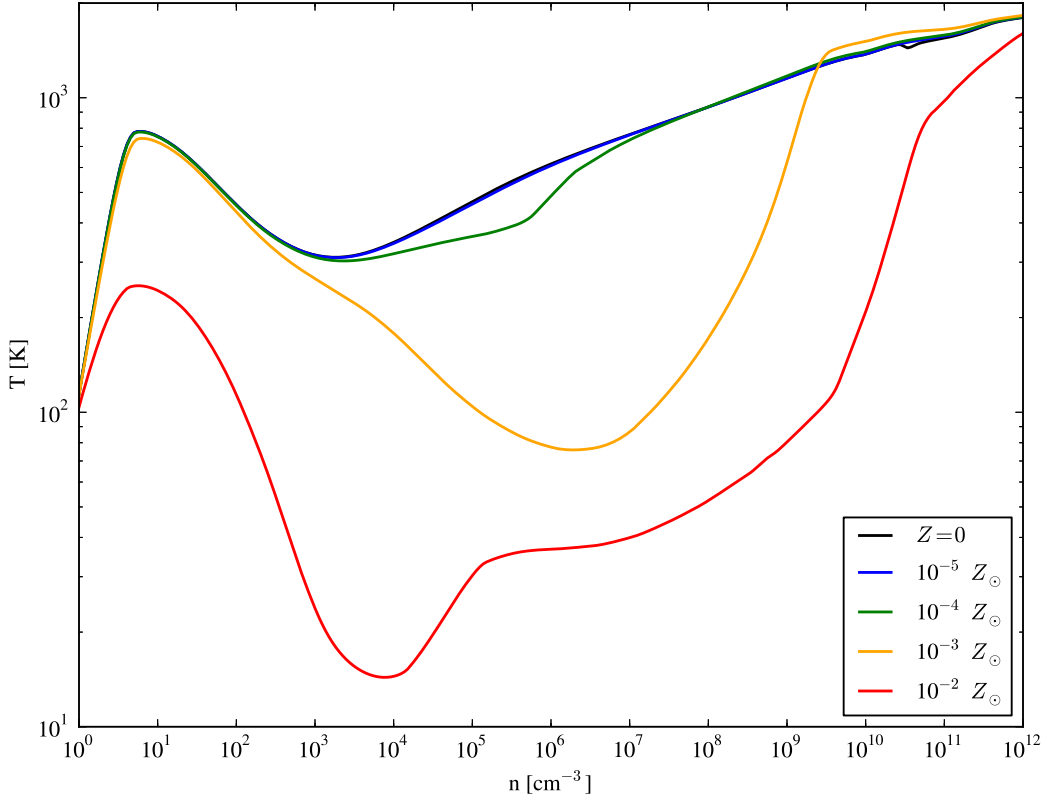}
  \end{center}
  \caption{Evolution of temperature versus number density for a
one-zone collapse test for gas at metallicities from 0 to 10$^{-2}
Z_{\odot}$.  This test is based on the results of
\citet{2005ApJ...626..627O}, and approximates the collapse of a
self-gravitating, cooling gas cloud.  This test problem uses \enzo's
primordial chemistry network with tabulated metal cooling rates
calculated with the \texttt{Cloudy} code, and compares favorably to
the results of \citet{2005ApJ...626..627O}.}
  \label{fig.onezone}
\end{figure}

The specific energy evolves as

\begin{equation}
\frac{de}{dt} = -p \frac{d}{dt} \frac{1}{\rho} - \Lambda,
\end{equation}

where $\Lambda$ is the cooling rate in units of erg s$^{-1}$ g$^{-1}$
and energy, temperature, density, and pressure are related by the
ideal gas law, including effects from molecular hydrogen as
appropriate.  Figure \ref{fig.onezone} shows an example of the
one-zone collapse test performed with an initial number density of 1
hydrogen atom per cm$^{-3}$ and an initial temperature of 100 K using
the 12 species chemistry network with H, D, and He species and metal
cooling rates calculated with the \texttt{Cloudy} code.  The effects
of metal cooling can be clearly seen; as the metallicity increases
from zero to $10^{-2}$~Z$_\odot$, the gas rapidly and significantly
deviates from the primordial result (black line).  Our primordial
results compare very well to those shown in
\citet{2005ApJ...626..627O}; however, we use a different metal cooling
method, so the lines describing the evolution of the metal-enriched
gas are not directly comparable.

%% file: tests-photoevaporation.tex
\subsubsection{Photo-evaporation of a dense clump}
\label{sec.tests.raytracing}

The photo-evaporation of dense clumps of gas is prevalent in radiation
hydrodynamics simulations, and this test problem examines the
ionization front propagation into a dense clump, shadowing effects
behind the clump, and the hydrodynamic response on the clump from
photo-heating, all using \enzo's Moray ray-tracing module.  The
problem setup is the same as Test 7 in the Cosmological Radiative
Transfer Comparison Project \citep{IlievEtAl2009} and
\citet{Wise11_Moray}.  The simulation domain is 6.6~kpc on a side with
an ambient medium of pure neutral hydrogen of density $n_{\rm H} = 2
\times 10^{-4}\; \cubecm$ and temperature $T = 8000 \unit{K}$.  We
place a spherical overdensity in hydrostatic equilibrium with the
ambient medium.  It has a radius $r = 0.8 \unit{kpc}$, hydrogen
density $n_{\rm H} = 0.04\ \cubecm$ (i.e., overdensity of 200), and
temperature $T = 40$ K, and is centered at $(x,y,z) = (5, 3.3, 3.3)
\unit{kpc}$.  In \citet{IlievEtAl2009} all of the codes used a fixed
$128^3$ grid to ease the comparison, but in this test to demonstrate a
higher resolution AMR solution, we employ a $128^3$ grid with two
additional levels of refinement by factors of 2 for cells with a
baryon mass greater than 1.5 (method 2 in
Section~\ref{sec:refinement_criteria}). This test is run for 15 Myr.

\begin{figure}
  \centering
  \includegraphics[width=1.0\textwidth]{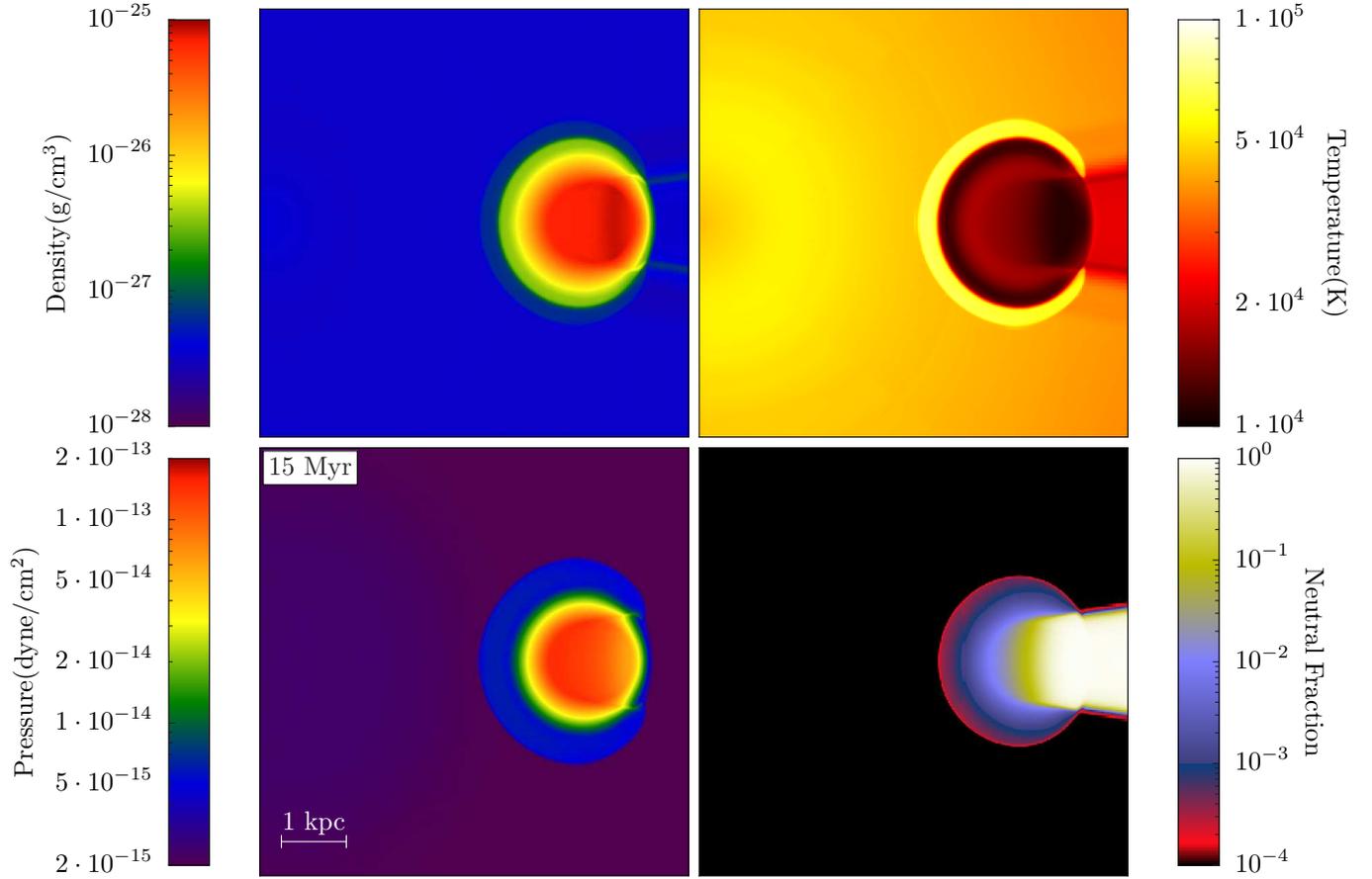}
  \caption{Photo-evaporation of a dense clump using the Moray
ray-tracing module.  Clockwise from the upper left: slices
through the clump center of density, temperature, neutral hydrogen
fraction, and pressure at $t=15$~Myr after the initialization of the
simulation.  A point source of radiation is at the center of the -x
boundary and illuminates the clump with a constant luminosity, casting
a clear shadow behind the clump, as seen in the neutral hydrogen
fraction.}
  \label{fig:shadowing}
\end{figure}

The cloud is subject to radiation from a point source at the center of
the $x=0$ boundary with an ionizing photon luminosity $\dot{N}_\gamma
= 3 \times 10^{51}$ photons s$^{-1}$, corresponding to a flux $F_0 =
10^6 \unit{photons s}^{-1} \unit{cm}^{-2}$ at the clump surface
closest to the radiation source.  The radiation source has a spectrum
of a $T = 10^5 \unit{K}$ blackbody, and we use four energy groups with
the following mean energies and relative luminosities: $E_i = (17.98,
31.15, 49.09, 76.98) \unit{eV}, L_i/L = (0.23, 0.36, 0.24, 0.06)$ that
are optimized to reduce errors in the solutions with a full spectrum
and energy discretization \citep{Mirocha12}.  (Note that this choice
of energy groups is different from those used in Wise \& Abel 2011.)
\nocite{Wise11_Moray} We use a minimum angular resolution of 10 rays
per cell and a constant radiative transfer timestep of 25 kyr.  Figure
\ref{fig:shadowing} depicts the clump at $t = 15$ Myr with the outer
layers expanding after being photo-heated.  It also shows the sharp
shadowing effects of the dense clump in the neutral fraction plot that
is representative of ray tracing techniques.

%% file: tests-fld-ifront.tex
\subsubsection{Cosmological I-front propagation}
\label{sec.tests.fld}

This test problem examines \enzo's flux-limited diffusion radiation
transport and associated ionization chemistry solvers on a
cosmological I-front test.  This corresponds to test problem 4.6 from
\cite{ReynoldsHayesPaschosNorman2009}, run using the cosmological
deceleration parameter $q_0 = 0.05$ and initial redshift $z_i=10$.  In
this test problem, the physics of interest is the expansion of an
\ion{H}{2} region in a uniform medium around a single monochromatic
ionizing source (with frequency $h\nu = 13.6$ eV).  The I-front
propagates rapidly at first, approaching 90\% of the Str{\" o}mgren
radius by a scaled redshift of $-\log\left[(1+z)/(1+z_i)\right]
\approx 0.15$, before cosmological expansion overtakes ionization,
pushing the Str{\" o}mgren radius outward faster than the I-front can
propagate.  The Str{\" o}mgren radius is analytically given by
\[
   r_S(t) = \left[\frac{3\dot{N}_{\gamma}}{4\pi \alpha_{\rm B}
   n_{\rm H}(t)^2}\right]^{1/3}, 
\]
where $\dot{N}_{\gamma} = 5 \times 10^{48}$ photon sec$^{-1}$ is the
strength of the ionizing source, $\alpha_{\rm B} = 2.52 \times 10^{-13}$
s$^{-1}$ is the \ion{H}{2} recombination rate, and $n_H(t)$ is the
proper number density of hydrogen.  We then define $\lambda =
\alpha_{\rm B}\, n_{H,i}\, /\, H_0\, / (1+z_i)$, where the subscript $i$
indicates quantities at the initial redshift.  The analytical
solution for the I-front position as a function of time is then

\begin{eqnarray*}
   r_I(t) &=& r_{S,i} \left[\lambda e^{-\tau(a)} \int_1^{a(t)}
     e^{\tau(\tilde a)} \left(1 - 2q_0
     + \frac{2q_0(1+z_i)}{\tilde{a}}\right)^{-1/2}\mathrm
     d\tilde{a}\right]^{1/3}, \quad\text{where} \\ 
   \tau(a) &=& \lambda\frac{6q_0^2(1+z_i)^2}{F(a)-F(1)}, \\
   F(a) &=& \left[2-4q_0 - \frac{2q_o(1+z_i)}{a}\right] 
      \left[1-2q_0 + \frac{2q_0(1+z_i)}{a}\right]^{1/2},
\end{eqnarray*}
and $a(t)$ is the cosmological expansion coefficient.

The simulation parameters are: box size $L_i\approx 27$ kpc, Hubble
constant $H_0 = 100$ km s$^{-1}$ Mpc$^{-1}$, and cosmological
parameters $\Omega_m = 0.1$, $\Omega_\Lambda=0$, and $\Omega_b = 0.1$.
The initial values for the simulation are a radiation energy density
$E_{r,i} = 10^{-35}$ erg cm$^{-3}$, temperature $T_i = 10^4$ K,
density $\rho_{b,i} = 2.35 \times 10^{-28}$ g cm$^{-3}$, and an
ionized hydrogen fraction $n_{\rm HII}/n_{\rm H} = 0$.

\begin{figure}
\begin{center}
\includegraphics[width=0.45\textwidth]{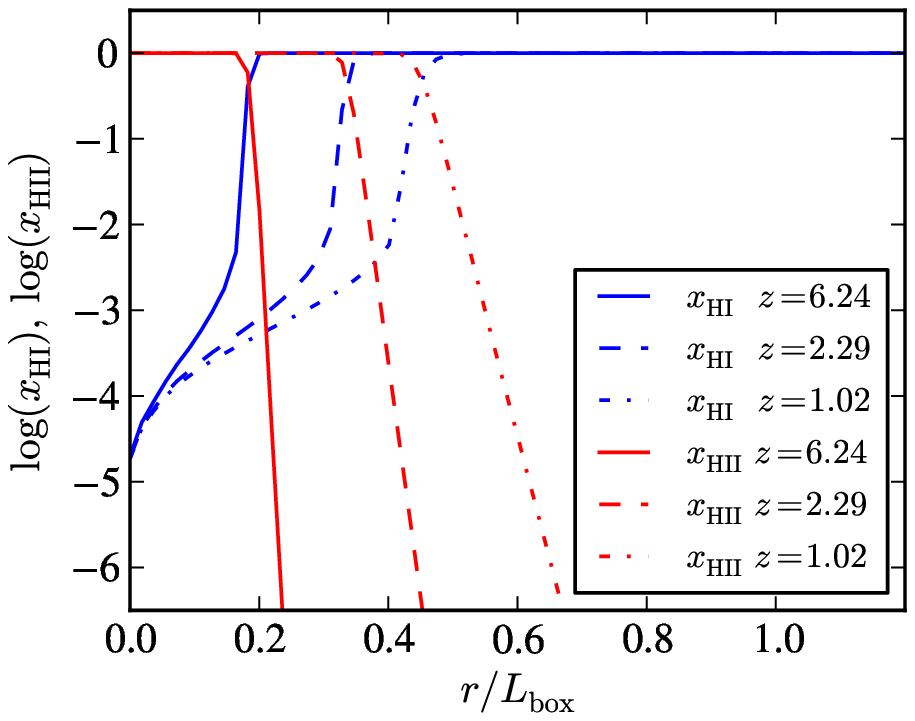}
\includegraphics[width=0.45\textwidth]{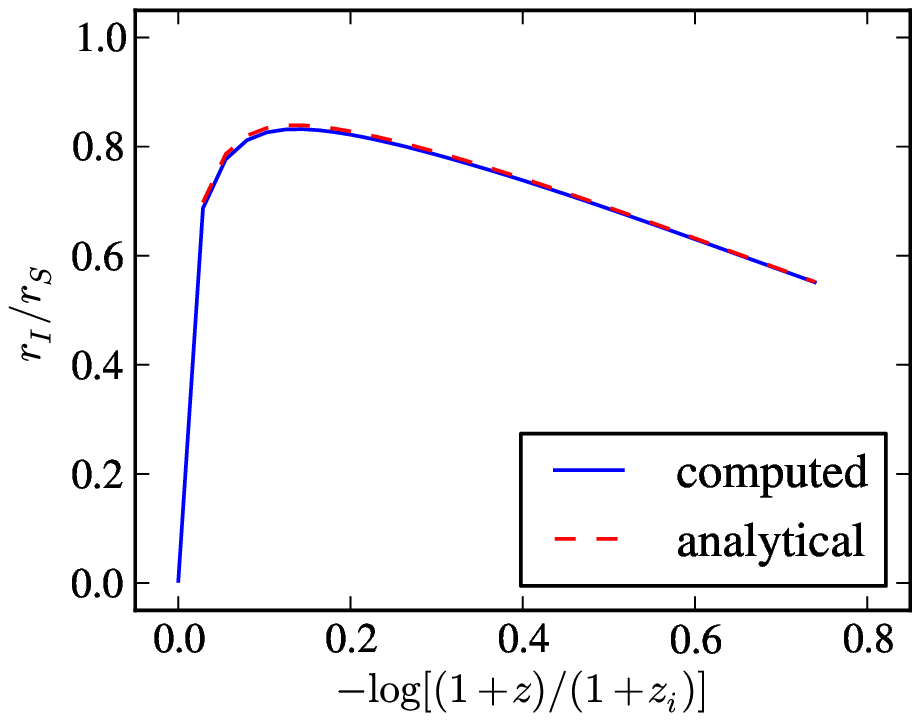}
\caption{Ionized fraction profiles and I-front location history for
the propagation of an ionization front in an expanding universe, using
\enzo's flux-limited diffusion radiation transport method.  Left
panel: profiles of the \ion{H}{1} and \ion{H}{2} fractions (blue and
red lines, respectively) at redshifts $z=\{6.24,\, 2.29,\, 1.02\}$
(solid, dashed, dot-dashed lines).  Right panel: ratio of the I-front
and Str{\" o}mgren radii as a function of scaled redshift in the test
problem (blue solid line) and analytical solution (red dashed line).}
\label{fig.fld}
\end{center}
\end{figure}

In Figure \ref{fig.fld}, we plot profiles of the \ion{H}{1} and
\ion{H}{2} fractions at redshifts $z=\{6.24,\, 2.29,\, 1.02\}$, as
well as the ratio of the I-front and Str{\" o}mgren radii throughout
the simulation.  As seen in the left panel (species fractions as a
function of radius), the I-front is initially quite narrow, but slowly
becomes wider as the simulation proceeds due to the diffusion
approximation used by the radiative transfer scheme.  However, as seen
in the I-front propagation history plot (right panel), the computed
I-front location very accurately matches the exact value.

%% file: tests-aniso-conduct.tex
\subsubsection{Anisotropic Thermal conduction}
\label{sec.tests.conduct}

\begin{figure}
\begin{center}
\includegraphics[width=0.42\textwidth]{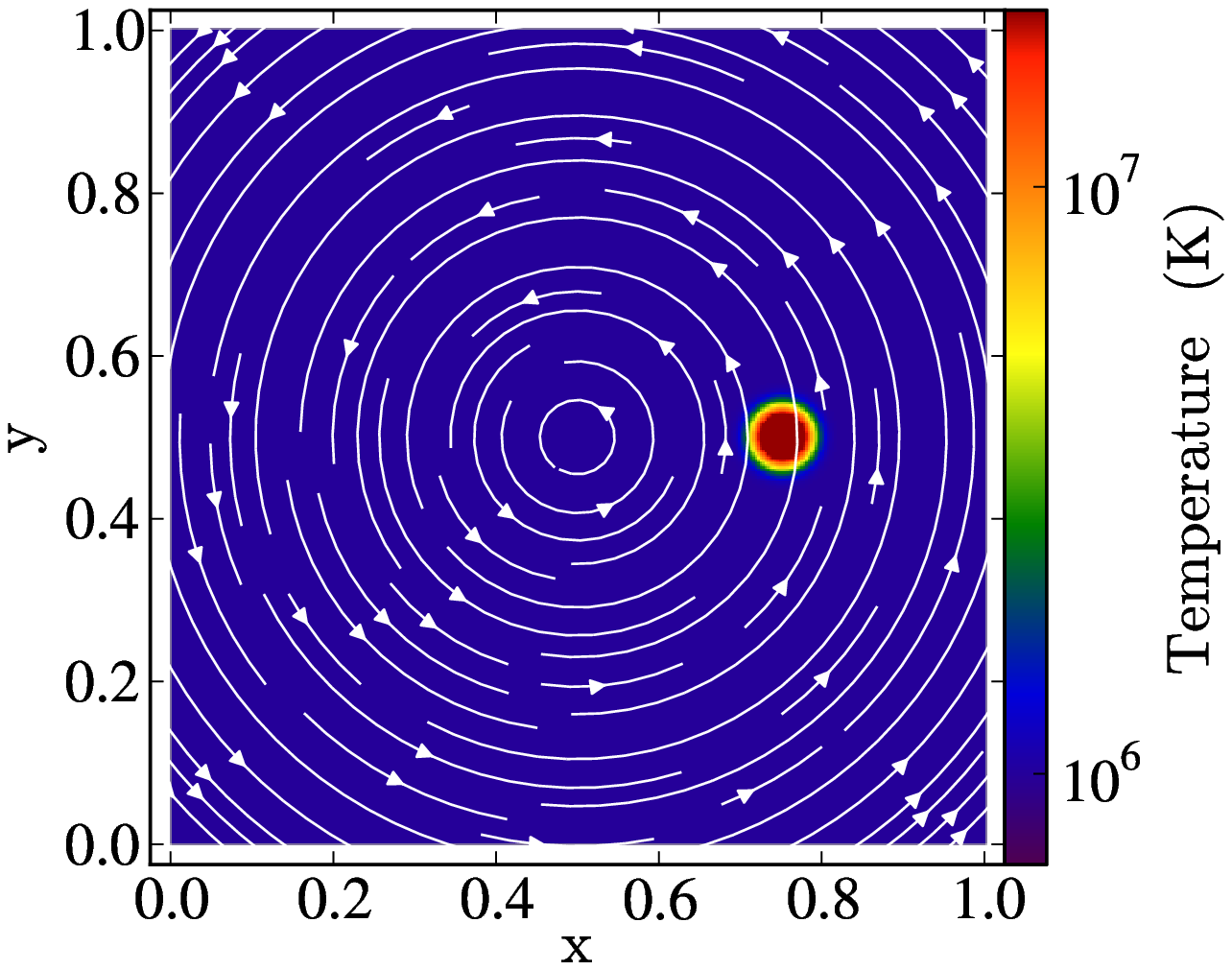}
\includegraphics[width=0.42\textwidth]{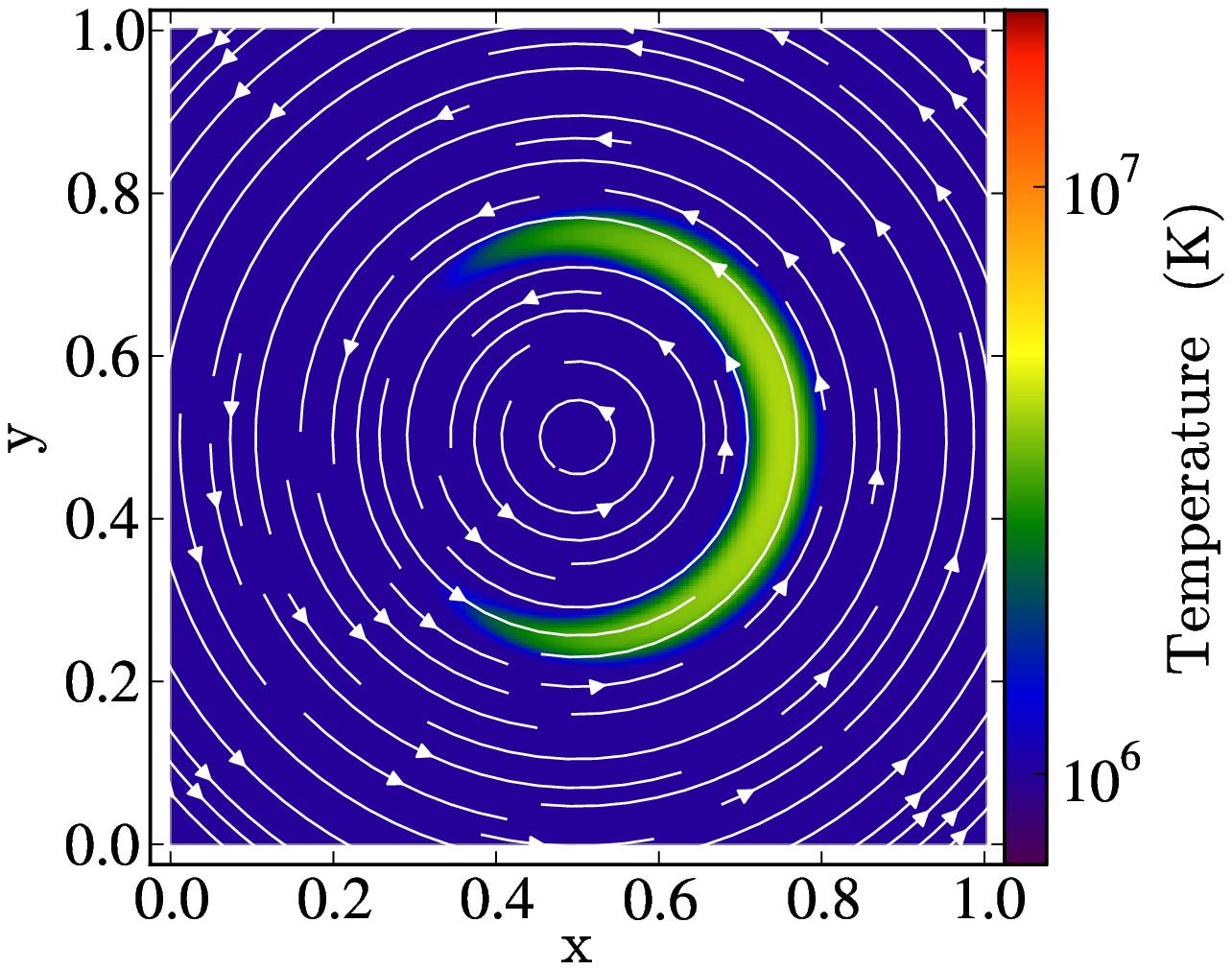}
\caption{Two-dimensional anisotropic conduction test in a uniform,
constant temperature background with circular magnetic fields
(indicated by white streamlines) centered on (0.5, 0.5) in a $256
\times 256$ grid. The background medium has a density of 1 proton/cc
and a temperature of $10^6$~K.  At t$ = 0$ (left panel), a Gaussian
heat pulse is injected at (0.75, 0.5) with a FWHM of $1/64$ (with all
numbers given in units of the box size) and a peak temperature of
$10^8$~K, and allowed to evolve without hydrodynamical motion (i.e.,
static gas) and no radiative cooling for 300 Myr.  At t$ = 300$~Myr
(right panel), heat has been transported along magnetic field lines
with no significant diffusion perpendicular to field
lines. Furthermore, there are no detectable oscillations in the
temperature in regions where the magnetic field is not parallel with
the grids.}
\label{fig.conduct}
\end{center}
\end{figure}

Figure~\ref{fig.conduct} shows a test that demonstrates the correct
behavior of anisotropic thermal conduction in \enzo. We initialize a
two-dimensional, $256 \times 256$ cell simulation having a physical
scale of 1 kpc on a side, a uniform density of 1 proton cm$^{-3}$ and
a background temperature of $10^6$ Kelvin.  Magnetic fields with a
strength of B$_0 = 1$~$\mu$G are initialized such that the field lines
form circles around the center of the simulation volume, such that
B$_x = -B_0\sin(\theta)$ and B$_y = B_0\cos(\theta)$, where $\theta$
is the angle measured from the $+x$ direction in a counterclockwise
manner.  A Gaussian temperature pulse is injected at $(0.75, 0.5)$ (in
units of the box size), with a peak temperature of $10^8$ K and a FWHM
of $1/64$ of the box size.  This initial setup is shown in the left
panel of Figure~\ref{fig.conduct}.  The simulation is then allowed to
evolve with \textit{only} anisotropic conduction turned on (e.g., no
hydrodynamics, radiative cooling, or cosmological expansion), and with
a Spitzer fraction of f$_{\rm sp} = 1$.

The right panel of Figure~\ref{fig.conduct} shows the state of the
simulation after 300 Myr.  Heat has clearly been transported only
along field lines -- there has been no diffusion perpendicular to the
magnetic field setup, which is critical for many studies involving
anisotropic thermal conduction.  No oscillations are seen in the
temperature field in regions where the fields are not aligned with the
grid, suggesting that the flux-limiter is operating as expected (see
discussion in Section~\ref{sec.num.conductions}).

%% file: parallel.tex
\section{Parallel Strategy and Performance}
\label{sec.parallel}

\subsection{Parallel Strategy}

The current version of \enzo\ has been parallelized for distributed
memory platforms using the Message Passing Interface (MPI).  This is
done using a single grid object as the basic unit of parallelization.
Each grid object -- including all cell and grid data -- is fully
contained on a single processor\footnote{In this context, we use the
word {\it processor} to mean a basic distribution unit; this could be
either a core or a node, depending on details of the system.  Note
that the current version of \enzo\ is not threaded, although a hybrid
MPI + OpenMP version is under development.}.  Parallelization is
accomplished by distributing grids amongst processors.  This is done
on the root grid using a simple tiling system, where the root grid is
split up into $N_{\rm root}$ tiles, with $N_{\rm root}$ typically
equal to the number of processors, $N_p$.

Load balancing on levels other than the root level (i.e., grids for
which the level $l > 0$) is different, as the refined patches are not
generally uniformly distributed.  Grids on refined levels are first
placed on the same processor as their parent to minimize
communication; however, this generally does not result in a
well-balanced computational load.  Therefore, the code has a number of
options for load balancing the grids on a given level. Each grid is
assigned an estimated computational load, generally equal to the
number of cells in the grid (which, empirically, is a good estimate of
computational cost).  The first load-balancing option is to move a
grid from the processor with the highest computational load to the
processor with the lowest load, with the proviso that only grids with
load factors less than half the difference between the highest and
lowest loaded processors will be moved.  This continues until the load
ratio between the most-to-least loaded processor is below 1.05 or
until no suitable grid can be found to transfer.  A second load
balancing option uses a space-filling Hilbert curve to order the grids
by their approximate spatial position.  Then, once the grids have a
specific one-dimensional ordering, we can divide up the grids into
$N_p$ groups (with the division taking place as equally as possible).
Load balancing is done separately for each level.  Clearly load
balancing is most successful if there are significantly more grids
than processors; however, small grids are less efficient because of
the large number of ghost zones compared to active zones, and so the
code uses a simple heuristic in order to split up grids until there
are of order 10 grids per processor.  This generally results in good
load-balancing while not producing grids that are wastefully small.

Communication between processors is done using a non-blocking
communication strategy that allows overlap of communication and
computation.  This can be done efficiently because each processor
retains a copy of the entire hierarchy of grids, except that grids
that do not `live' on a given processor only contain meta data
(essentially just the location and size of the grid; such grids are
denoted as `ghost' grids).  Replicating the hierarchy means that all
communication of data from one grid to another can be identified by
each processor independently.  The metadata for `ghost' grids are
quite small and so the extra memory required is generally not onerous
unless very large numbers of grids are used (more than a few hundred
thousand grids).  A schematic of this distribution is shown in
Figure~\ref{fig.amr_hierarchy}.

Data is transferred through a three-step procedure that takes
advantage of the capabilities of the MPI library: (i) as the code
progresses and data is needed from another grid on another processor,
the receiving processor posts an MPI non-blocking receive indicating
that it is expecting data; this outstanding receive is recorded in a
table, (ii) the sending processor calls the MPI non-blocking send
function, and then finally (iii) the receiving processor, after it has
carried out all the computation it can, waits for any MPI message to
arrive.  Each message is coded so that it can be matched with the
appropriate receive posted in the first step, and based on that, the
appropriate routine is called to processes the data.  Step (iii) is
repeated until there are no outstanding receives.


\subsection{Performance}
\label{sec.performance}

\subsubsection{Performance Measurement \& Instrumentation}

Because of the wide variety of simulations, methods, and uses of
\enzo, it is difficult to state in general terms which routines within
the code will be most costly during a given simulation.  As such, we
have designed a lightweight registering system that has been
implemented for the most commonly used routines (such as the
hydrodynamic and gravity solvers) as well as refinement level timers
that measure the time spent on each level.  Beyond this minimal set of
routines, we have designed a simple way for the user to modify the
source by adding \texttt{TIMER\_START("Your\_Routine\_Name")} and
\texttt{TIMER\_END("Your\_Routine\_Name")}.  These timers are created
and managed individually on each processor in an asynchronous fashion,
and contribute minimal computational, memory, and IO overhead.

At each complete root grid timestep (or less often if specified), each
timer is then communicated to the root processor where it calculates
the mean, standard deviation, minimum, and maximum for each of the
timers of that name across all processors.  For level timers, there
are additional attributes such as the number of cell updates, the
current number of grids, and the average cells updates per second per
MPI process.  This information is then output to a logfile.  This
provides a simplified interface to the user that can be used to
diagnose performance issues as well as estimate a given problem type's
scalability.  In addition to the logfile, we have developed a plotting
interface for quickly producing figures that process the data from the
logfile.  These capabilities are described in the online
documentation, along with further discussion of the performance
measurement implementation.

\subsubsection{Unigrid scaling}
\label{sec:weak_scaling}

\begin{figure}
\begin{center}
\includegraphics[width=0.6\textwidth]{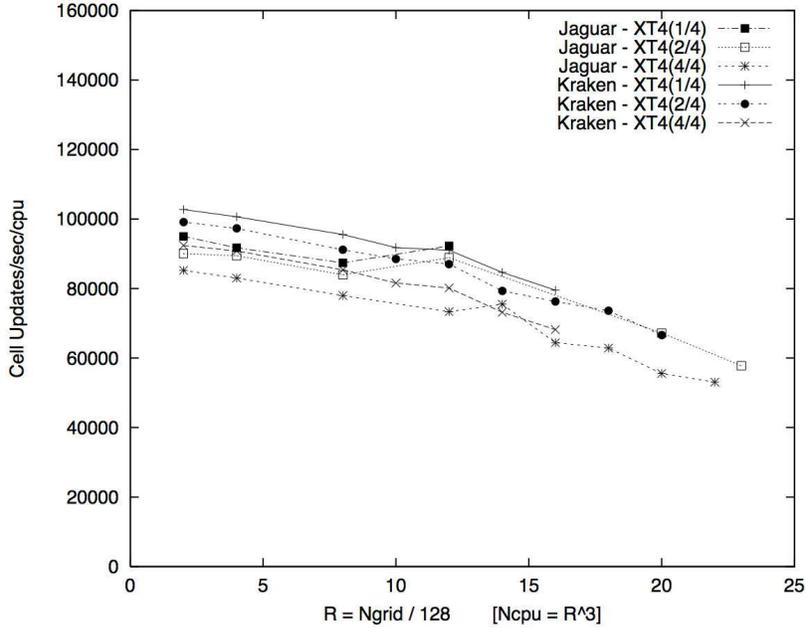}
\caption{\enzo\ weak scaling performance for a set of Lyman-$\alpha$
forest cosmology simulations with constant comoving spatial resolution
per grid cell, showing cell updates per second per processor plotted
as a function of the number of root grid tiles of dimension $128^3$
(R) in each dimension.  The number of MPI tasks is N$ = R^3$, so R$ =
16$ on this plot corresponds to a $2048^3$ computational mesh running
on 4,096 MPI tasks.  This plot goes from R$ = 2$ (8 MPI tasks) to R$ =
24$ (13,824 MPI tasks) on two supercomputers -- NICS Kraken and ORNL
Jaguar when they were Cray XT4 systems -- and using 1, 2 or 4 MPI
processes per node, where each compute node contained a single
quad-core AMD Opteron CPU with a speed of 2.1 GHz on Jaguar and 2.3
GHz on Kraken.}
\label{fig.uniscale}
\end{center}
\end{figure}

It is advantageous to use \enzo\ in its ``unigrid'' (i.e.,
non-adaptive mesh) mode for a variety of problems, including
non-gravitating turbulence
\citep[e.g.,][]{2002ApJ...569L.127K,Kritsuk04, 2007ApJ...665..416K,
2009ASPC..406...15K}, the Lyman-$\alpha$ forest
\citep{2005MNRAS.361...70J,2009MNRAS.399.1934P}, or feedback of
metal-enriched gas from galaxies into the intergalactic medium
\citep{2004ApJ...601L.115N,2011ApJ...731....6S}.  Achieving good
scaling of the code in unigrid mode is relatively straightforward --
upon initialization, unigrid simulations are decomposed such that each
MPI process has a roughly equal subvolume (and thus an equal number of
grid cells), meaning that work is evenly distributed among
computational elements.  Communication patterns for both the gravity
solve (which uses a fast Fourier transform) and the fluid solves
(which transfer boundary information between subvolumes) are
predictable and straightforward, and rebuilding of the grid hierarchy
does not take place, removing a substantial global operation and a
great deal of communication.

Figure~\ref{fig.uniscale} shows \enzo\ weak scaling results for a
sequence of scaled unigrid Lyman-$\alpha$ forest calculations. These
calculations include dark matter dynamics, hydrodynamics using the PPM
solver, six-species non-equilibrium chemistry and radiative cooling,
and a uniform metagalactic ultraviolet background.  In this sequence
of test calculations, we perform a weak scaling test on up to 13,824
MPI tasks on the NICS Kraken XT4 and ORNL Jaguar XT4
supercomputers\footnote{These simulations were performed prior to
conversion of both machines to the current-generation systems}.  In
this test, each MPI task was given a $128^3$ root grid tile (i.e.,
$128^3$ grid cells containing baryon quantities) and, initially,
approximately $128^3$ dark matter particles.  The number of grid cells
on each processor was constant throughout the calculation; the number
of dark matter particles on each processor varies as they are moved
from subvolume to subvolume as structure evolves.  The grid resolution
was kept at a constant comoving size of $\simeq 40$~kpc/h, and as the
core count was increased, so was the simulation volume.  On each
machine, a compute node contained a single AMD Opteron quad-core chip
(2.1 Ghz on Jaguar; 2.3 Ghz on Kraken) with 2 GB/memory per core (8
GB/total per node).  Both machines used the SeaStar2 interconnect.  In
the scaling study, calculations were run with 1, 2, or 4 MPI tasks per
node.  The figure shows cell updates per second per MPI process;
perfect scaling would be a horizontal line.

As can be seen in Figure~\ref{fig.uniscale}, the unigrid weak scaling
performance of the code is extremely good for this problem, with only
a 20\% decrease in cell updates per second per MPI task as the code is
scaled from 8 to 4,096 MPI tasks, and a 40\% decrease in performance
overall going from 8 to 13,824 (or $24^3$) MPI tasks.  We speculate
that this decrease is likely to be partially due to global MPI
communications used to, e.g., calculate the overall timestep of the
simulation, and also likely due to load imbalances due to increasing
cosmological power (and thus an increasingly uneven distribution of
dark matter particles between MPI tasks at late times) as the
simulation volume grows.  We also observe that a systematic difference
in speed can be seen between the two machines, which can be attributed
primarily to the slightly faster CPUs on Kraken at the time (2.3 Ghz,
vs. 2.1 Ghz on Jaguar).  The difference in speed when using different
numbers of MPI tasks per node can be attributed primarily due to
differences in competing usage of shared cache on the quad-core chips
used on this machine.

Broadly, excellent scaling in \enzo's unigrid mode is seen for a
variety of problems as long as each compute core is given an adequate
amount of work to do.  For cosmological simulations, this value has
been empirically determined to be roughly $128^3$ cells per core.  If
fewer cells per core are used, the CPU is essentially data-starved and
poor scaling is observed due to computing units being idle while
waiting for information to be communicated from other processes (for,
e.g., boundary information or gravity solves).  Substantially larger
cell counts per core would in principle help scaling by reducing the
amount of inter-process communication needed.

As a final point, we observe that scaling at larger core counts has
been measured, but only with an experimental hybrid-parallel (MPI +
OpenMP) version of \enzo.  Using this version, scaling comparable to
that shown in Figure~\ref{fig.uniscale} was seen on up to 98,304 cores
on the NICS Kraken XT5 (an upgraded version of the XT4 machine used
for the scaling study shown in the figure), using 2-8 OpenMP threads
per MPI process.  Hybrid parallelism has the potential to
substantially improve scaling by reducing the amount of communication
per grid tile, as described in the previous paragraph.

\subsubsection{AMR scaling}

\begin{figure}
\begin{center}
\includegraphics[width=0.48\textwidth]{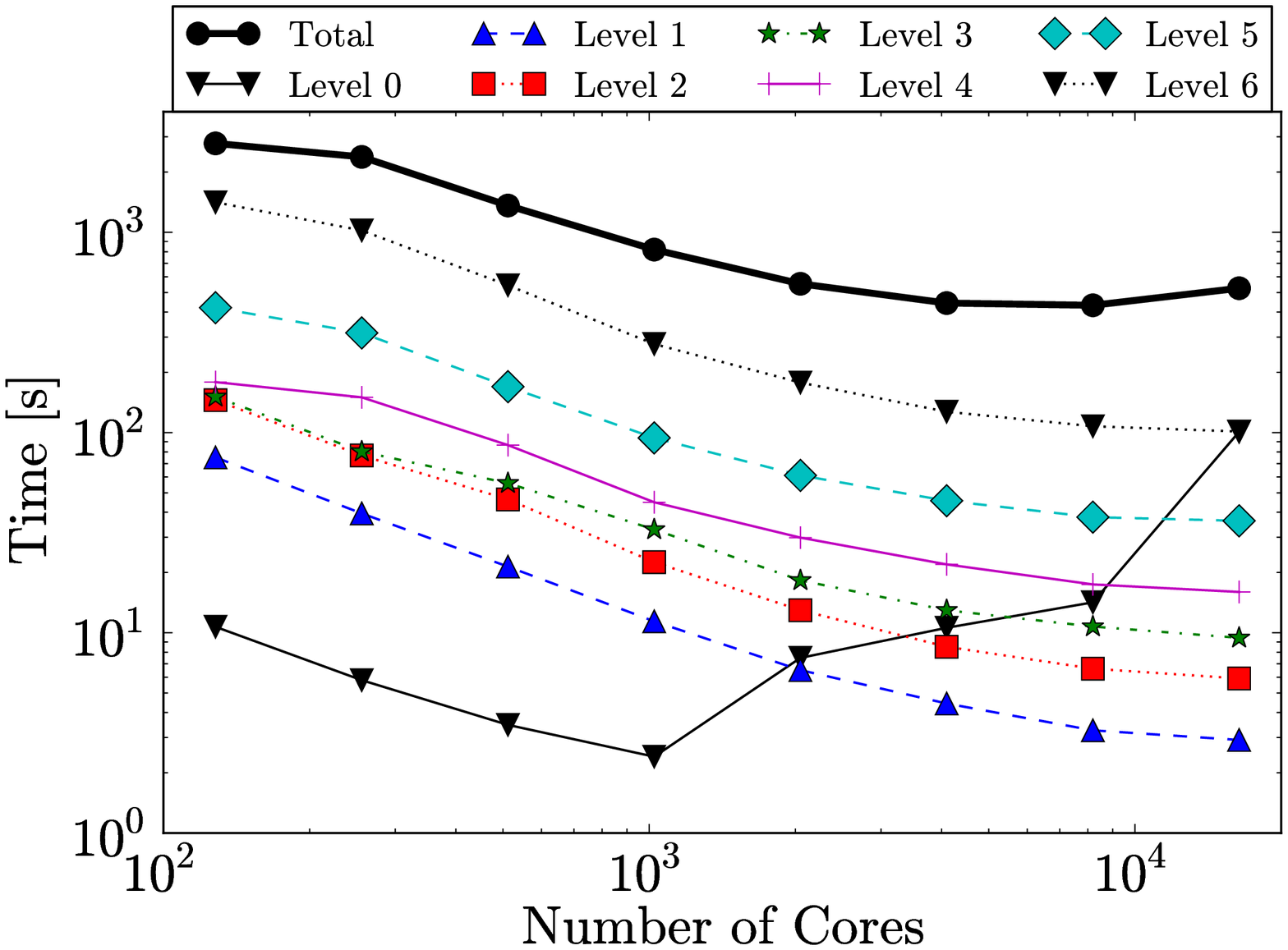}
\hfill
\includegraphics[width=0.48\textwidth]{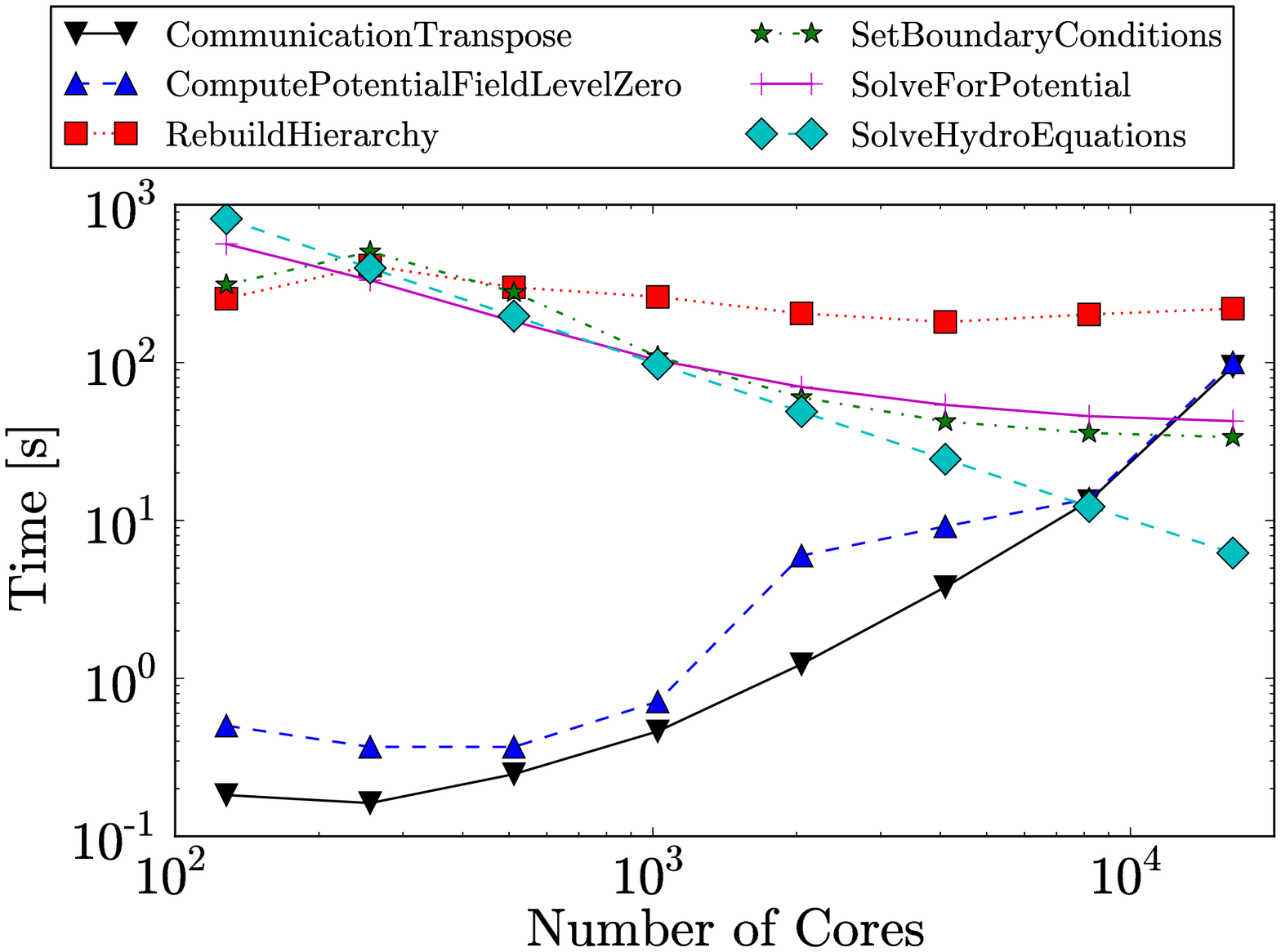}
\end{center}
\caption{\emph{Left:} Strong scaling test of a 512$^3$ AMR
cosmological calculation.  The root grid scaling is not representative
of the true strong scaling of \enzo\ because the root grid tiles are
not repartitioned when a simulation is restarted with additional
cores.  The weak scaling test in Figure \ref{fig.uniscale} is more
representative of the scaling on the root level.  The performance in
the refined levels show good strong scaling up to 2,048 cores.
\emph{Right:} Time spent in representative major routines in the same
AMR calculation.}
\label{fig:strong_scaling}
\end{figure}

Many astrophysical problems, such as cosmological galaxy formation
\citep{2012ApJ...749..140H}, high-resolution disk galaxy simulations
\citep{2011ApJ...738...54K}, high-redshift
\citep{2009Sci...325..601T}, and present-day star formation
\citep{Collins12a}, involve multi-scale physics that span several
orders of magnitude in both space and time.  In these situations,
using \enzo\ in its adaptive mesh refinement mode is beneficial.
Because the adaptive grid hierarchy is dynamic, grid boundaries and,
thus, communication patterns can be unpredictable, hindering strong
scaling to high core counts.

Figure \ref{fig:strong_scaling} shows strong scaling results from a
single 50 Mpc/h cosmology simulation run on $N_{\rm core}$ compute
cores, ranging from 128 to 16,384 cores at power-of-two intervals.  It
was run on the NICS Kraken XT5 supercomputer, which has two AMD
Opteron hexa-core processors with 16 GB of memory per compute node.
The simulations that utilized 128, 256, and 512 cores were executed on
128 nodes because of the memory requirements.  The higher core count
simulations were run with 8 cores per node.  This simulation would not
run with 12 cores per node because of the memory overhead associated
with the grid hierarchy being duplicated on each MPI process.
However, this overhead is greatly diminished if a hybrid-parallel (MPI
+ OpenMP) approach is used.

This simulation includes dark matter dynamics, hydrodynamics using the
piecewise parabolic method, six-species non-equilibrium chemistry and
radiative cooling, and a uniform metagalactic ultraviolet background.
The simulation uses the space-filling curve method for load balancing
the AMR grids.  It has a $512^3$ root grid that is divided into 512
tiles, and 6 additional AMR levels are used.  We perform these scaling
tests when the simulation reaches $z=4$, where the AMR grid hierarchy
is well-developed and is thus a reasonable representation of AMR
simulation behavior in general.  The results shown in Figure
\ref{fig:strong_scaling} come from a single root-level timestep of
$\Delta t = 2.1\, \textrm{Myr}$.  At this time, there are $3.04 \times
10^5$ AMR grids, $9.03 \times 10^8$ ($\sim 967^3$) computational
cells, and $1.34 \times 10^8$ dark matter particles in total.  The
breakdown of the number of AMR grids, cells, timesteps, and number of
cell updates on each level is shown in Table \ref{tab:amr_scale}.

\begin{table*}
  \begin{center}
  \caption{Strong scaling test computational details}
  \begin{tabular*}{0.9\textwidth}{@{\extracolsep{\fill}}c c c c c}
    \tableline\tableline
    {Level} & {$N_{\rm grid}$} & {$N_{\rm cells}$} & {$N_{\rm up}$} &
    {$N_{\rm timesteps}$}\\
    \tableline
    0 & 512 & $1.34 \times 10^8$ & $1.34 \times 10^8$ & 1\\
    1 & 61,868 & $4.01 \times 10^8$ & $4.01 \times 10^8$ & 1\\
    2 & 91,182 & $1.99 \times 10^8$ & $5.96 \times 10^8$ & 3\\
    3 & 59,932 & $7.62 \times 10^7$ & $5.34 \times 10^8$ & 7\\
    4 & 40,700 & $3.32 \times 10^7$ & $5.65 \times 10^8$ & 17\\
    5 & 28,346 & $2.76 \times 10^7$ & $1.36 \times 10^9$ & 49\\
    6 & 19,771 & $2.80 \times 10^7$ & $5.25 \times 10^9$ & 187\\
    \tableline
    Total & 302,311 & $9.03 \times 10^8$ & $8.83 \times 10^9$ & --\\
  \end{tabular*}
  \parbox[t]{0.9\textwidth}{\textbf{Note.} --- Data shown at $z=4$ for
    a root grid timestep of 2.1~Myr.  
    Col. (1): AMR Level. Col. (2):
    Number of grids. Col. (3): Number of computational
    cells. Col. (4): Number of cell updates. Col. (5): Number of
    timesteps.}
  \label{tab:amr_scale}
  \end{center}
\end{table*}


The left panel in Figure \ref{fig:strong_scaling} shows the
computational and communication time spent on each level.  In the AMR
levels, there exists good strong scaling up to 2,048 cores, and
marginal speed-ups are found at 4,096 cores.  On the root-grid level,
there exists good scaling up to 1,024 cores, but the performance
dramatically decreases at higher core counts.  This occurs because the
root grid is not re-partitioned into $N_{\rm core}$ tiles when the
simulation is restarted with a different core count.  This feature can
be easily implemented and is planned in the next major release of
\enzo, where scaling results would be similar to the weak scaling
shown in \S\ref{sec:weak_scaling}.  The right panel in Figure
\ref{fig:strong_scaling} shows the time spent in some representative
major routines in \enzo.  The local physics routines, for example
\texttt{SolveHydroEquations}, exhibit perfect strong scaling because
they involve no communication.  By investigating the scaling behavior
in each routine, it is clear that the communication in the
\texttt{SetBoundaryConditions}, \texttt{SolveForPotential} (the
multi-grid solver in AMR subgrids), and \texttt{RebuildHierarchy} are
responsible for the lack of strong scaling at $N_{\rm core} \ga$~4,096
in this particular simulation.  The transpose of the root grid tiles
are responsible for the performance decrease because it is not
optimized for situations where the number of tiles is greater than the
number of MPI processes.  These results are directly applicable to
simulations with similar computational demands.  In simulations with
fewer computational cells, strong scaling will cease at a smaller core
count because the CPUs will become data-starved more quickly, and the
opposite occurs with larger simulations.

\subsubsection{An Approximate Time and Memory Scaling Model}

In this section, we develop a simple, approximate model to estimate
the computational time and memory required to complete a given
calculation.  Note that we are {\it not} trying to model parallel
performance (which was discussed in the previous section), but have an
even simpler goal: to determine how to scale computational time
estimates as we increase the simulation resolution.  This is a
straightforward thing to do in codes that use static grids, as the
computational effort per timestep is constant.  However, for an AMR
calculation, the number of grids that will be generated during the run
is not known in advance, and so the CPU time per problem time can vary
drastically throughout the calculation.  For example, at the beginning
of a cosmological simulation, when the densities are nearly uniform,
only the static grid is required and the calculation progresses
rapidly.  However, as structure forms and dense clumps are generated,
the number of grid points swells by orders of magnitude (an increase
of $10^3$ is not uncommon) and most of the CPU time is consumed at
late physical time in the simulation.  Therefore, simply performing a
few steps at the beginning of the calculation does not produce a good
estimate of the required CPU time.

Nevertheless, we can try to determine how the compute time scales for
a given run as we increase the resolution.  First, we neglect
particles and concentrate on the time taken by the grids, which can be
justified both theoretically and empirically\footnote{Since the
potential is calculated on the grid, the only particle costs are:
depositing mass to the grid, interpolating accelerations, and updating
particle positions and velocities.  These are relatively
computationally inexpensive operations.}.  Second, we break the
problem down slightly, and examine the scaling over a short enough
period of time that the grid structure does not change significantly
(i.e. the number of grids at each level remains approximately
constant).  We then assume that the whole run scales in the same way,
or in other words, that changes in the resolution affect the
simulation in the same way at each timestep.  This is usually a good
approximation, as the most costly parts of the calculation typically
don't change their characteristics substantially between timesteps.

To make progress, we assume that the computational cost to advance a
single cell by one timestep is a constant.  This can be incorrect if
the chemistry solver requires many iterations, but is usually fairly
accurate.  For a unigrid calculation, the time would by proportional
to $N_{\rm root}^4$ since the number of cells scales as $N_{\rm
root}^3$ and, assuming the Courant condition is the factor that
controls the timestep, the number of steps to advance the calculation
over a given time interval is proportional to $N_{\rm root}.$
Therefore, to advance a hierarchy a given time interval, we find,
accounting for all levels and using a refinement factor of 2,

\begin{equation} 
t_{\rm SU} = C_1 \sum_l f_l N_{\rm root}^4 2^{4l}
\end{equation} 

where $C_1$ is a constant, which can be thought of as the time taken
to advance a single cell.  The factor $f_l$ is the fraction of the
volume on a given level that is actually refined.  By definition $f_0
= 1$, and $f_l \le 1$.  In writing down this equation, we neglect a
number of costs that are not directly proportional to cell count,
including communication between processors, the $\log{N}$ factor for
the root grid FFT, cache misses, optimizations, and other costs
associated with processing the hierarchy.  The first item on this
list, in particular, is clearly important for large processor counts;
however, we neglect parallel considerations in this section.

\begin{figure}
\centerline{\includegraphics[width=0.7\textwidth]{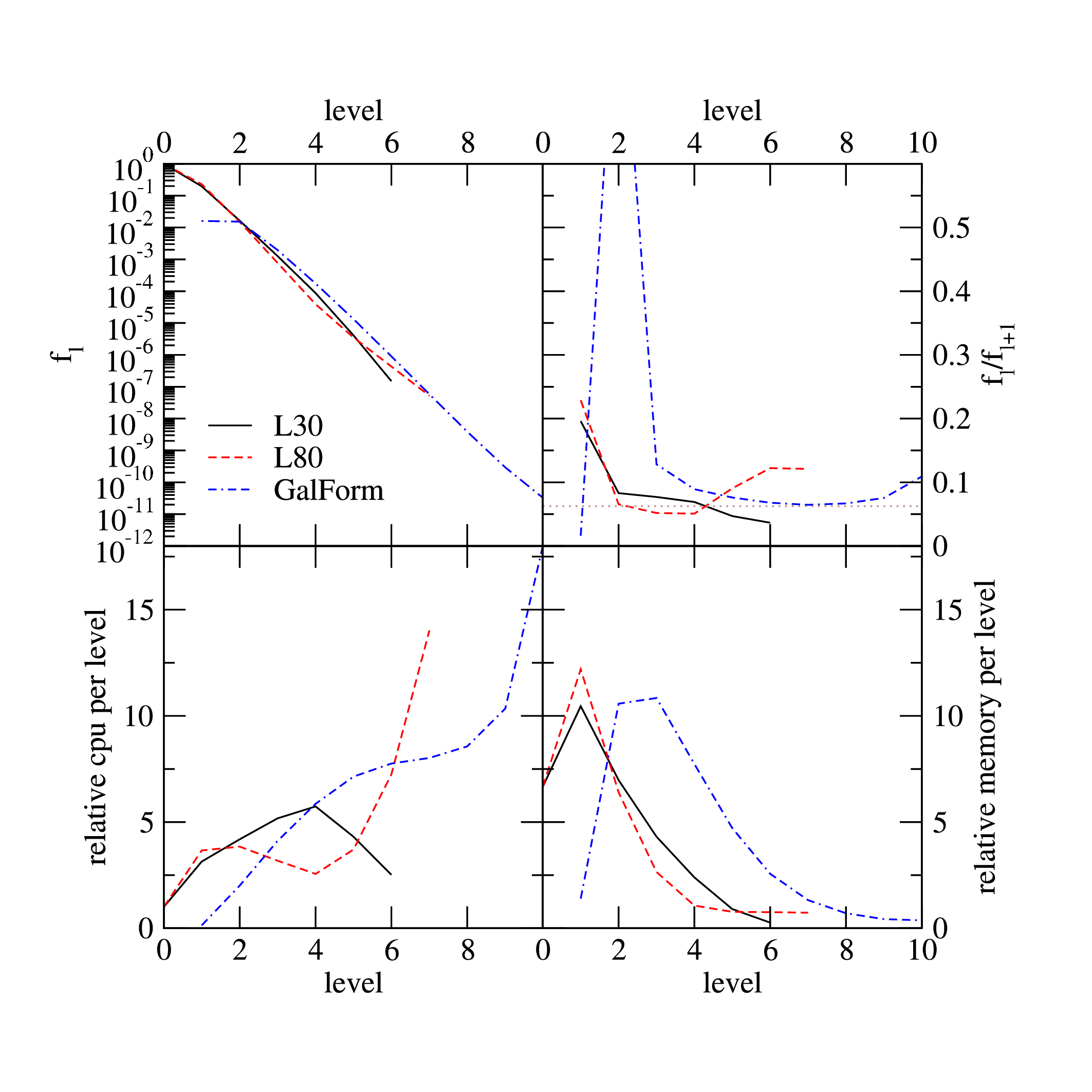}}
\caption{This figure shows (clockwise from upper left), the covering
fraction $f_l$ of the grids on a given level $l$, the ratio of
$f_{l}/f_{l+1}$, the relative memory usage of each level (normalized
by the memory usage of level 0), and, in the lower-left panel, the
relative CPU usage of each level, again normalized by $l=0$.  In each
panel, three curves are shown: the solid black and long dashed red are
two cosmological simulations with box sizes of 30 h$^{-1}$ Mpc and
80h$^{-1}$ Mpc respectively.  The dot-dashed blue line is for a
non-cosmological simulation of a ram-pressure stripped galaxy.  In the
top-right panel, there is a line at the critical ratio of 0.06, which
determines if the next finer level ($l+1)$ takes more (above) or less
(below) CPU time than level $l$.}
\label{fig:scaling}
\end{figure}

Note that unless $f_l/f_{l+1} < 1/16$ (i.e. if less then about 6\% of
a given level is further refined), the cost per level will increase
with level.  A key question, therefore, is the value of $f_l$ for each
level.  Unfortunately, this depends strongly on the simulation being
run.  In Figure~\ref{fig:scaling}, we show the values of $f_l$ for
three simulations: two cosmological runs with varying box size, and a
third simulation focusing on a single disk galaxy.

As, the figure demonstrates, the $f_l$ values all show a sharp decline
with level, dropping as power-laws with the level $l$.  Focusing first
on the ``refine-everywhere" cosmological runs, which both show similar
behavior despite the different box sizes and redshifts at which we
collect the data.  In fact, we find these results are very typical for
cosmological runs, with only slight variations depending on the exact
problem that is being simulated.  The top-right panel shows the ratio
of $f_{l}/f_{l+1}$ which, remarkably, hovers around the critical value
of 0.0625 determined earlier.  This implies that the relative CPU
usage of each level, shown in the lower-left panel, is nearly flat,
and that going to additional levels of refinement only increases the
amount of calculation by a factor of roughly $1/l_{\rm max}$.  Even in
the most extreme case, the L80 run, the last level adds only 25\% to
the computation time.

The memory used by each level, on the other hand, scales as

\begin{equation}
{\rm memory} = C_2 \sum_l f_l N_{\rm root}^3 2^{3l}
\end{equation}

and the terms in this sum are shown in the bottom-right panel of
Figure~\ref{fig:scaling}.  This demonstrates that the memory usage is
dominated by the top three levels, and adding additional levels 
adds minimally to the memory usage of the run.  We have empirically
tested this scaling for cosmological simulations and found it to be
reasonably accurate (although the increase is typically slightly
larger than found here, usually 30-50\%, probably because of
less-than-ideal parallel scaling).

The physical reason for the $f_{l+1}/f_l$ ratios found in the
cosmological simulations appears to be due to a combination of the
density structure of individual clouds and to the distribution of the
clouds themselves.  In particular, we note that for a $\rho \propto
r^{-2}$ density profile, the Lagrangian refinement criteria typically
used in such simulations produces an $f_{l+1}/f_l$ ratio of 1/8 for a
single, resolved cloud.  Of course, at some point we would resolve the
flat density of individual clouds and the ratio would climb, making
further refinement more costly; however, we are not yet in this regime
in this example.

These conclusions are, however, highly dependent on the type of run
being performed.  We contrast this cosmological case to the simulation
of a galaxy formation run, as shown by the dot-dashed curves in
Figure~\ref{fig:scaling}.  Note that grid levels 1 and 2 in this
simulation are statically refined to ensure that the Lagrangian volume
of the galaxy halo is resolved to at least level 2 at all times. More
importantly, the $f_{l+1}/f_{l}$ ratio of the highest levels are
systematically slightly larger than the critical 0.0625 value,
indicating that the CPU time is dominated by the most refined levels,
as shown in the bottom-left panel.  The memory usage is dominated by
levels 2 and 3, the static levels, as the bottom-right panel
demonstrates.  In this case, adding more levels of refinement would
substantially increase the cost of the simulation.

We have focused thus far on scaling when increasing $l_{\rm max}$, but
without changing the refinement criteria.  However, we can also keep
the levels fixed and modify the refinement criteria.  For cosmological
runs, this is typically done by increasing the mass resolution (and
simultaneously decreasing the dark matter particle mass), and adding
additional linear small-scale modes to the initial conditions.  The
effect of this is to boost the $f_l$ values at all levels (except for
the root grid where $f_0$ is already 1) by a factor linearly
proportional to the increase in the mass resolution.  This is because
the refinement criteria typically boosts the number of refined cells
on the first and subsequent levels by this factor.  Again, we have
empirically tested that this scaling is approximately valid, provided
that we keep the maximum level constant.

Putting this together, we find an approximate scaling for the
computational time required for cosmological simulations:

\begin{equation}
t_{\rm SU} \propto M_{\rm res}^{-1} l_{\rm max}.
\end{equation}

This is approximately accurate, and gives users some way to estimate
compute times for \enzo\ calculations.  

\subsection{GPU parallelization and scaling}

Many of today's leading supercomputers use a heterogeneous
computing platform:
on a single node of a distributed-memory platform, a multi-core CPU is
often paired with one or more many-core accelerators.  One programming
model that has been shown to successfully take advantage of these
hardware accelerators is to run the serial component of the algorithm
on the CPU, and the vector-parallel part of the algorithm on the
many-core hardware accelerator. Due to the much higher computational
performance of the many-core accelerators, if a code can be ported to
effectively utilize this heterogeneous architecture, massive speedup
in simulation performance may be achieved.  This heterogeneous trend
is likely to continue in future supercomputers, due to the relatively
low energy needs of hardware accelerators as well as other
factors. Thus, to ensure \enzo's efficiency on future high-end
computational platforms, some of the most time-consuming parts of
\enzo\ have been ported to many-core architecture.  Since GPUs are
currently the most popular many-core accelerator, NVIDIA's CUDA C/C++
programming technology was chosen to port some \enzo\ modules to
NVIDIA GPUs.

Thus far, the PPM and Dedner MHD solvers (described in
Sections~\ref{sec.hydro.ppm} and~\ref{sec.num.hydro-muscl},
respectively), have been ported to take advantage of GPUs using CUDA.
Since the CPU and GPU on a node have access to separate pools of RAM,
fields updated by other modules will be transferred to the GPU before
calling the GPU fluid solver.  Correspondingly, after calling the GPU
solver, the updated fields will be transferred back to CPU. This
ensures that the GPU-parallelized solvers can work correctly with
other parts of the code that have not yet been ported to the GPU, as
well as with the communication infrastructure within \enzo.  In
addition, the GPU solver supports SAMR, which is not a trivial task.
As discussed in Section~\ref{sec.amr}, one of the key steps in \enzo's
AMR implementation is flux correction, which is required when each
level of resolution is allowed to take its own time step.  In the GPU
version of the solvers, the fluxes are calculated on the GPU and only
the fluxes required for flux correction are transferred back to the
CPU.  This reduces data transfer overhead, which can be substantial in
a heterogeneous architecture of this sort.

The key step in porting to many-core architectures such as the GPU is
exposing the massive parallelism inherent in the algorithm. Due to the
explicit, directionally-split stencil pattern of both the PPM and
Dedner MHD solvers, they are inherently massively parallel and thus
should be a good fit for hardware acceleration. Both solvers essentially
contain two parts -- computation of fluxes, and a cell update.
When computing fluxes, the basic procedure is to compute the flux at
each cell-interface given the inputs from neighboring cells. In the
cell update part, the fundamental computation is updating the
cell-centered values using the previously-computed fluxes. In the CPU
serial code, a loop over the grid is used, where the loop body
contains these basic computations. In both parts,
algorithmically-different loop iterations are completely independent
of each other. Thus, the natural parallelization scheme is to map
one GPU thread to one iteration of the loop. However, the original CPU
code, which operates on each grid as a serial process, does not
completely expose this parallelism as some small temporary arrays are
reused among loop iterations.  This re-use of arrays introduces data
dependency among loop iterations, which is undesirable for GPU
parallelization.  Because of this re-use of arrays, the main change in
porting the serial CPU solvers to allow massively parallel GPU
computation was replacing those shared temporary arrays by larger
temporary arrays that are not shared among loop iterations. This
change exposed the massive parallelism in the algorithm, which could
then be accelerated in a straightforward manner using CUDA.

\begin{figure}
\centerline{\includegraphics[width=0.5\textwidth]{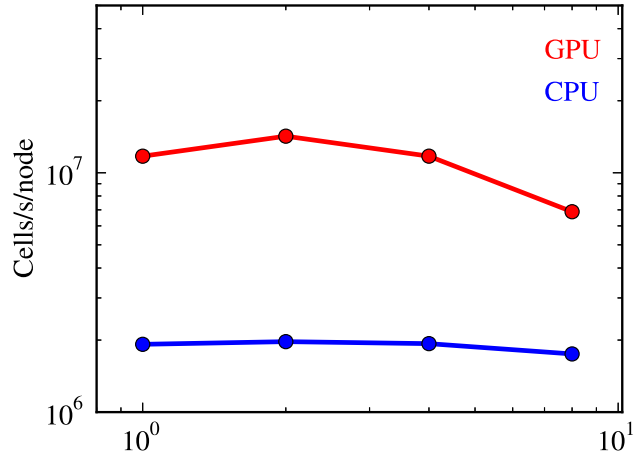}}
\caption{Weak scaling performance of MHD turbulence using the Dedner
MHD algorithm, which has both CPU and GPU solvers.  Each node of the
machine (a Cray XK6) contains one 16-core AMD Opteron 6272 CPU and 1
NVIDIA Tesla K20 GPU.  In this test, there are $256^3$ cells/node.
The y-axis shows cell updates per second per node, which for ideal
weak scaling is a horizontal line.  Blue line: CPU solver.  Red line:
GPU solver.}
\label{fig:gpu_scaling}
\end{figure}

To illustrate the speedup provided by porting two solvers to GPUs, we
show the results of a weak scaling test of driven MHD turbulence on a
uniform mesh in Figure~\ref{fig:gpu_scaling}.  This particular problem
type contains no physics other than the equations of ideal
MHD, and thus is representative of the type of calculation than can currently
benefit from the GPU-optimized solvers in \enzo.  In this scaling
test, we use the Dedner MHD solver, which runs on both CPUs and GPUs
and has been tuned to maximize performance on both platforms.  The
benchmarking platform is a Cray XK6 system with a Gemini
interconnect. Each node has a 16-core AMD Opteron 6272 CPU and a
single NVIDIA Tesla K20 GPU, which has 2,496 CUDA cores and a maximum
theoretical speed of 1.17 Tflops for double-precision computations.  For
the CPU run, all the 16 cores in each node are used by launching 16
MPI processes per node. For the GPU run, 16 MPI processes are also
launched per node, and all processes on this node share the single GPU
on that node. This can work because NVIDIA's MPS (Multi-Process
Service) technology allows multiple processes to concurrently use the
same GPU.  A series of weak scaling tests were run, varying from 1 to
8 nodes (16 to 128 MPI processes), with each node containing a cube of
$256^3$ cells (so, a 4-node computation would have a domain that is
$512 \times 512 \times 256$ cells).  Figure~\ref{fig:gpu_scaling}
displays the scaling results in terms of cell updates per second per
node -- as a result, ideal weak scaling is a horizontal line.  In
general, the simulations using the GPU-accelerated MHD solver perform
roughly 5 times better overall on this system than the equivalent
CPU-only calculation run on the same number of nodes.  We note,
however, that this somewhat mis-represents the speedup obtained by
porting the MHD code to GPUs -- in the GPU simulation case, the 16 CPU
cores on each node are mostly idle, as they are used only for
timestep calculation and boundary condition transmission (which are
both computationally inexpensive compared to the fluid solve).  This
means that, effectively, the GPU simulation is 80 times faster than a
single CPU core, so a user would likely see much greater effective
speedup on a system where the ratio of CPU cores to GPUs is lower.

%% file: development.tex
\section{Code development methodology}
\label{sec.development}

Over time, \enzo's development has followed a trajectory toward
increasing openness.  Started as the graduate research project of Greg Bryan at
the University of Illinois, it was subsequently stewarded by the
Laboratory for Computational Astrophysics (LCA) at the University of
California at San Diego, and has transitioned to a distributed,
completely open, and community-driven project.  Initially, \enzo\ was
versioned using a series of ``snapshots'' of the code base, usually
hand-created by the individuals doing development work.  These were
distributed to collaborators and colleagues, but the central ``trunk''
of development was updated primarily by a single person: while patches
and technology were accepted from external developers, the relatively
small number of individuals using the code resulted in a strong
centralization of development.

As the stewardship of the code passed to the LCA, the code was
released first to ``friendly users'' and then as a public open source
release.  However, while the code was made available with
documentation, technology developed in the broader community of users
was typically not re-integrated.  This led to a wide dispersal of
development, largely independent, by individuals who downloaded and
used the version of the code developed by the LCA.

Following the first public, open source release of \enzo, the code was
migrated to the Subversion version control system.  This is a
centralized version control system, and the ``stable'' \enzo\ source
code was made globally readable following the \enzo\ 1.5 release.
Access to the primary development tree required a password and login
for each user, and providing upstream changes either required this
password and the granting of write access or a sequence of patches and
manually-created diffs (much like the original development system).
The technical friction of manually contributing patches and
modifications, combined with Subversion's difficulty with tracking
merges, resulted in further fragmentation of the code base.

A version developed at Penn State and Stanford forked from a version
prior to the LCA version and was the one in which MHD with Dedener
divergence cleaning, the MUSCL hydro solvers, the ray tracing
radiative transfer module, relativistic fluid dynamics, the shearing
box boundaries and updates to the multi-species chemistry were
included. This version was the one that was eventually merged using
the distributed version control system (DVCS) Mercurial
(\url{http://mercurial.selenic.com/}) into a branch of the code known
as \texttt{week-of-code}, so named after the in-person development
sprint at KIPAC in June 2009 at which it was created.  The
fundamental, and transformative, distinction between the previous
\textit{centralized} version control system and mercurial's
\textit{distributed} version control system is the elimination of
gatekeepers.  While there still exists a canonical, central location
where stable and development versions of \enzo\ can be obtained,
changesets and versions can be exchanged between peers without the
intervention of designated gatekeepers.  This has the direct effect of
enabling local development to be versioned and its provenance ensured,
while still retaining the ability to benefit from ``upstream''
development.  An important, even crucial, side effect is that the
technology used for local versioning provides mechanisms for easily
submitting locally-developed modifications to the community source
location.  Mercurial internally represents all changes as nodes in a
directed, acyclic graph (DAG), which results in the natural ability to
more consistently and easily manage merging development streams.

Currently, \enzo\ is developed using the hosted source control
platform Bitbucket (\url{http://bitbucket.org/}) at
\url{http://bitbucket.org/enzo/}.  There are two mailing lists, one
for usage-focused questions and discussion, and another for
development discussion.  Both of these lists are open and publicly
archived.  Bitbucket provides mechanisms for inspecting source code,
hosting branches and forks of the primary source, and for code review.
All proposed source code changes for \enzo\ are subjected to a peer
review process, where experienced developers read, inspect, test, and
provide feedback on source code changes.  All developers, including
long-time \enzo\ contributors and developers, are subject to this
process before their code is included in the primary \enzo\
repository.  By using a remote, hosted system, \enzo\ is now
\textit{completely} open to contributions from the community.
Individuals, who may or may not consider themselves \enzo\ developers,
are free to ``fork'' the \enzo\ code base, develop changes (signed
with their own name), and submit them for review and inclusion into
the primary code repository.  In
contrast to the centralized, gatekeeper-focused technology used
previously, this enables anyone to contribute changes to be evaluated
for inclusion in the \enzo\ codebase.  One challenge that this
presents is that the \enzo\ code is a moving target -- to that end, we
recommend that users include the changeset hash of the \enzo\
repository that generated their simulation data (and also the version
of \texttt{yt} used to analyze the data) in their publications.

While peer review is able to catch many bugs and problems with source
code changes, \enzo\ is also subject to ``answer'' testing (described
in detail in Section~\ref{sec.tests.suite}).  We have created a set of
parameter files and problem types that exercise the underlying
machinery of \enzo.  These ``test problems'' have affiliated
``tests,'' which consist of scripts that use \texttt{yt}
\citep{2011ApJS..192....9T} to produce results such as mass
distribution, projections, profiles and so on.  The testing
infrastructure then evaluates whether the variation in the new results
compared to a ``gold standard'' of results has exceeded an acceptable
threshold, typically set to roughly the level of roundoff error in
single-precision floating point arithmetic.  Optionally, for those
test problems that are deemed unsafe to change to any precision, the
tests also produce hashes of the outputs; these hashes will only
remain unchanged in the event of bitwise identicality between results.
The results of the gold standard are versioned and stored in Amazon
S3, enabling remote testing to proceed.  While the testing process --
building, running, analyzing and comparing -- is not yet automated
against incoming pull requests, we hope to deploy that functionality
in the future.  The primary challenge is that of compute time; the
tests are organized into multiple categories, including by the
expected run time, but the full suite of tests can take several days
to run.

%% file: conclusions.tex
\section{Conclusions}
\label{sec.conclusions}

In this paper, we have presented the algorithms underlying \enzo, an
open-source adaptive mesh refinement code designed for
self-gravitating compressible fluid dynamics, including the effects of
magnetic fields, radiation transport, and a variety of microphysical
and subgrid processes.  In addition, we have described the \enzo\ code
development process, have shown the outputs of a set of representative
test problems, and have provided information about \enzo's performance
and parallel scaling on recent supercomputing platforms.  The \enzo\
code, its test suite, and all of the scripts used to generate plots
and figures for this paper are open source and are available at the
\enzo\ website, \url{http://enzo-project.org}.  Furthermore, the
\texttt{yt} toolkit, which is designed to analyze \enzo\ data (as well
as data from a wide variety of other simulation tools), can be found
at its website, \url{http://yt-project.org}.  Both of these codes have
active user and developer communities, extensive documentation and
user support, and strong mechanisms for users to contribute their
changes and fixes to the codebase.

The developers of the \enzo\ code are currently working on several
projects that will extend the functionality, scalability, or overall
performance of the code in the near future.  Projects that will appear
in forthcoming releases of the \enzo\ code include:

\begin{itemize}
\item The creation of a hybrid-parallel version of \enzo, combining
MPI for communication between nodes of a supercomputer and OpenMP for
thread-based parallelism within a node.  This will reduce on-node
memory usage and improve overall scaling behavior.
\item The restructuring of \enzo's treatment of particles to
accommodate a wider range of ``active'' particles that can easily
interact with each other and with multiple grids, and include sink,
source, and particle creation, destruction, splitting, and merging
functionality.
\item A new HYPRE-based AMR gravity solver that is faster, more
accurate, and more scalable than the current multigrid solver.
\item New infrastructure for problem initialization, enabling users to
more quickly and easily create new types of simulations.
\end{itemize}

With the continual rapid development of computer hardware, it makes
sense to not only review \enzo's current capabilities, but to look
toward its future development in view of predicted technological
trends. These trends in supercomputing hardware suggest that
substantial modifications to \enzo's core infrastructure, and very
possibly some of the core algorithms, will be required. More
specifically, the progression involves the usage of specialized
large-core-count, vectorized computing units such as graphics
processing units or chips like the Intel Xeon Phi, as well as
precipitously decreasing amounts of RAM per computing core.  The
former trend means that the amount of processing power per compute
node will continue to increase, likely much faster than the bandwidth
between nodes, and will require tremendous reduction in inter-node
(and possibly inter-CPU) communication in order to maintain code
scalability.  Also, much of the current code will need to be rewritten
to take advantage of the vector nature of these CPUs, making
assumptions that are quite unlike those made in much of the current
codebase.  The latter trend means that duplication of data -- for
example, the grid hierarchy -- must be effectively eliminated to save
memory, and all inter-core and inter-node communication must be
carefully thought through to minimize the amount of data moved.  An
additional challenge as one goes to core counts in the tens to
hundreds of millions (or more) is that the reliability of individual
computing elements will become much more of an issue, requiring
robustness to hardware failure to be built into the code at some
level.  Furthermore, we are nearing the physical limits of transistor
speed and interconnect latency~\citep{feynman1999feynman}, meaning
that simple hardware improvements will not make these challenges
disappear, and careful thought (and the rewriting of a great deal of
code) must take place!  These challenges are not unique to the \enzo\
code, and in fact are faced by effectively all applications that wish
to take advantage of new computational architectures. We therefore
anticipate that \enzo\ (or a code that has the capabilities of \enzo,
from a user's point of view) will continue to be usable at the largest
scales on such machines.

%% file: acknowledgments.tex

\acknowledgments

Development of \enzo\ has been ongoing since 1994 by a wide range of
agencies and institutions.  In all grants listed, we put the
initials of the PI (if an \enzo\ developer) or the \enzo\ developer
funded by the grant (if the PI is not a developer of \enzo).

This work has been supported by the National Science Foundation by
grants
AAG-0808184 (DRR),
AAG-1109008 (DRR),
ACI-9619019 (MLN),
ASC-9313135 (MLN),
AST-9803137 (MLN), 
AST-0307690 (MLN), 
AST-0407176 (RC),
AST-0407368 (SS, EJH),
AST-0507521 (RC), 
AST-0507717 (MLN), 
AST-0507768 (AGK),
AST-0529734 (TA),
AST-0607675 (AGK),
AST-0702923 (EJH),
AST-0707474 (BDS), 
AST-0708960 (MLN), 
AST-0807075 (TA),
AST-0807215 (JB),
AST-0808184 (MLN, AGK),
AST-0808398 (TA),
AST-0908740 (AGK, DCC),
AST-0908819 (BWO), 
AST-0955300 (NJG),
AST-1008134 (GB),
AST-1109570 (AGK), 
AST-1009802 (JSO), 
AST-1102943 (MLN),
AST-1106437 (JB),
AST-1210890 (GB),
AST-1211626 (JHW),
OCI-0832662 (BWO, MLN),
OCI-0941373 (BWO),
PHY-1104819 (MLN, JB),
the CI TraCS fellowship (OCI-1048505; MJT),
and the Graduate Research Fellowship program (NJG; SWS).

This work has been supported by the National Aeronautics and Space
Administration through grants
NAGW-3152 (MLN),
NAG5-3923 (MLN),
NNX08AH26G (MLN, TA),
NNX09AD80G (BWO),
NNX12AH41G (GB),
NNX12AC98G (BWO),
NNZ07-AG77G (BDS),
NNG05GK10G (RC),
ATP09-0094 (SVL),
Chandra Theory grant \#TM9-0008X (BWO),
Hubble Space Telescope Theory Grant HST-AR-10978.01 (BDS),
the Fermi Guest Investigator Program (\#21077; BWO),
and the Hubble Postdoctoral Fellowship through the Space Telescope Science
Insititue, \#120-6370 (JHW).

This work has been supported by the Department of Energy via the
Los Alamos National Laboratory (LANL) Laboratory Directed Research and
Development Program (BWO, DCC, HX, SWS), 
the LANL Institute for Geophysics and Planetary Physics (BWO, DCC, CP,
BC),
the Los Alamos National Laboratory Director's Postdoctoral Fellowship
program (No. DE-AC52-06NA25396;
BWO and DCC), and the
DOE Computational Science Graduate Fellowship (DE-FG02-97ER25308; SWS)

Additional financial support for the \enzo\ code has come from
Canada's NSERC through the USRA and CGS programs (EL) and through a 
Japan MEXT grant for the Tenure Track System (EJT).

We acknowledge the  many academic institutions that have supported \enzo\
development, including (in alphabetical order)
Columbia University,
Georgia Institute of Technology,
Michigan State University and the MSU Institute for Cyber-Enabled
Research, 
the National Center for Supercomputing Applications, 
the Pennsylvania State University,
 the San Diego Supercomputer Center (through the Strategic Applications
Partner program and the Director’s office),
Princeton University,
SLAC National Accelerator Laboratory,
the SLAC/Stanford Kavli Institute for Particle
Astrophysics and Cosmology,  
Southern Methodist University,
Stanford University,
the University of Arizona,
the University of Califoria at San Diego, 
the University of Colorado at Boulder,
the University of Florida,
and the University of Illinois.
We acknowledge support from the Kavli Institute for Theoretical
Physics at Santa Barbara, the Aspen Center for Physics, and the UCLA
Institute for Pure and Applied Mathematics, which have
generously hosted \enzo\ developers through their conference and
workshop programs.

Computational resources for \enzo\ development have come from the NSF
XSEDE program, the NASA High Performance Computing program, the DOE INCITE
program, and the DOE Advanced Simulation and Computing (ASC) program.

The \enzo\ collaboration would like to acknowledge the following
scientists, who have made contributions to the \enzo\ codebase at some
point during their research career: Gabriel Altay, Brian Crosby,
Elizabeth Harper-Clark, Daegene Koh, Eve Lee, Pascal Paschos, Carolyn
Peruta, Alex Razoumov, Munier Salem, and Rick Wagner.

The \enzo\ collaboration would also like to acknowledge the significant contributions to
\enzo\ development made by the late Dr. Robert P. Harkness.

%% file: appendix.tex
\appendix
\section{Interpolation methods}
\label{app:interpolation}

In this appendix, we provide details for the various interpolation
methods available in the code.  We assume throughout that we are
dealing with Cartesian coordinates and cells of equal sizes, although
we allow for an arbitrary refinement factor $r$ between cells at
different levels.

\vspace{0.5cm}\noindent
{\bf SecondOrderA} 

This interpolation algorithm is generally second-order, but has
monotonicity constraints as described below.
In one dimension, we define the parent values as $Q_{-1}$, $Q_0$, and
$Q_{+1}$, where the central parent cell has a left edge at $x_0$ and
width $\Delta x$. We first linearly interpolate the parent values to
the cell edges: $Q_{-1/2} = (Q_0 + Q_{-1})/2$, and similarly for
$Q_{1/2} = (Q_0 + Q_1)/2$, and then compute a monotonic slope: $\Delta
Q_0 = \minmod{Q_{1/2} - Q_0}{Q_0 - Q_{-1/2}}$ where

\begin{equation}
\minmod{a}{b} = \left\{ \begin{array}{ll}
0 & {\rm if} \quad ab < 0 \\
\min{ \left( | a |, | b | \right) } {\rm sign} (a) & {\rm otherwise}
\end{array}\right.
\end{equation}

This slope is then used to compute the interpolated subgrid values.
Defining $q_i$ to be the subgrid values at cell centers for the $r$
cells corresponding to parent cell $Q_0$ (for refinement factor $r$),
we can write:

\begin{equation}
q_i = Q_0 + \frac{i+(1-r)/2}{r} f_x
\end{equation}

where $f_x = 2 \Delta Q_0$ (this notation is used for consistency with
the 2 and 3 dimensional cases).

In two dimensions, the procedure is very similar in that we linearly
interpolate parent values to the four cell corners: $Q_{-1/2, -1/2}$,
$Q_{-1/2, 1/2}$, $Q_{1/2, -1/2}$, $Q_{1/2, 1/2}$ (by averaging parent
cell-centered parent values).  We then compute monotonic slopes across
the two diagonals, since for a linear function extrema occur at
corners:

\begin{eqnarray}
\Delta Q_0 & = & \minmod{Q_{1/2, 1/2} - Q_0}{Q_0 - Q_{-1/2, -1/2}} \\
\Delta Q_1 & = & \minmod{Q_{-1/2, 1/2} - Q_0}{Q_0 - Q_{1/2, -1/2}}
\end{eqnarray}

We then translate these slopes into the grid axes with $f_x = \Delta
Q_0 -  \Delta Q_1$ and $f_y = \Delta Q_0 + \Delta Q_1$ so that the
interpolation itself can be written simply as

\begin{equation}
q_{i,j} = Q_0 + \frac{i+(1-r)/2}{r} f_x + \frac{j+(1-r)/2}{r} f_y
\end{equation}

Finally, we write down the three-dimensional version -- unfortunately,
here the number of monotonicity constraints is four (the 4 diagonals
across the 8 opposing corners of the cube), while the number of slopes
is three, so the problem is over-constrained.  Somewhat arbitrarily,
we adopt the following procedure.  As before, we compute $\Delta Q_0$,
$\Delta Q_1$, $\Delta Q_2$, and $\Delta Q_3$ with the minmod limiter
across the 4 diagonals.  We then define

\begin{equation}
s =\frac{\Delta Q_1 + \Delta Q_2 + \Delta Q_3}{\Delta Q_0}
\end{equation}

which is the value of $\Delta Q_0$ that the other slopes ($\Delta
Q_1$, $\Delta Q_2$, and $\Delta Q_3$) imply, normalized by the desired
value of $\Delta Q_0$ itself.  If $0 < s < 1$, then no adjustment
needs to be made, as the monotonicity constraints are met.  If these
equalities are not met, then, if $s<0$ we define $\chi_n = 1$ if
$\Delta Q_n/\Delta Q_0 < 0$, and $\chi_n = 0$ otherwise (if $s > 1$, this is
reversed, so $\chi_n = 1$ if $\Delta Q_n/\Delta Q_0 > 0$, and 0
otherwise).  These weights are used to determine which slopes to
modify to match the $\Delta Q_0$ constraint.  We then compute the
amount of adjustment required:

\begin{equation}
f = - \frac{\Delta Q_0 s^\prime + \sum_n (1-\chi_n) \Delta Q_n}{\sum \chi_n \Delta Q_n + \epsilon}
\end{equation}

where $s^\prime = \min(\max(s,0), 1)$ and $\epsilon$ is a small number
to prevent numerical errors. We then use this adjustment fraction to
compute the new, adjusted $\Delta Q_n$ (for $n = 1, 2, 3$) that match
the $\Delta Q_0$ constraint as closely as possible with

\begin{equation}
\Delta Q_n = f \chi_n \Delta Q_n + (1-\chi_n) \Delta Q_n
\end{equation}

Finally, these are converted to the grid axes with $f_x = \Delta Q_2 +
\Delta Q_3$, $f_y = \Delta Q_1 + \Delta Q_3$, $f_z = \Delta Q_1 +
\Delta Q_2$.  We then do the interpolation itself with

\begin{equation}
q_{i,j,k} = Q_0 + \frac{i+(1-r)/2}{r} f_x + \frac{j+(1-r)/2}{r} f_y + \frac{k+(1-r)/2}{r} f_z
\end{equation}


\vspace{0.3cm}\noindent
{\bf SecondOrderB} 

We also provide a variant on the above procedure, with two changes.
The first is that the slopes across the diagonals ($\Delta Q_0$, etc.)
are computed directly rather than with the minmod limiter
(e.g. $\Delta Q_0 = (Q_{1/2,1/2,1/2} - Q_{-1/2,-1/2,-1/2})/2$).  To
ensure positivity in the resulting interpolation when applied to
positive conserved quantities, the slopes are limited so that the
smallest corner value is 0.2 of the cell center value.  The procedure
described above is then applied to turn these four slopes into a
linear interpolation.


\vspace{0.3cm}\noindent
{\bf SecondOrderC} 

This is a completely different second-order interpolation scheme,
which is based on Cloud-In-Cell (CIC) interpolation \citep{Hockney88}.
In one dimension, we define the parent values as $Q_0$, and $Q_{+1}$,
where the left parent cell has a cell center at $x_0$ and width
$\Delta x$.  Then, the interpolated value for a subgrid cell $q_i$
with a cell left edges at $x_i = x_0 + i \Delta x^p/r$, where $i$ runs
from 0 to $r-1$ for refinement factor $r$, is given by:

\begin{equation}
q_i =  \frac{2r - 1 - 2i}{2r} Q_0 + \frac{1+2i}{2r} Q_{+1}
\end{equation}

The extension to two and three dimensions is straightforward, with
weights in the other dimensions computed in the same way and then
multiplied to get a total of 4 and 8 weights for the 2 and 3
dimensional cases, respectively.  This scheme preserves monotonicity,
but is not conservative.


\vspace{0.3cm}\noindent
{\bf ThirdOrderA} 

This interpolation method provides third-order accuracy based on the
Triangular Shaped Cloud (TSC) methodology \citep{Hockney88}.  As
usual, in one dimension, we define the parent values as $Q_{-1}$,
$Q_0$, and $Q_{+1}$, where the central parent cell has a left edge at
$x_0$ and width $\Delta x$.  Then, the interpolated value for a
subgrid cell $q_i$ with a cell left edges at $x_i = x_0 + i \Delta
x^p/r$ where $i$ runs from 0 to $r-1$ for refinement factor $r$, is
given by:

\begin{equation}
q_i = a_i  Q_{-1} + b_i Q_0 + c_i Q_{+1}
\end{equation}

and the weights are given by:

\begin{equation}
a(i) =  \frac{(r-i)^3 - (r-i-1)^3}{6r^3}; \qquad c(i) = \frac{3i^2 + 3i + 1}{6r^3}
\end{equation}

and $b(i) = 1/r - a(i) - c(i)$.  The extension to two and three
dimensions is straightforward with the weights computed in the same
way as for the one-dimensional case but then multiplied to determine
the 9 and 27 factors necessary for the 2 and 3 dimensional cases,
respectively.

One problem with this interpolation technique is that it is not
conservative.  In particular, the sum of the interpolated subgrid
values:

\begin{equation}
\tilde{Q_0} = \sum_{i=0}^{r-1} q_i
\end{equation}
is not, in general, equal to $Q_0$.  We can retain conservation by
adding the factor $(Q_0 - \tilde{Q_0})/r$ to the interpolated values
(or by multiplying the interpolated values by the ratio
$Q_0/\tilde{Q_0}$).  Unfortunately, the result of this procedure does
not preserve monotonicity -- it can introduce local minima and maxima
at parent cell boundaries.  We can then attempt to correct that by
taking a weighted average between the interpolated values on the
subgrid and the parent value, with weights computed such that the
interpolated values at the edge of the parent cell are not local
maxima compared to the interpolated values in the neighboring parent
cell.


\vspace{0.3cm}\noindent
{\bf FirstOrderA}

Finally, for completeness, we include a first-order accurate piecewise
constant interpolator for which, using the same definitions as in the
previous case, we take $q_i = Q_0$, for $i = 0$ to $r-1$. 

\vspace{1cm}